\documentclass[a4paper,12pt,headsepline=true,dvipsnames,BCOR=9mm]{scrbook}
\usepackage[utf8]{inputenc}
\usepackage[T1]{fontenc}
\usepackage{graphicx}
\usepackage{amsmath}
\usepackage{amsfonts}
\usepackage{amssymb}
\usepackage{siunitx}

\usepackage{charter}
\usepackage[expert]{mathdesign}
\usepackage{changepage}
\usepackage{lscape}
\usepackage{mathtools}
\DeclarePairedDelimiter\ceil{\lceil}{\rceil}
\DeclarePairedDelimiter\floor{\lfloor}{\rfloor}

\usepackage[super]{nth}

\usepackage{color}
\usepackage[plainpages=false,pdfpagelabels=true,bookmarksnumbered=true,hidelinks=true,
			pdfauthor={Benedikt Werner Hosp},
			pdftitle={latent expertise gaze information in highly dynamic decision-tasks.},
			pdfkeywords={Eye tracking, expertise, cognitive models, visual search, expert eye movements, classification, deep learning, machine learning, medicine, sports, virtual reality}]{hyperref} 
\usepackage[english]{babel}
\usepackage{multirow}
\usepackage{bookmark}
\usepackage{booktabs}
\usepackage{threeparttable}
\usepackage{caption}
\usepackage{sidecap} 
\usepackage{wrapfig} 
\usepackage{subcaption}
\usepackage{siunitx}
\usepackage{tabularx}
\usepackage{xspace}
\usepackage{pdfpages}
\usepackage{tcolorbox}
\usepackage{makecell}
\usepackage{tabularray}

\usepackage{booktabs}
\usepackage{enumitem}

\usepackage{afterpage}

\usepackage{colortbl}
\usepackage{dcolumn}
\newcolumntype{K}{D{,}{,}{-1.0}}

\usepackage{lipsum}

\DeclareSIUnit{\px}{px}
\DeclareSIUnit{\byte}{B}
\DeclareSIUnit{\bit}{b}

\makeatletter
\DeclareRobustCommand\onedot{\futurelet\@let@token\@onedot}
\def\@onedot{\ifx\@let@token.\else.\null\fi\xspace}

\usepackage[colorinlistoftodos]{todonotes} 
\usepackage{xargs} 
\newcommandx{\change}[2][1=]{\todo[color=blue!40,#1]{#2}}
\newcommandx{\TODO}[2][1=]{\todo[color=red!40,#1]{#2}}
\newcommandx{\rewrite}[2][1=]{\todo[color=green!40,#1]{#2}}
\newcommandx{\elaborate}[2][1=]{\todo[color=purple!40,#1]{#2}}

\newcommand{\phdtitle}{Latent gaze information in highly dynamic 
decision-tasks.}

\newcommand{\phdyear}{2021}
\newcommand{\phdauthor}{Benedikt W. Hosp}
\newcommand{\phdplaceofbirth}{Klettgau-Grießen}

\newcommand{\phddean}{Prof.\ Dr.\ Thilo \ Stehle}
\newcommand{\phdreviewerA}{Prof.\ Dr.\ Enkelejda\ Kasneci}
\newcommand{\phdreviewerB}{Prof.\ Dr.\ Ansgar\ Thiel}

\graphicspath{ {images/}
 		     }

\usepackage[backend=bibtex,style=ieee]{biblatex}

\newcommand\blfootnote[1]{%
	\begingroup
	\renewcommand\thefootnote{}\footnote{#1}%
	\addtocounter{footnote}{-1}%
	\endgroup
}

\definecolor{PP-blue}{RGB}{50,65,75}
\definecolor{PP-lightBlue}{RGB}{50,65,75}

\begin{document}

\frontmatter
\begin{titlepage}
\pdfbookmark{Title page}{Title}
	\null\vfill
	
	\begin{center}
		
		\vskip 1cm
		{
			{\huge \phdtitle}\par
			\null\vfill
		}%
		{
			\large{\textbf{Dissertation}\par}
			\vspace{5mm}
			\normalsize{ der Mathematisch-Naturwissenschaftlichen Fakultät \par
			der Eberhard-Karls-Universität Tübingen \par
			zur Erlangung des Grades eines \par
			Doktors der Naturwissenschaften \par
			(Dr. rer. nat.)\par
			\vskip 3cm
			vorgelegt von\par
			\textbf{\phdauthor}\par
			aus \phdplaceofbirth			
			\vskip 3cm
			\textbf{Tübingen\\ \phdyear}
			}
		}%
	\vfill
	\end{center}
	\clearpage
	\thispagestyle{empty}
	\vspace*{\fill}
	\begin{tabular}{rllr}
	   \multicolumn{3}{l}{\parbox{16cm}{Gedruckt mit Genehmigung der Mathematisch-Naturwissenschaftlichen Fakultät \\ der Eberhard Karls Universität Tübingen.}} \\ 
	           &                   &               & \\
	 	 \multicolumn{1}{l}{Tag der mündlichen Qualifikation: }& 04.02.2022\\ 
            \multicolumn{1}{l}{Dekan: }& \phddean  &\\
            \multicolumn{1}{l}{1.  Berichterstatter: }& \phdreviewerA  &\\
            \multicolumn{1}{l}{2.  Berichterstatter: } & \phdreviewerB  &\\
	\end{tabular}
\end{titlepage}

\chapter[Acknowledgements]{Acknowledgements}
\label{chap:ack}

I would like to thank various people who supported my constantly new, crude ideas about gaze behavior and its exploration. First and foremost, my parents, Franziska and Werner, to whom I owe not only the fact that I was allowed to follow this path, but also a loving upbringing and a lot of motivation (which have made me the person I am today). Also, my siblings, Josy and Philipp, who have always been there for me and to whom I have always looked up to. \\ 

I would like to thank Enkelejda Kasneci, Oliver Höner, Florian Schultz and Shahram Eivazi from University of Tübingen and Peter Haddawy and Myat Su Yin from Mahidol University. In addition to heated technical discussions we were having from time to time, they were always open to my ideas and always on hand, even when things got tight.  \\ 

A big thank you also goes to my colleagues from the working group. Nora Castner, Wolfgang Fuhl, David Geißler, Thomas Kübler, Björn Severin, Efe Bozkir, and Yao Rong. Thank you very much for your support, your instructions, your help, and your humorous cooperation. \\

Last but certainly not least, I would like to thank Lioba, who has been with me through all the ups and downs.

\chapter[Abstract]{Abstract}
\label{chap:summary}

Digitization is penetrating more and more areas of life. Tasks are increasingly being completed digitally, and are therefore not only fulfilled faster, more efficiently, but also more purposefully and successfully. The rapid developments in the field of artificial intelligence in recent years have played a major role in this, as they brought up many helpful approaches to build on. At the same time, the eyes, their movements, and the meaning of these movements are being progressively researched. The combination of these developments has led to exciting approaches. In this dissertation I present some of these approaches which I worked on during my PhD.

First, I provide insight into the development of models that use artificial intelligence to connect eye movements with visual-expertise. This is demonstrated for two domains or rather groups of people: athletes in decision-making actions and surgeons in arthroscopic procedures. The resulting models can be considered as digital diagnostic models for automatic expertise recognition. Furthermore, I show approaches that investigate the transferability of eye movement patterns to different expertise domains and subsequently, important aspects of techniques for generalization. Finally, I address the temporal detection of confusion based on eye movement data. The results suggest the use of the resulting model as a clock signal for possible digital assistance options in the training of young professionals. An interesting aspect of my research is that I was able to draw on very valuable data from DFB youth elite athletes as well as on long-standing experts in arthroscopy. In particular, the work with the DFB data attracted the interest of radio and print media, namely DeutschlandFunk Nova and SWR DasDing. All resulting articles presented here have been published in internationally renowned journals or at conferences.

\chapter[Zusammenfassung]{Zusammenfassung}
\label{chap:zusam}

Die Digitalisierung durchdringt immer mehr Lebensbereiche. Aufgaben werden zunehmend digital erledigt und damit schneller, effizienter, aber auch zielorientierter und erfolgreicher erfüllt. Die rasante Entwicklung im Bereich der künstlichen Intelligenz in den letzten Jahren hat dabei eine große Rolle gespielt, denn sie hat viele hilfreiche Ansätze hervorgebracht, auf die immer weiter aufgebaut werden kann. Gleichzeitig werden die Augen, ihre Bewegungen und die Bedeutung dieser Bewegungen immer weiter erforscht. Die Verknüpfung dieser Entwicklungen hat zu spannenden Ansätzen in der Wissenschaft geführt. In dieser Dissertation stelle ich einige der Ansätze vor, an denen ich während meiner Promotion gearbeitet habe.

Zunächst gebe ich einen Einblick in die Entwicklung von Modellen, die mit Hilfe künstlicher Intelligenz Verbindungen zwischen Augenbewegungsdaten und visueller Expertise herstellen. Dies wird anhand zwei verschiedener Bereiche, genauer gesagt zwei verschiedener Personengruppen, demonstriert: Sportler bei Entscheidungsfindungen und Chirurgen bei arthroskopischen Eingriffen. Die daraus resultierenden Modelle können als digitale Diagnosemodelle für die automatische Erkennung von visueller Expertise betrachtet werden. Darüber hinaus stelle ich Ansätze vor, die die Übertragbarkeit von Augenbewegungsmustern auf verschiedene Kompetenzbereiche untersuchen sowie wichtige Aspekte von Techniken zur Generalisierung. Schließlich befasse ich mich mit der zeitlichen Erkennung von Verwirrung auf der Grundlage von Augenbewegungsdaten. Die Ergebnisse legen eine Nutzung der Modelle als Zeitgeber für mögliche digitale Assistenzoptionen in der Ausbildung von Berufsanfängern nahe. Eine Besonderheit meiner Untersuchungen besteht darin, dass ich auf sehr wervolle Daten von DFB-Jugendkader- athleten sowie von langjährigen Experten in der Arthroskopie zurückgreifen konnte. Insbesondere die Arbeit mit den DFB-Daten stieß auf das Interesse von Radio- und Printmedien, genauer, DeutschlandFunk Nova und SWR DasDing. Alle hier vorgestellten Beiträge wurden in international renommierten Fachzeitschriften oder auf Konferenzen veröffentlicht.

\tableofcontents

\mainmatter


\chapter{List of Publications}
\label{chap:listPub}

\subsection*{Published Articles}

\begin{enumerate}
	
	\item M. S. Yin, P. Haddawy, \textbf{B. W. Hosp}, P. Sa-ngasoongsong, T. Tanprathumwong, M. Sayo,and A. Supratak. "A Study of Expert/Novice Perception in Arthroscopic Shoulder Surgery." In Proceedings of the 4th International Conference on Medical and Health Informatics (pp. 71-77). August 2020.  \\https://doi.org/10.1145/3418094.3418135

	\item \textbf{B. W. Hosp}, F. Schultz, O. Höner, and E. Kasneci. "Soccer Goalkeeper Expertise Identification Based on Eye Movements.” PloS one, 16(5), e0251070. 2021. https://doi.org/10.1371/journal.pone.0251070

	\item \textbf{B. W. Hosp}, F. Schultz, E. Kasneci, and O. H{\"o}ner. “Expertise classification of soccer goalkeepers in highly dynamic decision tasks: A deep learning approach for temporal and spatial feature recognition of fixation image patch sequences,” Frontiers in Sports and Active Living, vol. 3, p. 183, 2021. \\https://doi.org/10.3389/fspor.2021.692526

	\item \textbf{B. W. Hosp}, M.S. Yin, P. Haddawy, R. Watcharopas, P. Sa-ngasoongsong, E. Kasneci. "States of Confusion: Eye and Head Tracking Reveal Surgeons’ Confusion during Arthroscopic Surgery." In Proceedings of the 2021 International Conference on Multimodal Interaction (ICMI ’21), October 18–22, 2021, Montréal, QC, Canada. ACM, New York, NY, USA. \\https://doi.org/10.1145/3462244.3479953

	\item \textbf{B. W. Hosp}, M. S. Yin, P. Haddawy, P. Sa-ngasoongsong, and E. Kasneci. „Differentiating Surgeons' Expertise Solely by Eye Movement Features”. Companion Publication of the 2021 International Conference on Multimodal Interaction (ICMI '21 Companion), October 18--22, 2021, Montréal, QC, Canada. ACM, New York, NY, USA. https://doi.org/10.1145/3461615.3485437
		
\end{enumerate}
 \newpage
\subsection*{Submitted Articles}

\begin{enumerate}
	
	\item \textbf{B.W. Hosp}, F. Schultz, O. Höner, and E. Kasneci. "In the Search of A Superior Gaze Behavior: Cross-Domain Shared Expertise-Related Gaze Features." 
	Submitted to: ACM Symposium on Eye Tracking Research \& Applications (ETRA '22).
	
\end{enumerate}

\subsection*{Additional Articles}

\begin{enumerate}
	\item \textbf{B. W. Hosp}, S. Eivazi, M. Maurer, W. Fuhl, D. Geisler, and E. Kasneci. „RemoteEye: An Open-Source High-Speed Remote Eye Tracker.” Behavior research methods, 1-15. 2020. https://doi.org/10.3758/s13428-019-01305-2	
		
	\item W. Fuhl, S. Eivazi, \textbf{B. W. Hosp}, A. Eivazi, W. Rosenstiel. and E. Kasneci.  "BORE: Boosted-Oriented Edge Optimization for Robust, Real Time Remote Pupil Center Detection." In Proceedings of the 2018 ACM Symposium on Eye Tracking Research \& Applications (pp. 1-5). June 2018. \\	https://doi.org/10.1145/3204493.3204558
	
	\item W. Fuhl, E. Bozkir, \textbf{B. W. Hosp}, N.  Castner, D. Geisler, T. C. Santini, and E. Kasneci. "Encodji: Encoding Gaze Data into Emoji Space for an Amusing Scanpath Classification Approach." In Proceedings of the 11th ACM Symposium on Eye Tracking Research \& Applications (pp. 1-4). June 2019.  \\https://doi.org/10.1145/3314111.3323074

\end{enumerate}

\newpage

\section{Scientific Contribution}

This work is divided into five chapters that focus on different aspects of latent gaze information in highly dynamic decision-tasks. The first chapter lists the publications that are related to this work and contains an overview of their scientific contribution. The second chapter introduces the necessary basics by starting with an introduction into the fundamentals and then proceeding to expertise and confusion research, which are two important perceptual-cognitive processes for this work. Chapter three sets the focus on the main contributions, which are the objective, robust, and reproducible linkage between gaze and expertise, which all three are exemplary shown on two data sets from different domains. Another contribution shows an approach to infer expertise-related features that are shared between different domains as a step towards general gaze expertise definition. One further contribution is an automation step to remove arbitrariness and manual feature selection in the process of building a model for expertise detection. The last contribution is an online system to detect states of confusion that can be used to temporarily schedule assistance options. The results of the mentioned contributions are discussed in terms of applicability and transferability in chapter four.  Chapter five focuses on ethical considerations regarding behavioral data and machine learning. This work is based on the papers from the upper publication list. The original publications are listed in the appendix.

\chapter{Introduction}
\label{chap:1}

While human beings might never be able to read another person's mind, science is already able to deduce certain information about cognitive processes based on user monitoring and data-driven methods. Eye tracking is one of the most promising emerging technologies that provides these insights, as the movements of our eyes reveal a lot of information about our cognitive states that are not obviously visible, but more subtle. Eye tracking as a method has already been used in ancient times, but the technology of video-occulography has its roots in the mid of the 20th century. It evolved from a mere laboratorial technique that included sophisticated ideas about optic systems to observe the movements of the eyes (i.e. Yarbus experiment 1967, \cite{yarbus1967eye}), to a wide field of ubiquitous and precise devices from handful vendors \cite{tobii2015,smart_eye}.

Lately, this technology has used video cameras to record the eyes and their movements to provide insights into perceptional processes. To date, several cognitive effects were researched by investigating movements of the eyes. 
Based on the cognitive load theory ~\cite{sweller2011cognitive}, i.e.  
scientists can reason the mental effort of a person during a task or understand the attention and mental effort while driving by examining the changes in pupil sizes ~\cite{palinko2010estimating,bozkir2019person,appel2021cross,appel2018cross,appel2019predicting,gao2021index}. Other studies, for example, deal with the influence of anxiety on cognitive load 
\cite{chen2009cognitive} or how online learning affects the mental load of 
students \cite{bradford2011relationship}. However, eye-tracking technology can be used to detect even more complex perceptual-cognitive processes. The detection of some of these processes can be advantageous for the optimized development of human behavior. Especially in fields related to subconscious behavior and high dynamics, behavior can be hard to describe. There are several aspects to be mentioned, but the two most interesting ones are expertise and confusional states. There is great interest in the recognition of these processes via visual perception ~\cite{castner2020gaze,tanaka1997expertise,murphy2021esport,castner2020pupil,eder2021support,geisler2020minhash,castner2018overlooking,kubler2015automated,ACTNEURO2017,eivazi2017towards,NSKE092017, kubler2017subsmatch,TC022017,kubler2017subsmatch,TCE092015,kubler2017,TCDRJACUWE122014,castner2018scanpath,castner2020deep}, as the application of eye tracking as a research method can provide objective measures, and likewise, can improve objective diagnostics of expertise and confusion in several fields, like soccer or medicine, and thus, enabling an objective way to understand and analyze subtle behavior better. \\

\subsection*{Hardware}

The most common types of eye-tracking devices are mobile (head-mounted) eye trackers and remote eye trackers. Mobile eye trackers are worn like glasses as where remote eye trackers are placed in front of or under a computer display facing the subject in front of the screen. 
Next to commercial systems, there are a plethora of open-source eye trackers 
available. Multiple devices with different sampling rates as well as different accuracy or precision values (important attributes that describe the performance of an eye tracker) are available. One of my first contributions to the scientific community was the development of "RemoteEye" \cite{hosp2020remoteeye}. With over 500 Hertz and an accuracy of < 1°, I 
developed a low-cost, high-speed, and open-source remote eye tracker for research. The eye tracker is easy to handle and uses a feature-based approach. 
The most important aspect of this eye tracker is, that it can be built by anyone. It's parts are either off-the-shelf LEDs and cameras, 3D printed boxes, or simple aluminum bars. Besides creating a basic understanding of eye-tracking methods and algorithms for my later research, it was important to me to provide open and low-threshold access to this technology. In fact, open-source is very important for the research community, because eye tracking is still a niche market. While Tobii \cite{tobii_gaming_2021} started to produce game ready eye trackers, that can be used in several computer games and Microsoft included an API for eye-tracking devices in 
Windows OS, the usability for research is still limited. Current state-of-the-art commercial eye-tracking devices are very expensive or restrictive regarding data accessibility and therefore hinder a great number of researchers to use eye tracking as a method in their research. With "RemoteEye" ~\cite{hosp2020remoteeye}, I developed a system that can easily be built and used for research even by small labs with a lower budget. 

\subsection*{Gaze Signal}

\begin{figure}
	\centering
	\includegraphics[width=1\linewidth]{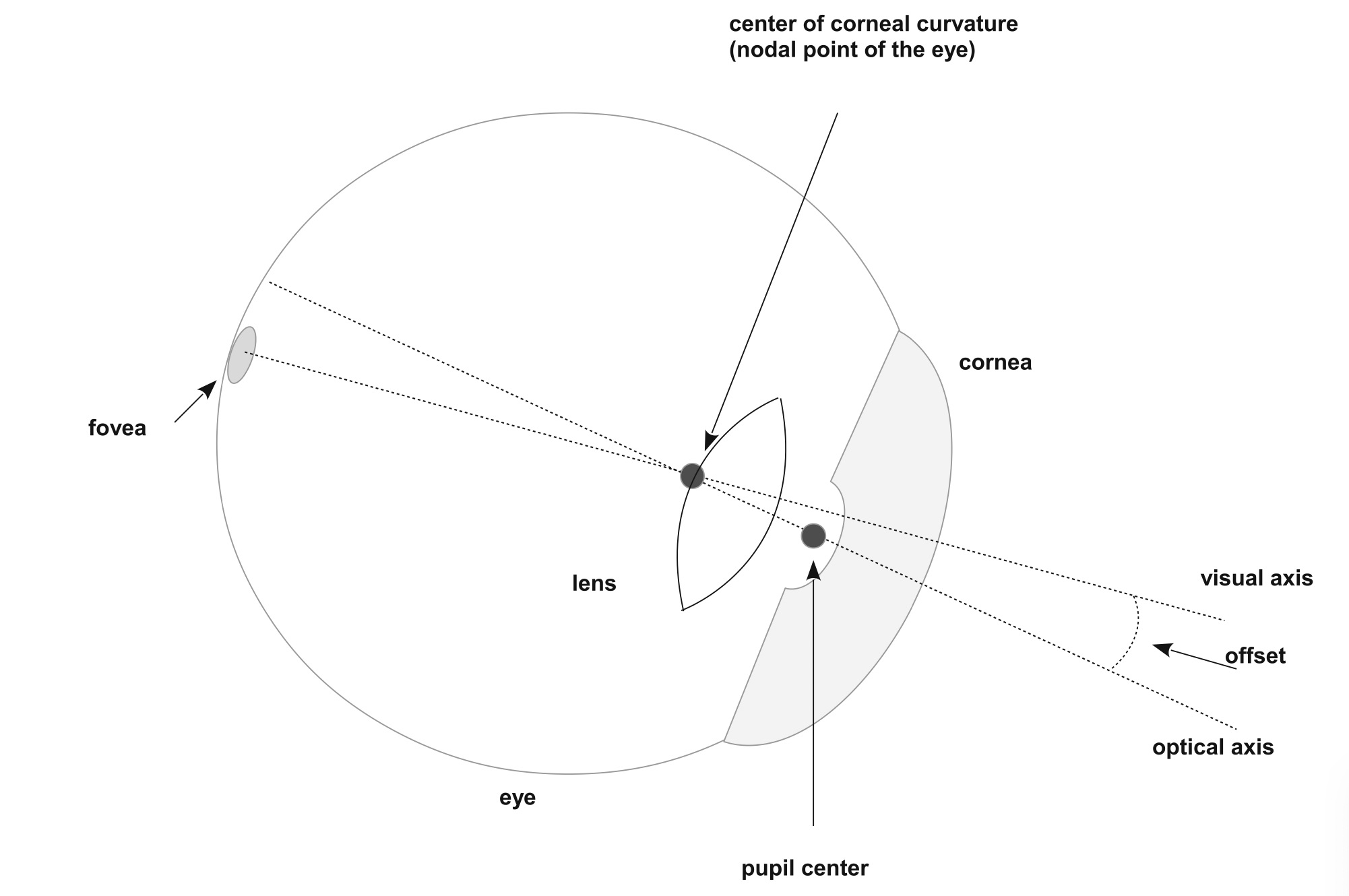}
	\caption{ Illustration of an eye showing important variables for gaze estimation. Source: Guestrin, Elias ~\cite{guestrin2010remote}}
	\label{fig:eye}
\end{figure}

One of the aims of eye tracking is to tell where the subject is looking at in spatial and temporal dimensions. For this purpose, one popular approach is to capture the image of the eyes and find the center of the pupil and/or glints (also known as \nth{1} Purkinje images, \cite{holmqvist2011eye}). Usually, glints are artificially created reflections on the cornea which come from infrared light-emitting diodes (LED) of the eye tracker. Infrared light is not visible to the human eye, so it does not blind or distract the subject. 
With the location of the pupil center and/or the glints, one can calculate the optical axis of the eye. To know the point of regard (POR) of the eyes one then needs to calculate either the visual axis of the eye or at least the relationship between POR and optical axis. This step is important, as the offset between optical and visual axis can deviate approximately 4-5$^{\circ}$ horizontally and 1.5$^{\circ}$ vertically \cite{bennett1962eye}, and differs from subject to subject. Thus, this is done by a person-specific calibration routine. While certain points in space or screen are shown to the subject, the locations of the pupil and/or glint centers are recorded. Subsequently, there are different options. One is modeling the offset between the optical and visual axis of the eye by computing a 3D model (illustrated in Fig.~\ref{fig:eye}) of the eye. Another one is to use some combinations of pupil centers and/or glint features and the POG to estimate the relationship of the two axis, represented by an equation of a higher-order polynomial. This allows interpolating the relationship between POG and pupil/glint from any other location, usually with differing accuracy. It is the same technique for head-mounted and remote eye trackers, except that remote eye trackers need to find the face and eyes in the captured image first and are further away from the eye. Thus, they have fewer pixels to cover each eye but are usually connected to a powerful desktop computer, allowing higher processing speed. In the latest years, the common method of feature-based eye trackers (that are based on features like pupil center or glints),  has slowly been replaced by uprising appearance-based methods. Appearance-based methods mostly take advantage of machine/ deep learning by taking images of the whole eye, finding and collecting features of the eye by themselves and relating them to certain points that are shown in space or screen during calibration. These techniques need a lot of computational power, but their accuracy is constantly improving and will run down the performance of feature-based methods anytime soon.

\subsection*{Eye Movements}
On average, humans make 3-5 eye movements per second \cite{holmqvist2011eye} 
which are needed to perceive visual information from our environment. Which eye movement events can be calculated resiliently, mainly depends on the speed of acquisition of the eye-tracking signal (in Hertz) \cite{andersson2010sampling}. In the center of our visual field, humans see the best when the object is imaged directly on the fovea which is an approximately 2° big area, slightly displaced at the back of the eye on the retina (as mentioned before, this is on the visual axis which can deviate up to 5° from the optical axis). Looking at the retina, one can see that the fovea is the point of highest acuity. Points that lie further away from the fovea, have less receptors. Therefore, the further away from the fovea, the worse vision becomes.

Fixations and saccades are seen as basic eye events. A fixation is theoretically a temporally and spatially limited accumulation of gaze points, which is calculated by an algorithm. It is assumed that visual information acquisition occurs during fixations. In detail, however, there are different methods with which a fixation can be computed. For low-speed eye-tracking devices (\textasciitilde 50Hz) threshold-based algorithms are mostly used (for overview see~\cite{salvucci2000identifying}). With higher speed, velocity-based algorithms are more often used ~\cite{holmqvist2011eye} as they earlier detect saccade launches. However, there are other approaches, too, e.g. using bayesian statistics ~\cite{tafaj2012bayesian,santini2016bayesian} or even machine learning techniques ~\cite{fuhl2018rule}.

Saccades are the jumps between fixations. That is, saccades are made when attention is drawn to another object. Thus, the attention-generating object is again on our fovea 
(area on the retina with the highest acuity). During saccades, humans are blind. However, the brain interpolates the images so that we are not aware of 
it. Saccades can be triggered consciously or unintentionally. However, for healthy people, both eyes usually move simultaneously and in the same direction. In addition to saccades, there are also micro-saccades, whose significance for perception has not been conclusively clarified. Eye-tracking software often calculates only fixations and considers all points between two fixations as part of a saccade.
Besides micro-saccades, tremor and drift are considered to be part of a fixation 
(so-called fixational eye movements). The prevailing opinion about their 
usefulness is that these minimal movements help the eye to stay on target and 
prevent the trigger from disappearing by constantly refreshing the potential 
on the cells of the retina. Other eye movements are named but difficult to calculate and detect. Vergence stands for the adjustment of focus on objects in different depths. Here the eyes move in opposite directions. Much more interesting movements are smooth pursuits. These cannot be triggered consciously and manifest themselves as a fixation on a moving object. However, if the movement of the object becomes too fast, smooth pursuits are no longer applied but instead small sequential saccades. The last known eye movement is the vestibular-ocular reflex. Here the head and eyes move in opposite directions. This is comparable to a fixation of the eyes while the head is turning.

Based on fixation and saccade calculations, second-order features can be derived. In this work, properties derived from and describing primary gaze events, like fixations, saccades, and smooth pursuits, are called second-order features. These derived properties are particularly important in expertise research ~\cite{hosp2021differentiating}, as simple comparisons between accumulated fixations or saccades offer little information ~\cite{klostermann2020fewer}. Much more meaningful are data on velocity, acceleration, deceleration, frequency, duration, dispersion, and amplitude. With the help of these properties, a much more precise picture of eye movement characteristics can be drawn. Most high-speed eye-tracking device vendors provide the calculations of these features as exportable CSV-files. These high-speed features are especially necessary for highly dynamic decision-tasks. The subject does not only need to be precise and correct, but also fast. This means there is little time to perceive important clues. In highly-dynamic situations, it is therefore important to record even volatile movements of the eyes. These volatile movements are particularly present in expertise classification.


\section{Expertise Research}
Expertise is a qualitative measure that describes a person's ability to solve a 
certain task or area of work. Therefore, beginners have less expertise than intermediates or even experts. Becoming an expert usually takes years of training and practice of purposeful procedures that contribute to the solution of the task. As a qualitative measure, there is a difficulty in measuring expertise quantitatively. A certain amount of hours or years of experience in the field is often used as a measure. However, this excludes talented individuals. Since everyone develops at their own pace, people with a lot of talent and relatively little experience can have a higher level of expertise than people with a lot of experience but little talent. 'How to measure expertise objectively and robust' is one of the central research questions in visual-expertise research. The diagnosis and differentiation of expertise at different levels are particularly important for understanding the factors underlying expertise and its development. In order to study the development of a person's cognitive and motor skills, the 'expert performance approach' ~\cite{ericsson1991toward} is often used. It states that expertise is best revealed in a laboratory situation when the conditions are kept as realistic as possible. In sports, for example, physical education and training often cannot be made more intensive. Therefore, emphasis has been placed on improving cognitive factors in recent years, which have lately been recognized as advantageous but have not been trained adequately so far. However, an objective diagnosis is needed first and foremost. 

In order to create a diagnostic model, usually known classifications of subjects are used first. This classification is done, for example through talent scouts or competitions (both are common in soccer), but also through the status in an educational program or the number of hours one has already invested in the solution of a task (e.g. in medicine). In any case, a known classification model is needed, to develop a diagnostic model. As a second step, for each expertise class, data is collected from as many subjects as possible that represent this class. This is important to provide the machine learning algorithm with enough examples from which decision boundaries can be derived. On the one hand, this allows a broad picture of characteristics for the respective class, but also a robust boundary to separate classes from each other. On the other hand, a typical problem in expertise research is that the number of experts is small. This leads to the fact that studies dealing with expertise and its research have a small number of experts or subjects, which makes this data extremely valuable, but of limited use in terms of generalization.
In addition to a small number of expert examples, an unnatural environment can lead to unwanted effects. For example, in sports psychology, more efficient gaze behavior during decision-making has already been linked to expertise. Unquestionably, experts show strengths in finding and interpreting relevant cues, but when it comes to the question of which gaze behavior features can describe the differences, the findings diverge. The results of some studies ~\cite{bertrand2009effects,williams1998visual,vaeyens2007effects,williams1994visual,roca2011identifying,roca2013perceptual} suggest that expert behavior (more experienced, more talented, or more successful performance) is associated with an increased number of fixations. On the other hand, ~\cite{nagano2006visual,perry20improvements,savelsbergh2005anticipation,krzepota2016gaze,williams1997assessing,williams1998visual,savelsbergh2006four,vaeyens2007mechanisms,canal2011visual,williams1997assessing,north2009perceiving} find no dependence of expertise on the frequency of fixations. Further, there are even studies that conclude that fewer fixations can be associated with expertise ~\cite{savelsbergh2002visual,woolley2015use,canal2011visual,helsen1999multidimensional}. A similar situation exists with the length of fixations and the reference to expertise. For example, ~\cite{bertrand2009effects,williams1998visual,williams1994visual,roca2011identifying,roca2013perceptual} linked shorter fixations with expertise and ~\cite{savelsbergh2002visual,woolley2015use,savelsbergh2006four,canal2011visual} linked longer fixations with expertise. Some studies even found no significant relationship at all ~\cite{savelsbergh2005anticipation,krzepota2016gaze,williams1998visual,vaeyens2007mechanisms,vaeyens2007effects,helsen1999multidimensional,north2009perceiving}.

The different results may lead to the conclusion that there is either no relationship between expertise and gaze behavior or that it was not found. However, this is a fallacy, because the studies mentioned did not pay much attention to the naturalness of the scene. The method of data collection plays an important role. For example,~\cite{mcguckian2018systematic} conclude after a closer look at these studies that when the demands resemble a realistic situation in a soccer game, expertise may be associated with shorter and more frequent fixations. Similarly, ~\cite{kredel2017eye} report that differences in expertise are much more emphasized when subjects are exposed to a highly realistic scene during data collection.
Thus, so-called internal and external validity are both of high importance.

Regardless of these contradictions, expertise has been formed over years of experience and practice. On the one hand, it is assumed that experts develop their own optimal methods of perception by solving highly similar tasks for many years and optimize their perception in the 
process. On the other hand, it is assumed that there are certain commonalities in the experts gaze behavior. In addition to these commonalities, the differences between levels of 
expertise are also of particular interest for research~\cite{klostermann2020fewer,panchuk2015eye,kredel2017eye,brams2019relationship,moran2019exploring,mann2007perceptual,harris2020eye,wilson2010psychomotor,manning2006radiologists,wilson2011perceptual,ericsson1991toward,mann2007perceptual,mcguckian2018systematic,kredel2017eye,bertrand2009effects,williams1998visual,vaeyens2007effects,williams1994visual,roca2011identifying,roca2013perceptual}. This interest stems from the possibility of deriving insights into visual perception at different stages of development, but also from the possibility of using the knowledge of the commonalities and differences to define unique expertise levels. Diagnosing the correct expertise level (based on findings of visual perception research) can function as a basis for the development of possible assistance options. Both, in turn, can be used for a perceptual-cognitive tutoring or training system. Diagnostics are needed to determine the correct level of expertise of a subject at any given point in time. Likewise, to define the assistance options to determine the necessary progress for each of these levels of expertise to grant the subject to move to the next higher level of competence by learning new aspects of visual gaze behavior.

Different expertise classes show different similarities in gaze behavior so that a beginner needs another kind of assistance than an advanced user ~\cite{hosp2021differentiating}. In recent years, the perception of experts has been investigated in various fields and tasks. Aspects of perception that allowed separation of expertise classes have often been found but were typically thought to be limited to a certain domain or task. While a look at the current research situation shows a mass of expertise research studies, only little inter-domain or inter-task work is done. Most work is somehow limited to a task or domain. For example, ~\cite{gegenfurtner2013transfer} shows that it is possible to transfer expertise from familiar tasks to semi-familiar tasks, but not to unfamiliar tasks. However, they took the same subjects for both tasks, which introduces a high risk of enabling recognition of subject-specific characteristics instead of expertise. Thus, while differences have often been found, only little is known about inter-task or at least inter-domain expertise that is transferable or generalizable. 
The problem of a missing generalizable feature set that works for more than one task or domain has yet not been addressed properly.
So far, no dedicated set of traits was found that is better suited to recognize expertise than others. Therefore, previous study results could hardly or not at all be transferred to other studies and were always limited to one field, task, or data set~\cite{gegenfurtner2013transfer}. 
However, since it is expected that experts in the same task exhibit certain commonalities regarding their gaze behavior, in a subsequent step, experts could also exhibit certain commonalities regardless the task or even domain. To investigate this hypothesis, studies are needed that evaluate the gaze behavior with the exact same methods, on different tasks or in different domains. 
The overall question is whether expertise-related features derived from the visual behavior are consistent across domains and whether experts of different domains share some visual strategy features. A superior set of perceptual properties would lead to a complete overturning of our understanding of perceptional expertise.\\

\section{States of Confusion}
Simultaneously, people's gaze behavior is sought to provide the deduction of even more latent characteristics. For example, it would be useful to recognize when a person needs assistance while accomplishing a task ~\cite{rong2020driver,braunagel2016necessity,ETCWWW032015,CEWW092015,TEKMWUE11111112015,EKKMWUE022014,kuebler2021etra,DDCBEKASNECI2020,Bozkir2019SAP,braunagel2017online,E102013}. In addition to mental load or expertise, eye movements might serve as a proxy to detect perplexity or confusion during the completion of a task. 
First, one needs to define what is meant by "confusion". In fact, there are a lot of different definitions. Laymen use it less specifically, thus, different than health workers. For example,~\cite{durward1979organic} say „symptoms and signs which indicate that the patient is unable to think with his customary clarity and coherence“ or „disorientation in time and place“ ~\cite{hamilton1985fish}. The use of the word is in fact so ambiguous, that in 1984,~\cite{simpson1984doctors} even conducted a study to find out how medical doctors and nurses define confusion and which symptoms they consider for it. Depending on the field of the health worker the definitions were quite different. In this work, confusion is considered as a temporal state of disturbance that inhibits the continuation of the task, the definition of the Faber Medical Dictionary stating confusion as „a condition in which among other things there is a disturbance of consciousness" ~\cite{wakeley1953faber} is used. 

Confusion can arise for a variety of reasons. For example, medical tasks are often lengthy and many steps have to be considered. However, especially in procedures such as arthroscopy, tissue can be very similar in many places, so navigating from the portal hole (entry point) to the surgical site can be extremely difficult and confusing. That means, confusion can also occur when the surgeon does not know where they are, how to proceed, or cannot recognize helpful visual cues such as specific bone formations. Even an incorrect rotation or orientation of the arthroscope camera can lead to confusion. In addition to the correct projection of the 2D output video from the arthroscope camera onto the 3D surgical area, the surgeon has to know where they are at all times in order to reach the correct surgical site in the body. During training operations on real human bodies, a supervisor always has to be present to show the trainee the correct way to proceed, in case of doubt. By automatically detecting confusion, adequate assistance can be provided digitally. In the best case, there is no need for a supervisor to stand next to the trainee, which not only saves money but also valuable time for the supervisor, who has typically little time for training, as they are usually also in charge of other operations, too. Enriching the arthroscopic camera image with helpful hints that enable the surgeon to progress without the need for a supervisor would be a simple, cost- and time-saving way of training young surgeons. For this process, however, the first step is to recognize when the surgeon needs assistance. In a second step, the necessary assistive steps have to be defined.

Although head and eye movement measurements are often used to detect 
cognitive processes, these measurements were never used in medicine 
to detect states of confusion or perplexity (as far as the author knows). Here, too, it can be 
assumed that the combination of eye (and head) tracking and machine 
learning can offer a suitable methodology to deal with such a problem.
With the knowledge of moments of confusion, person-specific 
assistance possibilities can be discussed, as next to expertise, there is high interest in understanding learning behavior ~\cite{koc2017visualizers,knoepfle2009studying,porta2012emotional,lai2013review}. As such, the absence of expertise during a task is of high interest, too. Though, learning behavior is typically highly subjective. This means that each subject has its own learning speed, 
its own ways of learning, and is at different levels of expertise. Here, too, many processes take place unconsciously so that verbal communication about the correct focus in the correct moment is particularly problematic and does not lead to the desired results. For example, subjects may be unsure how a task should be carried out further or are inactive during a task, which may indicate that the subject is not sure about their decision.
Especially for laypeople and beginners, complex tasks can cause perplexity or confusion. Therefore, beginners have to learn how to find suitable visual landmarks and use them optimally. This can be taught verbally only to a certain extent so that even just showing an expert viewpoint would be much easier to achieve with far greater added value. Novices in training need a teacher, but at least in some situations, the teacher can be replaced by a training system. Due to the complex nature of some real-world visual search tasks, it can be really difficult for the expert to describe visual points. An ordinal example gaze overlay of an expert could be recorded and shown asynchronously to several novices, simultaneously.

Confusion can be expressed in many ways, which makes it difficult to define a single measure that could be used to identify it. In contrast to expertise, however, class differences are less taken into account here than person-specific reactions are evaluated. Since people learn differently and at different rates, and since patients' tissues can differ greatly, there is always either some adaptation to the current situation or a highly abstracted, generalized approach needed. However, in order to define a uniform measure for recognizing states of confusion, certain characteristics must have their validity and be recognized in as many cases as possible. 
A few approaches were discussed and applied in the research of confusional states. For example, longer fixations on task-irrelevant areas were associated with confusion ~\cite{pachman2016eye}. Regarding input sensing, work has already been done with electroencephalogram (EEG), but mainly with so-called think-aloud protocols, ~\cite{di2019curiosity}. Little research has been done on the detection of confusional states by using eye-tracking data, which might also depend on the high complexity of the gaze signal.

\section{Data-Driven Analysis}
Luckily, for a few years now, machine learning is a rising field, which 
helps to deal with highly complex data such as gaze data. Machine learning and deep learning techniques found their way into the analysis of eye-tracking studies. Like a perfect fit, eye tracking creates a lot of data and machine learning usually works better with more data, which results in synergetic effects. Next to the analysis of eye-tracking data, these synergetic effects can be advantageous in the search for features that are thought to reflect expertise but are too complex to analyze with traditional methods. In fact, machine learning or, more precisely, deep learning simplifies the recognition of expertise, cognitive load, or, as already mentioned, confusion. Two areas of artificial intelligence are applied in this work: machine learning and deep learning. On the one hand, there are classical machine learning methods (shallow classifiers) like Support Vector Machines (SVM) ~\cite{noble2006support}, Random Forest~\cite{pal2005random}, or Logistic Regression~\cite{maalouf2011logistic}. On the other hand, there are deep learning techniques like artificial neural networks, which are considered to be a type of machine learning, while both are part of artificial intelligence.
Deep learning and machine learning are hard to differentiate, but one of the main differences between machine learning and deep learning is the ability to process unstructured data through artificial neural networks (ANNs). This is because deep learning through ANNs is able to convert unstructured information such as texts, images, sounds, and videos into numerical values. This extracted information is then used for pattern recognition or further learning. Among others, both are part of the field of artificial intelligence.

Nowadays, different forms of machine learning are used in different scientific fields. For example, there is a lot of research using traditional machine learning methods for expertise recognition, such as in dentistry ~\cite{castner2018scanpath,castner2020gaze,castner2018scanpath}, microsurgery ~\cite{bednarik2013computational,eivazi2011predicting,eivazi2012gaze,eivazi2017towards} or in sports~\cite{hosp2021soccer,murphy2021esport}. Likewise, the number of applied deep learning methods is slowly  increasing, too~\cite{sims2020neural,hosp2021expertise,castner2020deep}. However, there are multiple challenges that need to be addressed when developing an approach to artificial intelligence. To find the right approach, the following points should be considered.
Basically, the type of available data plays a role. If the data is available in unstructured form, deep learning methods can be used. Machine learning requires a certain structure to be available. Either the developer puts the data into a structured form or lets a neural network do this work. If the data is too complex to find patterns and relationships between them, deep learning is more likely to be applied. For example, images contain complex information. Here, information is organized in an unstructured way represented by vertical and horizontal pixels. For a classical machine learning approach, it is necessary that each sample is described by defined attributes. Classically, tables are created that show the expression of certain specified features for each sample. Thus, machine learning needs a lot of pre-processing to work on images, which means that machine learning needs more intervention of the developer, whereas deep learning has a certain autonomy. The approaches also differ in the amount of time they require. Machine learning methods can be set up and executed quite quickly, but their expressiveness can be limited. Deep learning methods require more time to set up, but can usually deliver better results with more time as more data becomes available.

If one wants to apply these methods in eye tracking, the eye-tracking signal can be used directly in order to generate a structured representation to use classical machine learning methods. The eye-tracking signal is stored in a file (or can be retrieved online) that shows a timestamp and the current gaze point at that time on the stimulus. Based on this gaze signal, higher layer features like fixations and saccades can be computed. Several derivations like the frequency of the fixations and saccades or the velocities can then be calculated, too. The computation of such features allows the structured description of individual samples, which can be the statistics of the features for instance over a trial or a stimulus. Such a structured representation could contain features as columns and the respective expression of the features of a sample as rows. In order to train the algorithm, each sample of the training set is assigned to a class, too. For such a structure, classical machine learning methods can be applied. Usually, the goal is to obtain a representation of the gaze behavior that describes the behavior as good as possible in a structured way. The machine learning algorithm tries to identify similarities between samples of the same class and differences between samples of different classes and thus to separate the classes, by defining a decision boundary between them. Based on this, a model is created that learns the separation of classes as robustly as possible that new unlabeled samples are correctly classified. However, gaze behavior analysis can be done in many ways. For example, if the gaze behavior is not only in the form of a gaze signal but also in the form of images (e.g. the sequence of a scan path represented by AOIs or image portions of the stimulus), deep learning approaches can be used, since images are unstructured data, too. Such approaches have found application in current research. For example, if one assumes that expertise results from the optimal perception of helpful visual cues ~\cite{mann2007perceptual}, it may be more useful to analyze behavior based on the sequence of visual cues. In the past, next to traditional algorithms such as Needleman-Wunsch ~\cite{likic2008needleman} or Smith-Waterman ~\cite{nakshathram2021sequence}, partly borrowed from bio-informatics, several types of scan path comparison algorithms have been developed ~\cite{kubler2017subsmatch,geisler2020minhash, kubler2015automated,TCE092015,TC022017,goldberg2010visual,TC022017,castner2020towards,NSKE092017,castner2018scanpath,castner2020gaze,castner2020deep,ACTNEURO2017,fuhl2019ferns}. However, deep learning methods show a particular strength here. Which features in the images or videos (which are ultimately only a sequence of images) are used for classification can be determined by applying different layers in a neural network. Likewise, different filters can be used to determine any latent features in the image sequences. Whether this is based on the saliency of parts of the image, particular edge detection, or even object detections, is part of investigations of current research and is therefore left to the developers. For example, CNNs have shown that the convolution operation, after which CNNs are named, can extract an extremely large amount of information from an image by interleaving several operations. In most cases, the first layers of a neural network are designed to recognize edges, corners, patterns, and objects. At the time of writing this thesis, research on the optimal use of CNNs and optimization by residuals, and 3DCNN, was in full swing.

\newpage

\chapter{Major Contributions}

This chapter summarizes the main contributions during my PhD work. For each of the papers presented in the following, I will describe the motivation for the research question and give a summary of the main findings. The full text of the papers can be found in the appendix. Figure ~\ref{fig:tree} provides an illustration of the several pieces of work and how they are connected.\\


\begin{figure*}[h]
	\centering
	\includegraphics[width=0.9\linewidth]{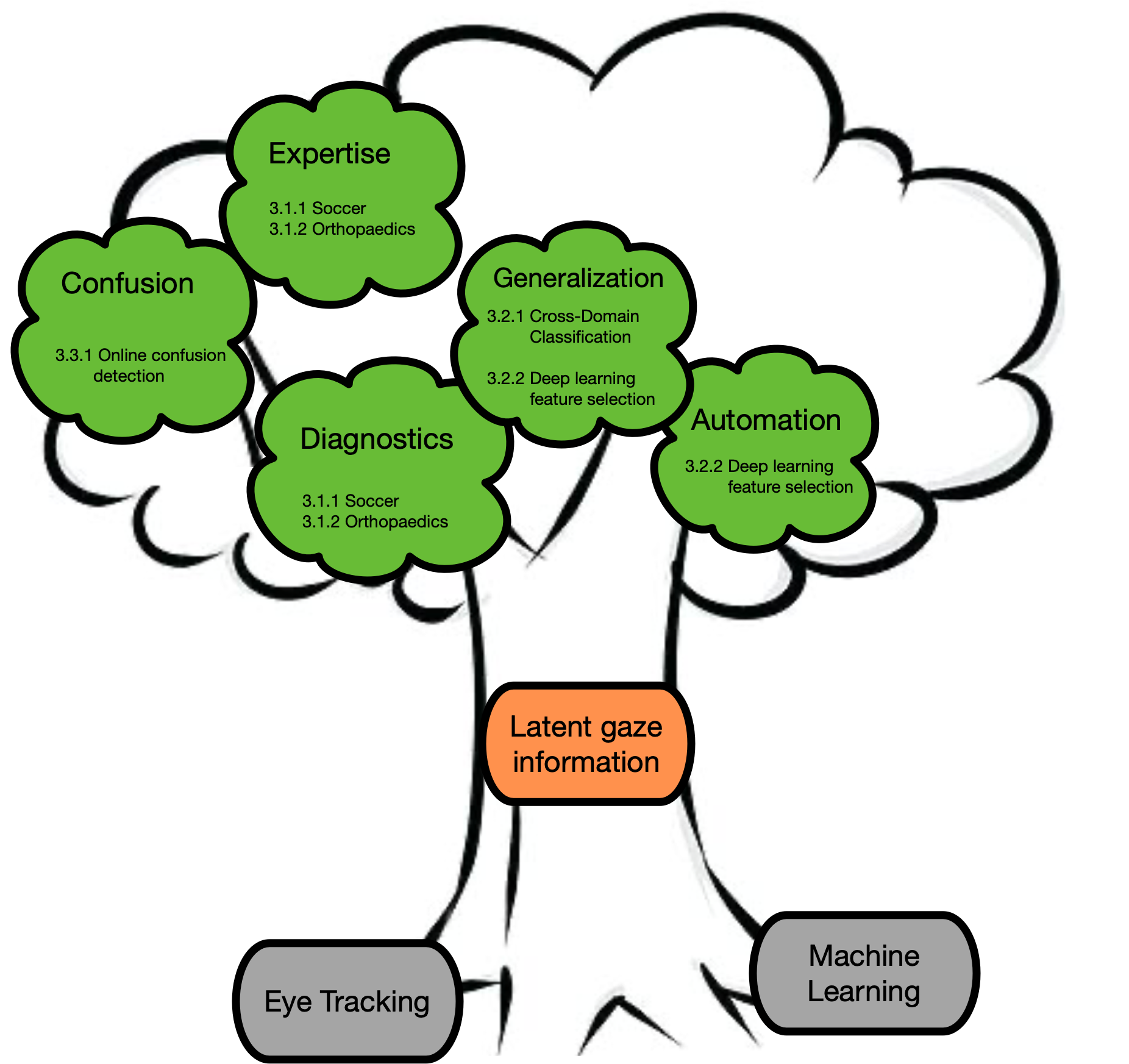}
	\caption{Overview and interrelationships of the presented papers.}
	\label{fig:tree}
\end{figure*}

As foundational to this work, eye tracking and machine learning can be seen as the roots of a tree. Eye tracking allows vital access to conscious as well as subconscious gaze information. Machine learning opens up novel ways of dealing with its complexity in order to infer further, deeper knowledge from it. The quality of the gaze signal as well as the robustness and applicability of the machine learning methods are essential for research on latent gaze information. The more these two research fields grow (higher recognition rate, more robust results, new methods), the more possibilities for researching on latent gaze information are provided. As the foundation grows, it delivers more essential knowledge and procedures and thus allow to infer even more latent gaze features. 
Research on latent gaze information can be considered as the main trunk of the tree. Since it combines the powers of both root technologies, eye tracking and machine learning, it allows to build models to answer a pool of research questions. In turn, the answers to research questions can lead to new applications (new branches of the tree). With growing roots, the trunk of the tree can grow bigger and thus enable more new branches to grow. Each branch symbolizes applications one can build upon the research on latent gaze information. One of them is diagnostics. Diagnostics are essential for further work like training or tutoring systems. Aspects that are typically diagnosed are expertise (covered in section 3.1.1 and 3.1.2) and confusion (covered in section 3.3.1). By combining both aspects in a training system, not only the expertise level, and thus, the level of assistance needed (expertise detection), but also the correct timing for assistance can be found out (confusion detection). Another branch is leading to a unified design process which can be achieved by a certain degree of automation (covered in section 3.2.2). By focusing on the removal of manual selection, the arbitrariness is cut out of the process, which enables some kind of automation. For generalization, some kind of robust, objective, and reliable cross-domain diagnostics is needed, that is based on the same, unified and automated processes.
This work includes first steps towards a general perspective on latent gaze information, their relationship and dependencies to diagnostics (covered in section 3.2.1).  \\

\newpage

\section{Gaze Expertise Linkage}

In certain research areas, gaze behavior has already been linked to expertise. However, this was mainly done manually, visually, or based on statistical methods ~\cite{blascheck2017visualization,kubler2015automated,TCE092015,TC022017,goldberg2010visual,geisler2020minhash,TC022017,castner2020towards,NSKE092017,ACTNEURO2017}. Since a few years, machine learning techniques are commonly used to infer knowledge about scan paths ~\cite{castner2018scanpath,kubler2017subsmatch,castner2020gaze,castner2020deep,eivazi2017towards,}. In fact, machine learning methods can remedy this linkage by using supervised methods that lead to a uniform approach on the one hand, and to explainable results on the other. Building uniform methods to analyze gaze-based expertise is an important step for comparisons of expertise and its definition. This first contribution is an objective, reproducible machine learning approach, that results in a model with explainable features. Likewise, this model helps to understand the evolution of perceptional features in several stages of expertise. This approach is applied to two different groups of subjects - containing expert, intermediate, and novice subjects - from sports and medicine. The analysis of both data sets shows that it is possible to recognize expertise based on a few gaze features presented as a ternary classification problem, with high accuracy (78.2\% and 76.46\%, respectively). Further, the high influence of idiosyncrasy of human gaze behavior on classification is shown and, likewise, which features describe the differences in expertise the best.\\ \\


\subsection[Soccer Goalkeeper Expertise Identification]{Soccer Goalkeeper Expertise Identification based on Eye Movements }
~\label{sec:soc1}

\textbf{B. W. Hosp}, F. Schultz, O. Höner, and E. Kasneci. "Soccer Goalkeeper Expertise Identification Based on Eye Movements.” PloS one, 16(5), e0251070. 2021.

\subsection*{Motivation}

While connections between expertise and gaze behavior have been made multiple times in different fields, little is done in sports, that 1) shows high accuracy on the classification of gaze behavior of three classes of expertise, 2) uses objective, reproducible, and state-of-the-art methods for classification, and 3) lead to explainable features. The following paper describes how to find an optimal set of features and feed it to a supervised machine learning algorithm to express the commonalities and differences of expertise-dependent gaze behavior in a robust, objective and reproducible way. The use of a classification model as an online diagnostic system is one of the far-reaching aims of this work. Therefore, how these findings can be used in the future is discussed in chapter ~\ref{sec:discGazeLink}.

\subsection*{Methods}

\begin{figure}
	\centering
	\includegraphics[width=1\columnwidth]{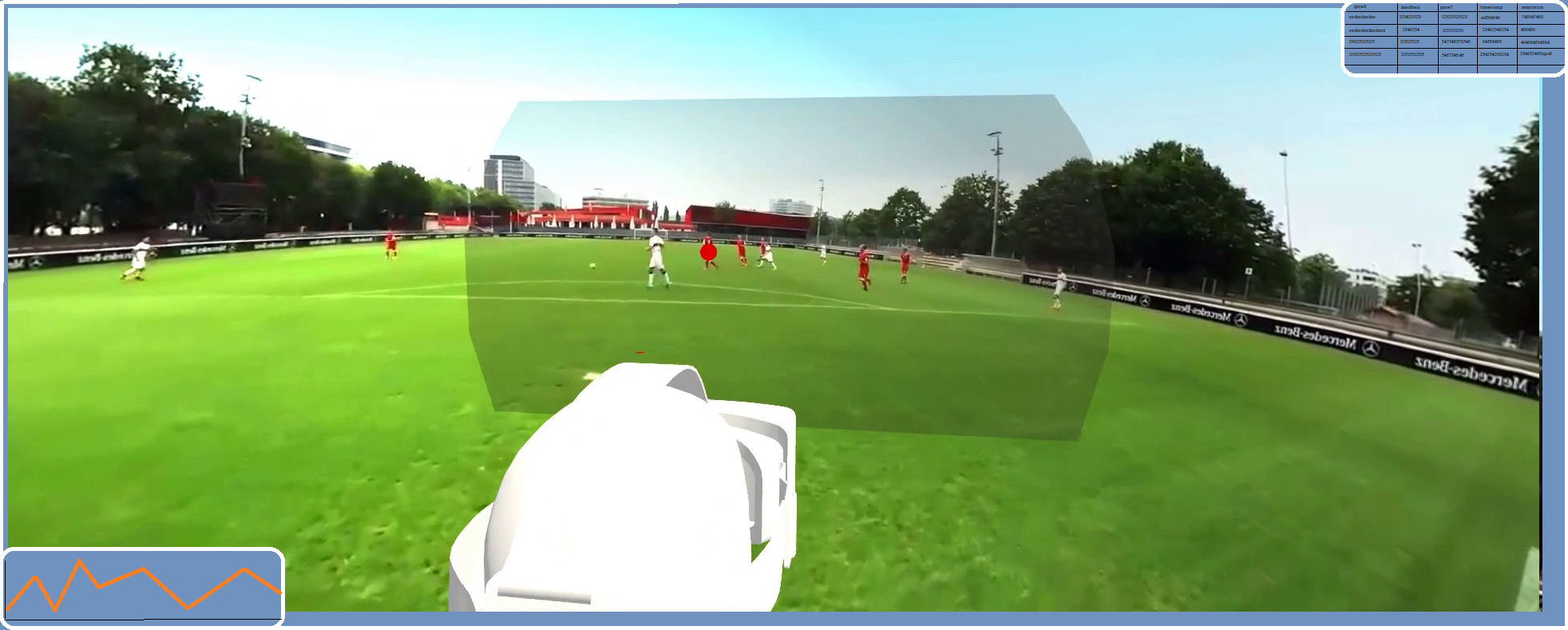}
	\caption{Zoomed out example of subjects' perspective during data collection.}~\label{fig:soccerPerspective}
\end{figure}

To infer the connection between gaze and expertise, a 360° camera was placed on the soccer field while soccer players physically replayed a defined common scene (Fig.~\ref{fig:soccerPerspective} shows a zoomed-out perspective view of the subjects). This omnidirectional video footage was then presented to our subjects on virtual reality (VR) glasses. Each scene shows a build-up situation that ended by the return pass to the position of the subject. Afterward, the subjects had to tell how to continue the scene. In total,there are 33 subjects from three different expertise classes. While the subjects were watching the stimuli on VR glasses, their eye movements were captured with 250 Hz. In the first step, the main model with all eye-tracking features of the eye tracker available is constructed. In a second step, a subset of features was defined that increases the accuracy of a test set while reducing the number of features dramatically. Three different methods were investigated. All features that have the highest p-values, thus, significant differences between the three classes of expertise (significant features = SF), all features that show the highest frequency by ranking them with a maximum relevance minimum redundancy (MRMR) algorithm and chi-square-test (most frequent features = MFF) and for comparison a model with all features available from the eye tracker (all features = ALL).

For all of the approaches, one first needs all available features from the eye tracking device. All of them are taken into account to build a first model. The effective aim is to find an optimal yet small set of features that has a high impact on the classification. A smaller feature set means shorter computing time and therefore better usability in an online diagnostic system. Starting with all features available, an SVM model is build to classify the data into three groups. The model reaches a certain accuracy $Acc_m$, which is the performance metric that is used to compare the impact of the features sets that were defined. Though the complete elimination of statistical errors is impossible, SF and MFF approaches need to be applied in a high number of runs, to lower the probability of statistical errors occurring. 

\subsection*{Results}

\subsubsection*{Feature set evaluation}

One of the main findings is the comparison of the different subsets of features. The subset of features chosen by the MFF approach (MRMR and chi-square-test) shows better performance in prediction accuracy (78.2\%) than a model with ALL(75.08\%) or SF (73.95\%). When looking at the 75th percentile, the differences are better noticeable (ALL: 80.989\%, SF: 79.25\%, MFF: 85.44\%). A classification model is considered as well performing with an accuracy of over 70\%. Thus, all three models can be considered as a classification model, but by looking at the recall of all three models, the MFF is the best performing, again (ALL: 71.87\%, SF: 73.19\%, MFF: 76.18\%).

\subsubsection*{Idiosyncrasy}

\begin{figure*}[h]
	\centering
	\includegraphics[width=1\columnwidth]{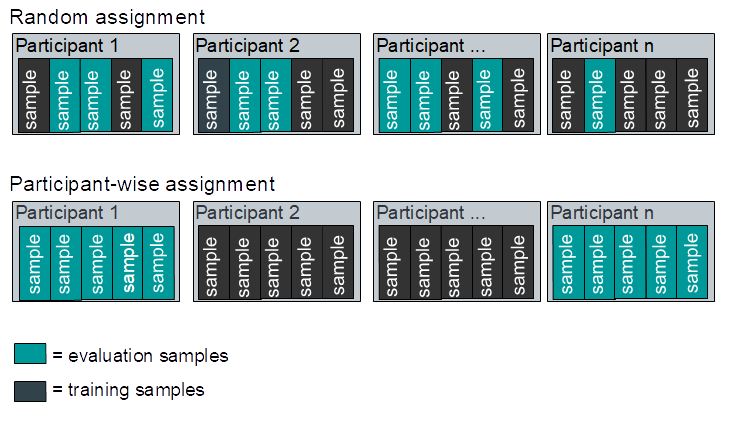}
	\caption{Random and subject-wise (idiosyncratic proof) sample assignment. }~\label{fig:idiosyncratic}
\end{figure*}

When assigning samples to the training and evaluation set, one has to consider an important point. Most eye movement features are idiosyncratic ~\cite{holmqvist2011eye}. Actually, a large portion of eye movements have already been proven to be idiosyncratic, like fixation duration, blink duration and rate, pupil diameter, saccade acceleration and deceleration, and saccade amplitude. During the model training step, these findings could be approved. By randomly assigning all samples of all subjects to the training or evaluation set, samples of each subject end up being distributed on both data sets (Fig. ~\ref{fig:idiosyncratic}). This leads to an unexpected learning behavior of the model, as the model rather matches the origin of the sample to a specific subject. Thus, it is not classifying a sample's class directly, but rather through its belonging to a certain subject. Such a model would estimate all samples of the evaluation set nearly perfectly, as the training set already contained highly similar samples of the same subjects. However, the model would fail to predict the belonging of new samples from new subjects correctly, as it has never seen data of this subject before. This behavior of idiosyncratic eye movement features reveals that the differences between subjects are much bigger than the differences within subjects. As such, a classification model would learn a subject-specific, bio-metric relationship instead of a correct class representation.

\subsubsection*{Expert variation}

A similar approach has been used to infer the classifiability of a certain group of subjects. Half of all experts were taken and switched with the same amount of intermediates by intentionally labeling them with the other class label. The accuracy is expected to drop under chance level, which would prove, that the differences between experts are smaller than the differences between experts and intermediates. The assumption was correct, as the model could not anymore differentiate between true experts and fake experts and intermediates vice versa. Defining a robust decision boundary was not possible anymore, which allows the statement to be defined: Differences between experts are smaller than differences between experts and intermediates.

\subsubsection*{Classification}

In a further approach, the expert and intermediate samples are considered to build a first SVM model with all features, that is able to predict the affiliation of the samples. A classification accuracy of 88.1\% was achieved. This means the model is able to estimate the affiliation of a sample of an expert or an intermediate correctly, with a probability of 88.1\%. There were 31 samples out of 260 falsely classified and the performance on intermediate samples was better than on expert samples. With a low miss rate of 
11.9\% the model shows great results. In a second approach, a better-performing model that needs fewer features was investigated. As the MFF set showed superior performance, this set is used in a ternary classification. As stated earlier, the ternary classification with the MFF set peaked at 78.20\%. Compared to the chance level of guessing, this model can be considered to be performing 
outstandingly.

\subsubsection*{Latent expertise features}

Next to the applicability of the MFF set as a foundation for a high-performing 
ternary classification, the MFF model revealed a certain amount of latent gaze information that is reflected by characteristics of the features. Most of the found features are typically not used as expertise markers. This might come from their difficult interpretability, as there is no obvious and simple characteristic behind these features. One first difference is found in the saccadic movements. Experts, as well as novices, tend to have a more homogeneous saccade behavior as the standard deviation of their saccade lengths is much smaller. However, novices have similarly long saccades as where experts have similarly short saccades. Apart from that, 
this allows proving the statement of~\cite{mann2007perceptual}, that experts have fewer but longer fixations. Their behavior is usually based on longer fixations to avoid saccadic suppression, as there is no information intake during a saccade. In this work, differences in fixation length between expertise groups were not directly found. This might be based on the split between short fixations and smooth pursuits or from the age difference between the single expertise groups. 
Conversely, further differences are found in the maximum deceleration of the saccades. 
In line with~\cite{zwierko2019oculomotor} deceleration behavior was found to be an adequate marker for expertise detection, too. There is a continuous increase in the maximum deceleration speed of the saccades. Novices are much slower than intermediates and experts.

Another quite interesting observation during data collection was the gaze behavior when the ball is passed around in the stimulus. Experts tend to only look at the ball shortly before and after a teammate is in possession of the ball. Novices tend to follow the track of the ball a much longer time. This is an important behavior as there are optimal times when the player can seek an overview over the scene to be able to react appropriately when getting into possession of the ball. These times are when the player is not in play, when the ball has been passed and cannot change its track, and when the line of sight to the ball is blocked (the subject is not playable). The values of the smooth pursuit dispersion vigorously prove such behavior. Experts have a small window between minimal and maximal smooth pursuit dispersion. Their maximum 
is less than half as long as the novices' and their minimal value is still 
$1/3$ shorter than for novices. Intermediates are placed between experts and novices. Thus, there is a continuous decrease visible. Likewise, the average smooth pursuit, as well as the maximum and standard deviation of the smooth pursuit dispersion correlate negatively with the classes. The classes differ significantly, which is also reflected in the average, 
minimum, and maximum smooth pursuits, with a p-value of $p < $\num{1e-12}. Novices show much longer smooth pursuits than intermediates and experts. Likewise, the 
shortest smooth pursuits of the novices are longer than the intermediates' 
and experts'. The same patterns can be observed in the maximum values of the smooth pursuits, as novices have a higher maximum than the intermediates and the experts.  The standard deviation of the lengths of the smooth pursuits shows a highly similar pattern but is statistically not significant. Novices' smooth pursuit scatter much more than intermediates or experts.

\newpage

\subsection[Differentiating Surgeons' Expertise]{Differentiating Surgeons' Expertise Solely by Eye Movement Features}
~\label{sec:linkSurgeon}


\textbf{B. W. Hosp}, M. S. Yin, P. Haddawy, P. Sa-ngasoongsong, and E. Kasneci. „Differentiating Surgeons' Expertise Solely by Eye Movement Features”. Companion Publication of the 2021 International Conference on Multimodal Interaction (ICMI '21 Companion), October 18--22, 2021, Montréal, QC, Canada. ACM, New York, NY, USA. \\
\\
and \\

M. S. Yin, P. Haddawy, \textbf{B. W. Hosp}, P. Sa-ngasoongsong, T. Tanprathumwong, M. Sayo,and A. Supratak. "A Study of Expert/Novice Perception in Arthroscopic Shoulder Surgery." In Proceedings of the 4th International Conference on Medical and Health Informatics (pp. 71-77). August 2020.

\subsection*{Motivation}

To infer how surgeons perceive their environment and especially how they navigate through tissue during arthroscopic surgeries, we investigate the distribution, analysis, and comparison of gaze behavior with a mobile eye tracker during live surgery. We want to know if, and when which, features in their gaze behavior can be found that differentiate different classes of expertise. Both papers are based on the same study. The work of~\cite{yin2020study} can be seen as a pilot study, where we focus on typical measures like area of focus distribution, cognitive load, and the classifiability of expertise with common gaze features. In~\cite{hosp2021differentiating} we focus more on the classification as we dig deeper into the classification of three groups of expertise. To further understand perceptual differences we focus on the developmental steps between the expertise groups, too. In both studies, we use the same recordings of the subjects as the data source. These findings are meant to be a base diagnostic system for a future training system that helps to understand the gaze behavior of surgeons in general and improve the education of young surgeons.

\begin{figure}[ht]
	\centering
	\includegraphics[width=1\columnwidth]{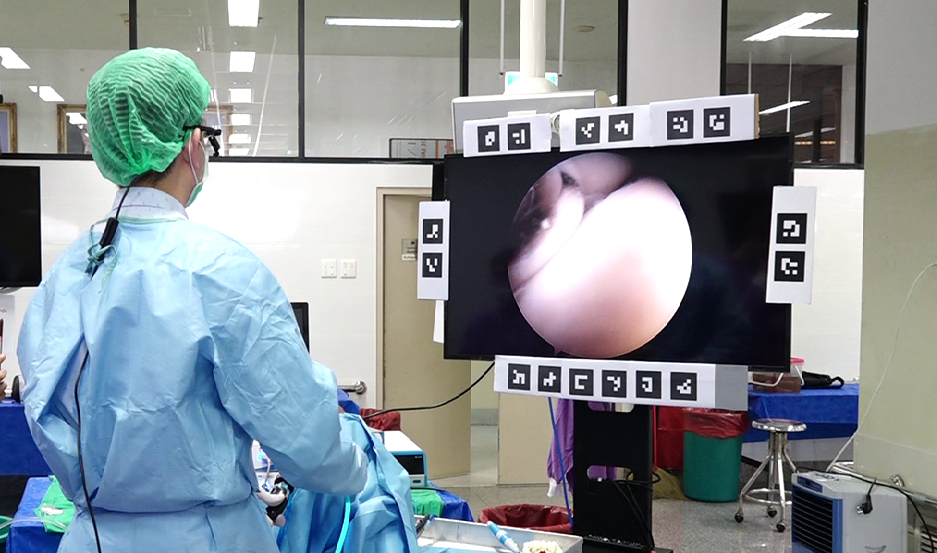}
	\caption{Surgeon during data collection looking at the output screen of the arthroscope. }~\label{fig:surgeonPerspective}
\end{figure}

\subsection*{Methods}

\begin{figure}[ht]
	\centering
	\includegraphics[width=0.45\columnwidth]{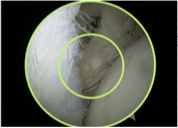}
	\caption{Visualization of inner and outer circle on the arthroscope's output screen. }~\label{fig:surgCircle}
\end{figure}

The gaze signal of 15 surgeons was recorded during a surgery on a soft-cadaver at the Ramathibodi Hospital in Bangkok, Thailand. There were n=5 experts with multiple years of experience, n=5 4th-year residents and n=5 3rd-year residents. All of them had to navigate from the portal on the shoulder to the operating site near the tendon of the shoulder, using an arthroscope (see Fig. ~\ref{fig:surgeonPerspective} for an overview of the scene). The surgeons had to tell when they were reaching one of 12 landmarks, which are placed on the way to the operating site. Starting from a rather general perspective, the visual attention of surgeons from different expertise levels was first focused on. ArUco markers are attached around the output screen of the arthroscope (4k, 52-inch, placed 4 feet away from the surgeon) to detect it and the circular video feed of the arthroscope therein, from the field camera video of the eye tracker (Tobii Glasses 2). The fixation patterns of experts and novices in the inner and outer circles of the arthroscope video are investigated, in order to state differences in the distribution of focus (Fig.~\ref{fig:surgCircle}).  In the next step, the importance of confusion detection during navigation is emphasized. While there are plenty of visual clues within the joint to detect landmarks, novices often miss to diagnose the correct target landmark and thus, report times of confusion. Previous studies have already proven the connection between confusion or disorientation with change in pupil diameter and head movements. With the percentage of change in pupil diameter (PCPD) an objective measure of cognitive load is used. For example,~\cite{kruger2013measuring} found higher cognitive load associated with higher PCPD values. Higher cognitive loads are suspected in times when a novice is confused, too, and considered the end of the last landmark to the beginning of the search for the next landmark as the baseline with low cognitive load, as the surgeon is not navigating during this time phase. The PCPD is computed by subtracting the average diameter of the confused time from the baseline diameter and dividing it by the baseline diameter. 
In a further investigation, whether there are differences in expertise visible in common eye movement features will be inferred. For a deeper analysis, the importance values of each feature is used to create a ranking (MRMR and chi-square-test), which tells us about their impact on the accuracy of new data. To infer more information about the differences of expertise reflected by their gaze signal which are considered to help to improve the understanding of the differences, the subset of features that ranked the highest are looked at and their characteristics are focused on as an explanation for evolutionary steps between classes of expertise.

\subsection*{Results}

\subsubsection*{Area of focus}
80\% of the navigation process from one landmark to another is considered to belong to a general area search. The remaining 20\% are considered to belong to a zeroing in or fine adjustment process. During general area search, experts, as well as novices, tended to focus on the outer circle by a ratio of 2:1. The differences can only be seen in the fine adjustment phase of navigation, where experts shifted their focus to the inner circle by a ratio of 2:1, but novices still focalize on the outer circle area with a roughly similar ratio as before. This finding is interpreted as an indication that experts adjust their attention according to the portion of the navigation task, as they know how close they are to the desired landmark, while novices might not be able to tell precisely.

\subsubsection*{Cognitive load}

Confused novices took up to two times longer than other novices to diagnose landmarks. In terms of time taken to accomplish the surgery, experts needed the least amount of time with the least standard deviation in task times. The pupil diameter changed during times of confusion by 1.02\% (left eye) and 1.12 \% (right eye) to the baseline. 
A much more extensive change can be seen in the gyroscope and accelerometer values.

\subsubsection*{Classification}

For a classifiability test typical metrics are used that were often used in medical studies like fixation rate (Hz), saccade rate (Hz), fixation duration (ms), saccade duration (ms), and average time to first fixation (ms) ~\cite{kocak2005eye,tien2015differences,richstone2010eye,merali2017eye,atkins2013surgeons}. In total, there are 12 metrics used (inclusive AOI intersections) that are extracted from the gaze data. Out of 15 subjects, two had an erroneous gaze signal from time to time, which would contaminate their overall statistics. Especially absolute statistics like fixation and saccade rate are not any more representative features of the samples of such subjects. Therefore, these samples were not used. 

Depending on their gain ratio, the features with the five highest values for our classification model are picked. Our logistic regression model shows a great accuracy of 84\% in classifying experts and novices. Only one expert has been misclassified as a novice, as their fixation rate is highly similar to that of novices and one novice has been misclassified due to similar characteristics to the experts in time to the first fixation.

Since the common eye-tracking features from the binary classification did not work for a ternary classification, alternative approaches for feature selection were searched for. For the further investigations on the classification of three groups of expertise, the importance of each of the features is calculated (with MRMR/chi-square-test) during classification and thereby get their impact on the accuracy. With a small amount of four features, one can classify the three classes of expertise with 76.46\% accuracy. Why a reduction of features is meaningful, is extensively discussed in Section~\ref{sec:disc_soc1}.

\subsubsection*{Latent expertise features}

Finally for the sake of explaining differences between expertise groups the following features were found most important to differentiate the two groups of novices and experts.

\begin{itemize}
	\item Average time to first fixation (ms)
	\item Fixation rate (Hz)
	\item Fixation rate AOI inner circle (Hz)
	\item Average fixation duration (ms)
	\item Average saccade duration (ms)
	
\end{itemize}

For the deeper analysis of the classification of three groups of expertise, the following features are found to be of interest:

\begin{itemize}
	\item standard deviation of saccade peak velocity
	\item minimum saccade amplitude
	\item total saccade amplitude
	\item min gyroscope z 
	
\end{itemize}

The binary model as well as the ternary model perform well and can therefore be well used to classify expertise. A deeper look at the characteristics of the four most frequent features from the ternary model shows that experts tend to have a more uniform distribution of peak velocities of the saccades. Thus, one interpretation of this result is that experts have a more structured speed behavior, which is more like a fixed scanning behavior. 
Higher values in the variance of the saccade peak velocity could signal a more chaotic gaze behavior, but it is hard to draw a conclusion. 
When looking at the minimal saccade length, a similar ratio is visible. Experts tend to have bigger minimal saccade lengths compared to intermediates and novices. In both features, the novices are between the experts and intermediates, which is atypical. Especially because the total amount of saccade amplitudes shows a reasonable evolution. Experts have a lower value than intermediates (who do more than twice the experts) and novices. Novices' scan path is more than five times higher than the scan path of experts and nearly double the scan path of intermediates. The last important feature is the gyroscope minimum measure on the z-axis. Z-axis can be seen as the movement from left to right of the head. Here, again, one sees an atypical behavior as intermediate subjects have the lowest minimum value followed by the experts and the novices.

\subsection{Conclusion}

To conclude the chapter about gaze to expertise linkage, the previous two papers are looked at individually and combined. 

The work on soccer data set shows how important idiosyncrasy is, when distributing samples of subjects on training and testing data set. We could confirm that a mass of gaze behavior features underlies a certain idiosyncrasy, which leads to overly positive results, when not taken into account properly. Further, the results could show, that the differences in gaze behavior between subjects from the same class are much smaller than the differences between subjects from different classes. In the work on surgeons, indicators of different search behaviors were found. While experts tend to have an optimized search behavior, novices seem to have more problems, thus, their behavior is more chaotic and less precise. This corresponds to the cognitive load values of novices and the completion times, which were two times higher than for experts.

Both data sets allow an objective, reproducible and robust classification of expertise, which can be seen as the basis for a diagnostic system. To robust such a system, more data of more subjects is needed. While the results of the work on soccer mainly highlighted the importance of different features describing the smooth pursuits, the results of the work on surgeons found a high influence on the length of the saccade amplitudes. However, the feature sets of both have one feature in common. Both pronounced the difference in the standard deviation of the saccade peak velocity. This feature can be understood as the variation of peak velocities of the saccades. A small value would show a homogeneous behavior, while a high value would show more chaotic behavior. In both data sets, one could see that experts have a much smaller standard deviation of saccade peak velocities. This means their behavior is less chaotic. This difference leads to the question of whether there are more commonalities between experts of different fields or if it was found by chance. This will be illuminated in the next chapter.

\newpage

\section{Cross-Domain Generalization}

When talking about perception and perceptual expertise, one usually talks about it in certain limits like domain or task. So far, there has been no proof, that perceptual expertise is restricted to a domain. Thus, perceptual expertise can also be some kind of domain-independent talent, which is considered to be one aspect of a successful generalization. In the first work presented in this chapter, the question asked was whether a subset of features can be found, that - applied in different domains - elicits expertise domain independently. The same feature set has been used to infer expertise classes by training a model with one data set of one domain and testing the model with unknown data from the data sets of a second and third domain.

However, generalization has multiple aspects. Another investigated research question was whether one can remove arbitrariness out of the way of finding spatial and temporal features by focusing on simple features like fixation image patches. This would allow a certain degree of automation from which generalization could benefit, too. It should be possible to apply our approach to any other domain where some kind of scan path can be created to infer expertise classes and differences.

\subsection{Cross-Domain Expertise-Related Gaze Features}

\textbf{B.W. Hosp}, F. Schultz, O. Höner, O. and E. Kasneci. "In the Search of A Superior Gaze Behavior: Cross-Domain Shared Expertise-Related Gaze Features." 
Submitted to: ACM Symposium on Eye Tracking Research \& Applications (ETRA '22).

\subsection*{Motivation}

So far, there is no proof that a superior set of shared features exists that explains expertise on more than one domain or task.  The following work focuses on the research of commonalities between different domains. To prove that there is a superior gaze behavior that is valid in multiple domains, there needs to be more research on cross-domain expertise, but one step into this direction has been done in this work by using a uniform, objective, and robust way of the feature selection process and apply this approach to at least three data sets of different domains or tasks. Such a subset of features would allow general statements about perception to be made, independent of domain. This approach needs to be defined and applied to find cross-domain but class-related expertise differences, reflected by a set of features. The current scientific understanding of perceptual expertise is 
mainly domain- or task-related, but there is no proof that these limits exist. 
The contribution of this paper contains the investigations on the generalization of 
perceptual expertise, where indicators are presented that cross-domain 
commonalities exist.

\subsection*{Methods}

In the first step all the gaze data from three studies that are accessible are collected. Data set A contains the samples of 33 soccer goalkeepers from the study in Section~\ref{sec:soc1}. Data set B contains all the samples from 15 subjects from the study in Section~\ref{sec:linkSurgeon}. Data set C is coming from a more static task. In data set C data of 58 dentists was collected during an OPT analysis. 
The fourth data set D contains data of 28 subjects, that are similar to data set A. Instead of a goalkeeper perspective, subjects from study D were virtually placed in the center of a soccer field and had to remain overview over 360° in virtual reality. Data set A and D are combined to A* since it contains data of highly similar tasks. Each of the data sets contains subjects that were defined as experts (based on years of experience or being picked by talent scout), novices (beginners in the field or no experience in the task), and intermediates (loosely defined as in between, with more experience than novices but way less than experts). 
As not every data set was captured with the same model of eye tracker nor vendor, all the features from all data sets are looked at and a subset of features that are shared by all of the data sets is defined. In the next step, the data is split to experts, intermediates, and novices in each data set. A balanced training set of a randomly picked data set $x \in \{A^*, B, C\}$ is defined and a bagged tree model is trained. In this first model, the feature selection from Section~\ref{sec:soc1} is used and the features are ranked by their importance for the model during the cross-validation. With this new subset of features data sets $ y \in \{A^*, B, C\} \setminus x $ are used as the test set.

\subsection*{Results}

\subsubsection{Classification performance}

After the feature selection process, the following features remain that have the highest impact on the classification and are therefore picked as candidates for the subset of features that might be shared by all the data sets.

\begin{itemize}
	\item maximum saccade peak velocity
	\item maximum fixation dispersion 
	\item standard deviation of saccade peak velocity
	\item maximum saccade amplitude
	\item minimum smooth pursuit dispersion
	
\end{itemize}

With the mentioned features an accuracy performance of 58\% was achieved. This sounds quite low, but this is a three-class problem. Thus, the chance level of picking the right class is 33.33\%. With 58\% the accuracy is slightly worse than the doubled chance level. An accuracy of over 66\% would lead to the fact, that single samples might be classified incorrectly, but the majority is classified correctly. Therefore, also the majority of a subject's samples are classified correctly and subsequently the subject in total, too.

Looking at the two data sets that were classified, the dentists' data set had a total classification accuracy of only 29\%. The intermediates were classified with 7.7\%, the experts with 35\%, and the novices with 45 \%. Thus, the dentists' data set is slightly worse than the chance level, and therefore, the most optimal features for that data set might not be found. Another reason for this classification might also be the totally different task of static diagnostics. There were restrictions on head movements as this study used a remote eye tracker while stimuli had been shown on a screen.
On the soccer data set, which task was much more similar to the surgeons, accuracy for the novices reached 83.4\%, 0.5\% for the intermediates, and 82\% for the experts. Again, because of the misclassifications of the intermediates, the average accuracy is at 60\%. All in all, with the mentioned features a model is trained with one data set and the two other data sets are classified with an accuracy of 58\%.
On a deeper look at the single classes on the combined test set (soccer and dentists), one can see that the novices were nearly optimally detected (92\%), the intermediates with 3.2 \% not at all, and the experts still with an accuracy of over 79\%. From 100 runs 34,700 samples were correctly classified as novice and 3,000 incorrectly as an expert. This is no problem, as it is known in expertise research there are subjects acting better than their initial classification. More problematic is the amount of samples that belong to the expert class but is classified as novice or intermediate. In 100 runs 30,000 samples were classified correctly as expert samples. 4,300 samples incorrectly as a novice, and 3,400 samples as intermediate.  At first, these results look complex to understand, but a closer look at how the intermediates are defined reveals the ambivalence of these results and a weak point in the classification. This will be discussed in the corresponding discussion section of this paper (Section~\ref{disc:cross1}).

\subsubsection{Shared, latent expertise features}

Having a deeper look at the features and their characteristics, one can see three important correlations. As the data was normalized based on each data set individually, the values can be positive as well as negative. For comparison, this is important, as the correlations are only visible there.
The surgeons' experts e.g. have a maximum saccade peak velocity of -221.560 °$/s$, followed by the intermediates with -0.7267 °$/s$ and the novices with 222.287°$/s$. Comparing the values with those of the soccer players, one can see that the experts also have a highly negative value of -593.31 °$/s$ followed by a high value of 234.56 °$/s$ and an even higher value of 1211.377 °$/s$ for the novices. Soccer players show more or less the same trend between the expertise classes. In the data set of the dentists, this trend is not visible. Only experts and novices show similar values, thus, intermediates will be misclassified as novices (their values correspond much closer to the novices). For the dentists a correlation between the trends of the standard deviation of the peak velocity was found. The dentists as well as, the surgeons follow the same trend (experts: ca. -15°$/s$, intermediates: ca. 6.5 °$/s$, and novices: ca. 10°$/s$). Here the data of the soccer players do not fit at all. A feature whose values correlate with both other data sets' experts and novices, is the minimum smooth pursuit dispersion. The values for the expert groups are slightly positive (0.019 to 2.25 pixels), while the values of the novices are slightly negative (-4.8 to -0.15). Only, again, the soccer players' intermediates correlate with the surgeons by being negatively close to zero.

\newpage
\subsection[A Deep-Learning Approach for Feature Selection]{Expertise Classification of Soccer Goalkeepers in Highly-Dynamic Decision-Tasks:  A Deep-Learning Approach for Temporal and Spatial Feature Recognition of Fixation Image Patch Sequences}

\textbf{B. W. Hosp}, F. Schultz, E. Kasneci, and O. H{\"o}ner. “Expertise classification of soccer goalkeepers in highly dynamic decision tasks: A deep learning approach for temporal and spatial feature recognition of fixation image patch sequences,” Frontiers in Sports and Active Living, vol. 3, p. 183, 2021.

\subsection*{Motivation}

Although recent research focuses on behavioral features, there is a lack of understanding of the underlying cognitive mechanisms. First and foremost because of missing adequate methods for the analysis of complex and high-speed eye-tracking data that go beyond accumulated fixations and saccades. The latest research signifies that, until now, there is no feature set that allows general statements to be made, not even in the same domain. In fact, if there is no manually picked superior feature set that yields high-performance results, a rational step would be to use a learning algorithm to find features automatically, without a human in the loop. Hence, we investigate a different way of spatial and temporal feature recognition by using fixation image patch sequences. This approach removes arbitrariness and manual feature selection totally out of the process of defining predictor variables as the foundation of classification. A comparison of the automated feature selection versus a manual feature selection is done in the discussion of this work (Section~\ref{disc:gazePatchnet}).

\begin{figure}[h]
	\centering
	\includegraphics[width=1\columnwidth]{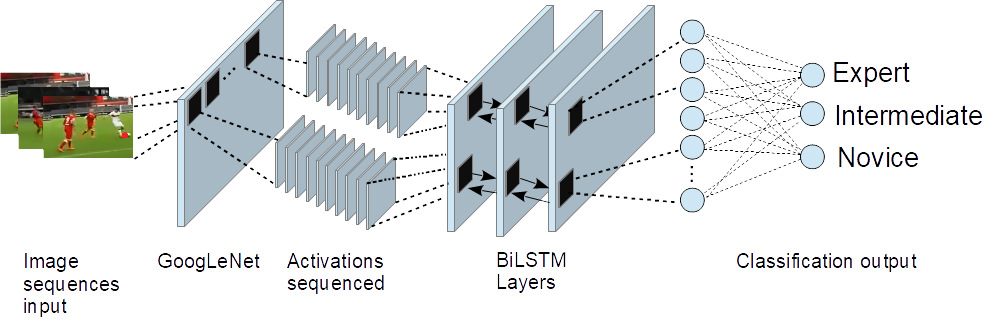}
	\caption{Pipeline model of the classification network. }~\label{fig:dnn}
\end{figure}

\subsection*{Methods}

Our method includes the finding of latent features (hidden in the image patches gazed at during fixations) and subsequent classification of these patches as consecutive sequences. The aim is to predict three different expertise classes. For that, the same data as in Section~\ref{sec:soc1} and all the fixation data is used to cut out the image patches, the subjects were gazing at during a fixation. For training, the images were augmented in several steps to adapt training to a realistic range of samples. The normal and the augmented samples are fed to a CNN (GoogLeNet) to find latent spatial features in the fixation patches. The procedure is illustrated in Fig.~\ref{fig:dnn}. To be able to do this, transfer learning is done on the GoogLeNet network, as this network is trained to recognize over 1000 classes of objects. The last layers after the last pooling layer is removed and added our BiLSTM network as well as a final three-class classifier to it. In each run 70\% of the data belong to the training set and 30\% to the validation set. All data of one subject is totally held out (hold-out-validation) to test our model with unseen data. One sample represents one video trial of one subject. Thus, the classification of expertise of subjects is looked at indirectly, instead, each sample is focused at. This means that some samples of the same subject can be detected to belong to different expertise classes.

\subsection*{Results}

The average classification accuracy reached 73.11\% over 33 runs. The accuracy of predicting a novice correctly as a novice is at 55.5\%. While only 166 samples out of 1,816 samples in total were classified as expert samples, 650 novice samples were classified as intermediate samples. For the intermediates, there is a similar data situation. Out of 1,605 samples 119 were classified as expert samples and 372 as novice samples. 1,114 samples were correctly classified as intermediate samples, which corresponds to an accuracy of 69.4\%.
The best recognizable group is the experts. With 15 samples being classified as novices and 30 samples as intermediate, a very large majority is classified correctly as expert samples. The average accuracy of detecting expert samples correctly peaks at 93.4\%.

\subsection{Conclusion}

The previously presented papers showed two different views on generalization. The first research question asked was whether one can find a subset of features that elicits expertise in a domain-independent manner. With the three data sets A* (containing gaze behavior of soccer goalkeeper and field player perspective in omnidirectional videos), B (containing gaze behavior of surgeons during a live arthroscopy), and C (containing gaze behavior of dentists during OPT analysis in 2D), indications were found that the similarity of the task seems to be important. Our investigations showed that data of a static visual search task can not be properly classified, as where the classification of data from a highly similar dynamic task shows much higher classification rates. The surgeon and the soccer data set had much more in common. Dynamics (video), kind of eye tracker (able to look freely around or limited by a screen), and task (find a proper way to continue) are the most important. Therefore, with the current data situation, perceptual expertise cannot be stated to be domain-independent. The results only allow stating that expertise might be domain-independent as long as the task is familiar. This is well in line with~\cite{gegenfurtner2013transfer}, who showed that it is possible to transfer perceptual expertise from familiar tasks to semi-familiar tasks but not too unfamiliar tasks.

Regarding the data of the two familiar tasks, one can see that there are strong correlations. Obvious trends across the classes for both data sets are the smallest maximum peak velocities of saccades of the experts, followed by intermediates, and then novices with the highest value. For the unfamiliar task (dentist data set), one sees correlations between the experts and the novices, in the maximum peak velocity of the saccades, as well as in the standard deviation of the saccade peak velocity, but they were not strong enough to build a robust classification. Regarding a correlation between all data sets, one can see that the minimum smooth pursuit dispersion might be an indication that perceptual expertise can be domain-independent.

Our second research question was to investigate whether one can remove arbitrariness out of the way of defining features by focusing on simple features like image patches. It can be stated that deep learning, especially the combination of a CNN and an LSTM network is able to define and find spatial and temporal features independently, without the need for manual work. Our approach on the classification of fixation image patches shows high accuracy values, that can compete with traditional machine learning methods. The results might also be improved by adding more data on more subjects.

\newpage
\section{Gaze-Based Assistance Timing}

\subsection[States of Confusion during Arthroscopic Surgery]{States of Confusion: Eye and Head Tracking Reveal Surgeons' Confusion during Arthroscopic Surgery}
~\label{sec3:conf}

\textbf{B. W. Hosp}, M.S. Yin, P. Haddawy, R. Watcharopas, P. Sa-ngasoongsong, E. Kasneci. "States of Confusion: Eye and Head Tracking Reveal Surgeons’ Confusion during Arthroscopic Surgery." In Proceedings of the 2021 International Conference on Multimodal Interaction (ICMI ’21), October 18–22, 2021, Montréal, QC, Canada. ACM, New York, NY, USA. \\ 

\subsection*{Motivation}

The use of eye-tracking methods to detect other cognitive processes besides expertise and cognitive load may be of particular importance for training systems that reduce training time. Online detection of confusion can contribute greatly to this. This is because if confusion can be detected in real-time during a task, then based on this, targeted assistance options can be applied to facilitate or even enable the continuation of the task. Targeted temporal detection of confusion can thus provide a basis for a training system. In particular, in arthroscopy, distracting information should not be shown on the screen during normal operation. Thus, a precisely timed view of auxiliary possibilities is directional with respect to perceptual-cognitive training systems. Young surgeons can thus shorten their training, which until now always required a supervisor to provide assistance in case of confusion. Optimal recognition of the right time for assistance can thus, on the one hand, accelerate and improve training and, on the other hand, save hospital resources through digital possibilities. To find an optimal timing for supportive measures, one has to focus on the detection of confusion. Confusion needs to be detected quickly and relatively accurately to signal the time when supportive actions need to be taken to assist non-experts in their learning process. The following work focuses on an online, machine learning-based approach to confusion detection in surgery. The findings of this work can be used to create training scenarios that not only optimize training for novices but also reduce the number of training hours that an expert must instruct. In addition to confusion detection, our goal is to create a fast model that can be used online. We want to predict confusion in real-time using a minimal set of features obtained from an eye tracker during surgery with an arthroscope. A long-term goal is to apply this method in an intelligent training environment that provides optimally timed assistance through temporally and spatially placed visual cues.

\begin{figure}
	\centering
	\includegraphics[width=1\columnwidth]{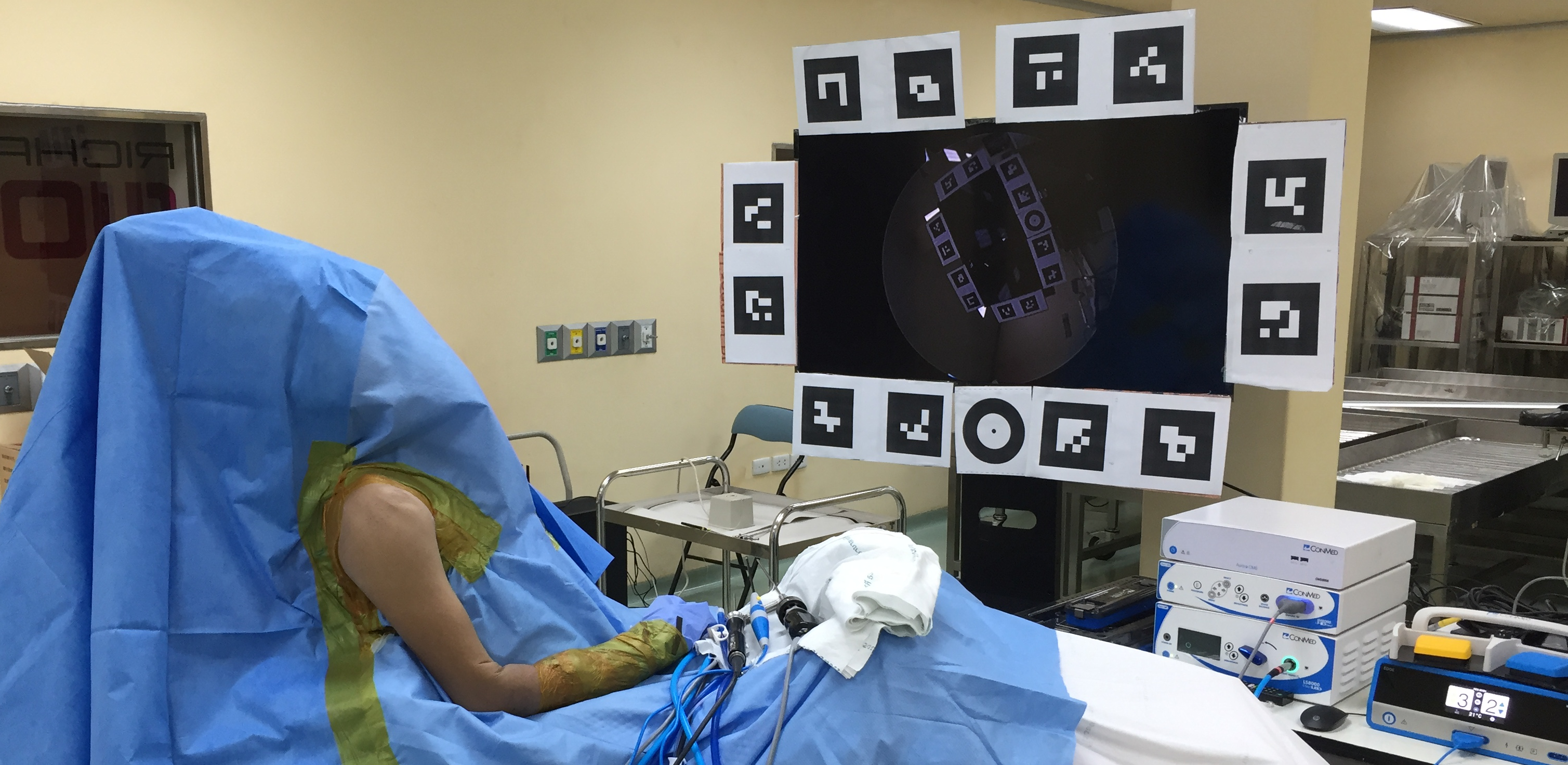}
	\caption{Side view of the operation side with covered cadaver (left) and arthroscope output screen (right). }~\label{fig:surgeonSetup}
\end{figure}

\subsection*{Methods}

For the purpose of confusion detection, the data from Section~\ref{sec:linkSurgeon} were used. There are have six novices that reported confusion during the surgery. For each of the moments, the samples from one second before until one second after the reported event were picked and were manually labeled as "confusion event". All other samples were labeled as "no event".  
Each sample contained the following features:
\begin{itemize}
	\item point of regard (x, y)
	\item pupil position (average of both eyes)
	\item gyroscope (x, y, z)
	\item accelerometer (x, y, z)
\end{itemize}

To build a random forest model, the samples are split into training and testing data set. Out of 1,266,758 samples, there were had 7,103 samples with a confusion event and 1,259,655 samples with no event. Out of these samples, 7,103 confusion samples and 7,103 no confusion samples were collected. In every run, 1,000 samples of both were randomly picked to predict their class. The other samples were used for training.
2/3 of the subjects were randomly picked for training and counted the number of confusion event samples for each. Afterward, the same amount of "no event" samples from the same subjects was collected. This means for our training set there was the same amount of confusion event samples as no event samples. This leads to a balanced training set (50\% confusion event samples and 50\% no event samples) and to a random baseline classification accuracy of 50\%, which allows for easy interpretation of the results later.
As the model is planned to be used in an online fashion, the classification accuracy (with unseen data) needs to be tested, as well as the classification speed. For that, a queue of n=2,000 samples is created. As the system was developed on saved data, the data were constantly read row by row. In each moment, there were n =2,000 samples in the queue, which represent one sequence. Every time a new sample is added to the queue, the oldest sample gets kicked out.
Subsequently, the average of the features of the samples inside the queue is computed. These values represent the current content of the queue, which is called delta sample. This delta sample is now given to the trained random forest model to classify it as a "confusion sample" or a "no confusion sample". To infer the average performance time, the computation time of 100 single runs is measured and the average performance time is calculated.

\subsection*{Results}

The average accuracy of the random forest model is 94.2\%. According to the accuracy, the average misclassification cost/loss is 0.088. The optimal loss value for the test approach is reached at ~49 trees with a misclassification cost of ~0.085. In total, there are ~50,000 samples for each class. Of class 0 (no event), 47,016 samples (93.8\%) out of 50,136 samples were predicted correctly and 3,120 (6.2\%) incorrectly. Similarly, for class 1 (confusion event), the model predicted 47,023 (94.3\%) samples correctly as confusion events and 2,841 (5.7\%) incorrectly. To measure the performance speed of the model, the computing time of each of the 100 runs is measured. On average each prediction takes 0.039 seconds. This corresponds to a frame rate of ~25 fps.

\subsection{Conclusion}

With the available data, one is able to reach high accuracy on the detection rate of over 94\%. This value and the sufficient speed of 39 ms are high enough to use this model in an online training system for young surgeons. In the next step, these assistive options need to be evaluated. 
Arthroscopy is one of the tasks that could benefit a lot from such a training system. In arthroscopy perceptual problems occur often, that are hard to explain verbally but are easy to solve digitally on the output screen. With this online system, these digital solutions can be applied directly. As such, there is no need for a supervisor to stand next to the trainee, as a correctly timed detection can remedy.

\chapter{Discussions \& Outlook}
\label{chapter:disc}


This chapter deals with the discussion of the findings of the papers discussed in this work with regard to application in practice and generalizability. To allow to expand a diagnostic system to an application and thus as a foundation for a training system in practice that is able to improve the perceptional skills of a subject correctly, there are several important aspects that have to be considered beforehand. A diagnostic model must not only correctly classify the right skill level of the subjects, much more important is a low false-negative rate. In the case of low-ranked subjects, those will become better at some point and therefore, rank higher as before. Thus, the diagnostic, which is necessary after each training run, needs to be sensitive to such changes in expertise. However, experts that have been classified as experts already, usually stay in the expert class. Sometimes even experts perform worse than their class average, but this is assumed to happen rarely. Thus, the false-negative rate  (expert being classified lower as intermediate or novice) is an important metric that needs to stay low and has to be observed. The main aspects of a diagnostic system are:

~\begin{itemize}
	\item \textbf{the classification accuracy is as high as possible.} Based on findings from statistics, a system that has a prediction rate of 100\% is not possible, as up to a point, human gaze behavior underlies idiosyncrasy~\cite{holmqvist2011eye}, which implies a certain degree of difference. Therefore, the aim is to get as high as possible. This will only be able at a certain point where most of the variations within an expertise class have already been fed to the training of the model. Otherwise, the diagnostic system might fail to at least some new data. However, there is no security as when this point is reached, as even a human classification cannot guarantee 100\% accuracy. Therefore, even after a long time of testing, the system will still be at a certain risk of misclassification which is tolerable but needs to be kept in mind.
	
	\item \textbf{false-negative rate needs to stay low.} This is much more possible and is based on the sharpness of the decision boundary between the classes, at least up to a certain degree. It is unlikely that every subject shows consistently high performance. In fact, it is assumed that there are some false-negative predictions. However, a proper diagnostic system can bypass the problem of rare false negatives by simply averaging the results of all runs for one subject. Thus, a system needs to tell the current classification of the current training run (the current stimulus) and averaged metric overall training runs (all stimuli), which whitewashes some rare false-negative results, that describe the total performance of a subject better.

	\item \textbf{continuous re-training of the model.} After a diagnostic run, the data of such need to be fed to the diagnostic model to help to improve the sensitivity of the decision-boundary. This shrinks the risk of misclassifications very much.
	
\end{itemize}

The classification accuracy, as well as the false-negative rate, can already be described with the current state of the diagnostic models addressed in this work. The aspect of continuous re-training is not, as for this the system needs to be used frequently. As such, it is the task of future users. Therefore, the presented papers will be discussed on the first two points. As soon as these aspects are sufficiently addressed, the next step will be to define support options. 

\section{Gaze-Based Expertise}
\label{sec:discGazeLink}

\subsection{Soccer}
~\label{sec:disc_soc1}
The results of the expertise detection in soccer goalkeepers lead to several conclusions. Firstly reached several milestones from the aforementioned aspects of a diagnostic system were reached, and secondly the process of understanding what it takes to perform well was simplified. Algorithmically the process of feature selection was improved, which is important at the beginning of every classification procedure. One important note is that it is vital to assign samples in a person-specific manner. Idiosyncrasy is a dead end, as it suggests high-performance classification, but fails completely in the prediction of new unknown samples. This is fundamentally based on the relationships of the differences between different groups. Samples of the same person have the highest similarity (idiosyncrasy). Samples from the same group come next as the differences across experts are shown to be smaller than the differences across groups of expertise (expert variation). The accuracy of the detection rate allows - with over 78\% - an application in practice. There is still some scope for optimization, but at its current state, the model is already quite performant. It is common in expertise research that subjects, assumed to rank low, might surprise with high talent and therefore get ranked in higher classes. Foremost, this can be counteracted by relying on strict rules when defining each of the classes. This is especially true for the novice class. Since our assignment is based on a relative rule (no experience in competitions and no training on a regular basis), which allows the feature characteristics of the novices to spread on a wide area but lead to strict decision boundaries for the intermediates and experts. A strict differentiation within the novices might reveal further differences in gaze behavior between novices with no experience at all and novices with experience being a long time ago.
With the current status, the expertise of novices has never been assigned professionally. Thus, the recognition of the novice group might be harder than the alternatives. This is especially evident in the false-negative and false-positive rates. 18.6\% of the novice samples were classified as intermediate samples, but only 1.6\% of the intermediate samples as novice samples. A portion of low performers usually can be found in higher-performing classes. And as there is no ground truth for novice expertise, the classification of a novice is considerably more difficult. A much more important finding is, that our model fundamentally has an extremely low false-negative rate, thus, only a few subjects are wrongly switched from higher to lower expertise classes. Which is one of the main aspects of a diagnostic system. 

Further, for a reproducible and objective way, the way how features get selected as predictor variables is essential. Features can have different influences on accuracy. Feature selection is done in order to optimize the accuracy of the predictions and by selecting only a subset of features, a dimensionality reduction is achieved, too. With less dimensionality, the problem is becoming less complex, and computational power and speed can be saved. But the methods of feature selection can have other purposes, too. A reduced feature set can help to avoid overfitting of the model. Using fewer features reduces the risk of the model memorizing certain training examples. Likewise, fewer features can improve the interpretability of the model, as the affection of certain features can be identified. Thus, our model reached another milestone: real-time operability. With a low number of features, the computation time is held low, which speeds up the whole process. For the use as a diagnostic system, we assume that only the classification accuracy needs to be improved further. This can be achieved by collecting more data from more subjects to robust the model against outliers. 

The simplification of the understanding of gaze behavior is based on the average characteristics of latent gaze features that represent each of the expertise classes. Novices tend to do a lot of small movements with their eyes. We interpret that behavior as a signal of tension or nervousness, as they have less experience and try to perceive every possible change in the scene around them as fast as possible. These results are not final but can be used to teach a more planned scanning behavior to novices. The characteristics might even change as soon as there is enough data collected from further subjects, because the lower the number of subjects, the more outliers make a difference. 

In summary, the current state of the model allows the usage as a diagnostic model in practice, because both aspects have been addressed sufficiently. The accuracy will usually grow higher as soon as more data is available. Thus, continuous re-training will considerably be important. For the use in a training system, one first idea is to use the general rules from the physical training of the specific domain. At different states of expertise, there are different aspects of perception that are being trained. This can be the perfect timing for shoulder glances, finding and utilization of free spaces, or just an optimal decision after a pass.

As the whole system is data-driven and only based on the gaze signal, the model can be used in other domains, too. The way how features are calculated and evaluated to be part of the classification, as well as the classification process is highly automated and as such, the pipeline to build the model provides high generalizability. The only requirement is that the gaze signal can be obtained online or at least from a file. Further details about the generalizability of this approach can be found in the work in Section~\ref{sec:discGeneral} and Section~\ref{disc:cross1} where the same approach is applied to data sets from different domains.

\subsection{Orthopedics}

In the field of arthroscopic surgery in orthopedics, several indicators were found that allow the separation of experts, intermediates, and novices. First of all, novices need much more time (up to two times) to solve the same tasks as experts. This is also reflected in their gaze behavior. During the initial search phase where subjects navigate to the main area where the next landmark is placed, novices were acting the same as experts. The difference is in the last 20\% of the search, where novices had problems in fine-tuning the arthroscopic camera to the correct place. Thus, they needed more time to reach the landmark. Novices were also the only group that reported confusion. During such a state of confusion their cognitive load grew, which could be shown with the PCPD of about 1\% compared to baseline. The typical metrics of gaze behavior could only lead to an understanding of the differences between the experts and the novices, but not any further. Since a three-class classification is considerably more complex, investigations in another way of feature selection are needed. With the use of MRMR methods and chi-square-test from Section~\ref{sec:soc1}, differences between the three groups were found and at least 76.46\% of the differences could be explained.

Regarding the two points raised earlier about the application of the model in practice, the model is quite powerful, but to improve the training time of young surgeons the performance needs to be improved further. A miss rate of 23.54 \% is still too high for an application in practice. On the plus side, this is only based on 15 subjects, which means there might be even other feature combinations that explain differences much better, but for a pilot study, the values are quite promising. To continue further, there needs to be more data collected from more subjects. A finer graded classification would help to understand the differences even further. With this study, the same challenges as with the soccer goalkeepers study are present, but as there is a really low number of subjects, fewer aspects could have been addressed. The only milestones that were reached with the current state of this work are to prove that there are differences between the classes, which can be found by looking at their gaze behavior, and a fast computable subset of features, which allows an online application of the system. 

Luckily, the four features are easily calculated, which would allow the usage of the classification as an online classification system. Though, the classification needs to be done segment-wise after a certain period of time, which needs to be investigated first. Further steps are to add more subjects to each class and refine the number of classes. This allows a much finer classification and therefore a better understanding of the differences between the classes. A finer classification is important to robust assumptions made by the model about gaze behavior and optimizes the recognition of class-specific weak spots to be used in a training system.

A high miss rate of 23.54\% symbolizes also a high false-negative rate, which would violate the second requirement for the application in practice. Similar to the model in Section~\ref{sec:disc_soc1}, continuous re-training of the model with much more subjects is essential here. However, until more data is available, no robust statement can be made about the application in practice. Again, the procedure to develop this model is highly generalizable as the same method are used for feature calculation, as well as feature selection and model creation as in the model of Section~\ref{sec:disc_soc1}. As far as generalizability is concerned, the same conclusion can be drawn as in Section~\ref{sec:disc_soc1}. However, due to the typical character of a pilot study which is based on a small number of subjects, there are limits to the informative value of the features used in this model in terms of their generalizability to a larger data set from the same domain and task. The features may change completely as more data is available.

\subsection{Outlook}

Although only one model from Section~\ref{sec:soc1} meets the requirements for use as a diagnostic system, the findings of both papers are far-reaching. Indeed, it can be said that eye tracking is in many ways well suited not only to detect perceptual differences but can even provide more profound information that different classes can be inferred from different gaze behaviors. Eye tracking is thus an essential part in the study of human behavior and will very likely continue to be able to provide important insights from the aspect of multimodal interaction between humans and machines. As mentioned before, one of these aspects could be a training system based entirely on eye tracking. Now, on the one hand, the models have to be made more robust by adding more data to learn from, and, on the other hand, the first steps have to be taken into the application. Since especially in the fields of soccer and surgery new ideas for improving training are constantly sought, such training systems meet a relatively large market, as soccer clubs and hospitals are increasingly turning towards digital possibilities.
Also in view of current trends in human-computer interaction, which aim at a personalizable self-diagnosis, this system would be an important building block, as the procedure performed in the two sections can be applied to other perceptual-cognitive processes as well since the procedure works independently of the training data labeling. To adapt the model to another perceptual-cognitive process, only the training data must be marked correctly. 

\section{Cross-Domain Generalization}

The understanding of expertise is mostly limited to a certain domain or task. Thus, it is assumed that perceptional expertise is it, too. However, what is if this is not the case? If perceptional expertise is something everyone can learn, cross-domain training systems can be build that help subjects to improve in several domains simultaneously. An ophthalmologist can provide perceptional expertise tests. They might be able to diagnose different levels of perceptional expertise and thus, the aptitude of a subject for a certain task or even diagnose diseases that hinder one to apply an optimal gaze behavior. Simple tests can be provided that can be included in several fields where perception plays a central role. Also, our understanding of perception might change completely. As visual perception can play a decisive role in soccer and medicine, highly specified training should become an essential part of education. The first steps in the direction of general perceptional expertise detection are discussed in the following.

\subsection{Expertise-Related Features}
~\label{disc:cross1}

One of these first steps is the search for shared features that describe the same expertise levels of subjects from different domains. With a model that is able to find commonalities between data sets from different domains, and thus, differences between expertise classes that are valid across domains, there is an important starting point for the search for generalized expertise. 

When looking at the single values of the features used of the different classes and domains, it can be stated that there are correlations between the classes and between a subset of domains. Correlations have been found in the maximum saccade peak velocity, the standard deviation of the peak velocity, and minimum smooth pursuit dispersion between subsets of the domains included. As not all domains share the same features equally, generalizability is hard to state. To answer the question about generalizability, first, the model needs to be trained with more data of the known domains (but equally from each class) and of foreign domains. However, with the current work, it is possible to find commonalities between different domains that reflect expertise in a general way. The findings suggest that it is worth continuing the search for general expertise features. There are two optional ways to go now.  If the search for commonalities between domains is considered as a ternary problem, the detection of the single classes, especially the intermediates, needs to be optimized first, as it is close to 0\%.  Therefore, in its current state, the model can not be used to build an application. The findings are too weak to function as a basis. If only novices and experts are considered and intermediates are disregarded, the model is performing quite well already. Both expertise classes show accuracy values more than doubled chance-level. Thus, they are sufficient to classify the majority of samples of a subject correctly and therefore reach a low false-negative rate, too.  So, by defining domain-independent visual expertise as a binary problem, the model can be used in an application to detect general visual expertise, but solely classified as novice or expert (If there is any use case where a binary differentiation is desirable).

However, in any case, it is astonishing that such fundamentally different domains and tasks have such a load of common perceptional features. In both cases, to robust the model, first of all, there need to be strict definitions of the single classes that allow a proper classification. In soccer, this can be the years of experience or hours practiced. It just has to be more precise than "has never participated in a competition". This definition is too vague.
The classification result of 3.2\% shows that intermediates from one data set are not equally skilled as intermediates of the other data sets. Thus, intermediates of one data set are considered as being better/worse and end up being classified as expert/novice. Only if the classes are clearly separable, the machine learning classification can perform well.  
Another step towards an application is to train the model with more data. This means more subjects from different expertise groups, but also more subjects in total. At the moment the classification accuracy is 58\%, thus,  slightly worse than double the chance level, which indicates there is still enough uncertainty that too many samples of a subject will get classified incorrectly. As soon as the model reaches over 66\% it is strong enough to classify the majority of a subject correctly. This would be sufficient to continue the search on general expertise features. So far, it is only know that there are common perceptional features between surgeons doing an arthroscopic surgery, soccer players in decision-making situations, and hints about some commonalities to dentists on a visual search task. The next major step would be to investigate even more domains and see whether the features found in this work can be transferred to even more domains. Our findings suggest considering some aspects. The task and the requirements for the technology (same kind of eye tracker, similar dynamic scenes, etc.) should fit each other. We assume the better the technology and the scene and task fit together, there are more commonalities visible in the gaze signal. 

\subsection{Deep Learning Feature Selection}
~\label{disc:gazePatchnet}

Another important step in the direction of generalized perceptional expertise is the way the data received are examined. So far, manual selection of features that are thought to be helpful is the main approach. Subsequently, machine learning techniques are more often used. One specific advantage is that a machine is not at risk of arbitrariness and as such, the generalizability of such techniques might be higher than with manual feature selection. Usually, machines select features because of their importance based on a calculation. Thus, removing the arbitrariness out of the process of feature selection.

In our current model, the accuracy of predicting an expert correctly is at 93.4\% as this class is the easiest to detect. The prediction rate of the intermediate class is much lower with an accuracy of 69.4\% because this class is supposed to be the hardest to detect. The accuracy, however, is more than double the chance level with about two-thirds of the intermediate samples being classified correctly. Much lower than the intermediates, the novices are predicted with an accuracy of 55.1\% which is nearly two times as high as the chance level but still ~11\% lower. The expert group is a pretty well recognizable group. The intermediate and novice groups are more heterogeneous as there are subjects that have more/less experienced than others. Another reason for this could be the missing metrics needed to divide between the two classes properly. This question is typically addressed with the availability of more data. The problem may stem from the small sample size of intermediate subjects as this group could be too small for the model to define robust decision boundaries. The fact, that expert samples were barely (15 samples) predicted to be samples of an intermediate player, shows that there are clear decision boundaries for the intermediate and expert classes.
Nevertheless, a long-term goal is to optimize the training for young players, whereas This study is the first step in that direction. For that, one needs to know which behavior is optimal and how one can design training steps for young players to reach this optimal behavior. The difference in active years/training, and therefore experience, between intermediate and expert subject, is much smaller and needs to be finer graded. As soon as the detection of the novices is more robust, the model is likely to become an application that is used in practice. 

Especially instead of providing a description of the behavior of different classes, this model describes a pipeline to find latent features by itself. This circumvents one problem: handcrafted features. The characteristics of handcrafted features may be difficult to teach a user in the form of new behavior based on feature values. Even if the optimal set of features is found, it is difficult to incorporate the findings into a training system. 

Conclusively, one can state that a certain degree of automation has been achieved in the process of feature selection. This improves the whole process as now features are used that are not only said to be meaningful, but that can algorithmically be calculated and explained. As such, the model shows high generalizability, as an only requirement the scan path of the subjects (independent of the task) as fixation image patches needs to be provided. The remaining pipeline is automated totally. However, a comparison between the manual selection in Section~\ref{sec:soc1} and the automatic one of this work, shows that manual selection performs superiorly. One explanation might be, that for the current state of the data set with a still-low number of subjects, the essence of what is important can not be robustly depicted in the fixation sequences. It might also be that the error of the eye tracker has a higher impact on the absolute values of the fixations than on the relative features picked in Section~\ref{sec:soc1}. At the moment, the manual selection is better suited, but sooner or later, at least when there is much more data available for training and eye tracker errors can be excluded, machine learning methods will pass manual selection. The correct data representation to feed them with are just not found.

\subsection{Outlook}

From these works, one learned that there are differences and commonalities in the classes across the data sets. Similar to the papers on classification one faces the same problem of defining expertise. Commonalities between experts and novices were found, which is a first step in the direction of understanding how expertise develops, but an understanding of intermediate subjects' visual perception lacks. Thus, one need to focus on strict definitions of each expertise class properly. The current definition is too vague. In fact, the cognitive factor is only one of the several factors that contribute to expertise. For goalkeepers, for example, it is still most important to be able to block shots on goal. If a goalkeeper can do this extremely well, they may be invited by the DFB and classified as an expert), even though they could make “worse” decisions after return passes. Conversely, it can also be the case that intermediate subjects are very good decision-makers, but did not hold as many balls, which is why they are not invited by the DFB. As a result, it is very important to not just test players from different classes but to test players with the assumed highest decision-making skills. The same situation exists in surgery and dentistry. The pre-classification as ground truth might classify a 4th-year resident as an intermediate because they are in the 4th year of the residency, but how much practice this surgeon has already had is not taken into account. In dentistry, the exact same problem is faced. As such, it is useful to search for commonalities and especially for general features of perceptional expertise, independent of the status of their education. By doing that, assistance options can be simply provided to everyone that wants to optimize their perception.

In fact, this model offers a different way of teaching a subject a new behavior by visualizing the test person what and when has to be observed. Therefore, a model should be created that in the best case, finds an optimal behavior. Based on such information, an optimal behavior for each class can be created and artificially extracted to create information that can be taught to users. A prerequisite will be the analysis of single scan paths, which can be accessed by looking at the fixation image patches.  Currently, as the fixation point is temporally and spatially averaged, another improvement might be achieved when optimizing the input layer by using an object detection beforehand. Especially when counting in the error rate of the eye tracker and early fixations, some samples might end up directly next to an object and some directly on it. In this case, the CNN will return different shapes. By using the object as an area of interest (AOI) and taking the intersection as input, this behavior can be unified as one can assume that the subject is perceiving the same object in both cases. The CNN can also be optimized. At the moment this CNN is trained on ImageNet to classify about 1,000 classes. By retraining the CNN on a set of 360° videos, with manually labeled teammates, opponents, goals, the ball, and free spaces, the intersections of the gaze with AOIs can become advantageous and result in higher classification rates.

\section{Gaze-Based Assistance Timing}

With detection of a gaze-based assistance timing, one can for example help surgeons to proceed, by either pointing out visual clues, which may be used by expert surgeons to navigate or drawing arrows on the output of the arthroscope which tells the surgeon where to navigate next (examples are shown in Fig.~\ref{fig:surgeonSetup}). Another possible usage of the knowledge of the correct timing for assistance can be to augment the whole output by describing the scene by segmenting and labeling each bone or tissue. Or simply name the shown parts in the output. There are multiple ways of supporting a surgeon. Depending on the state of expertise, the level
of assistance may be chosen, to allow different skilled surgeons, to train their different weaknesses. 

\begin{figure}[ht]
	\centering
	\includegraphics[width=0.8\columnwidth]{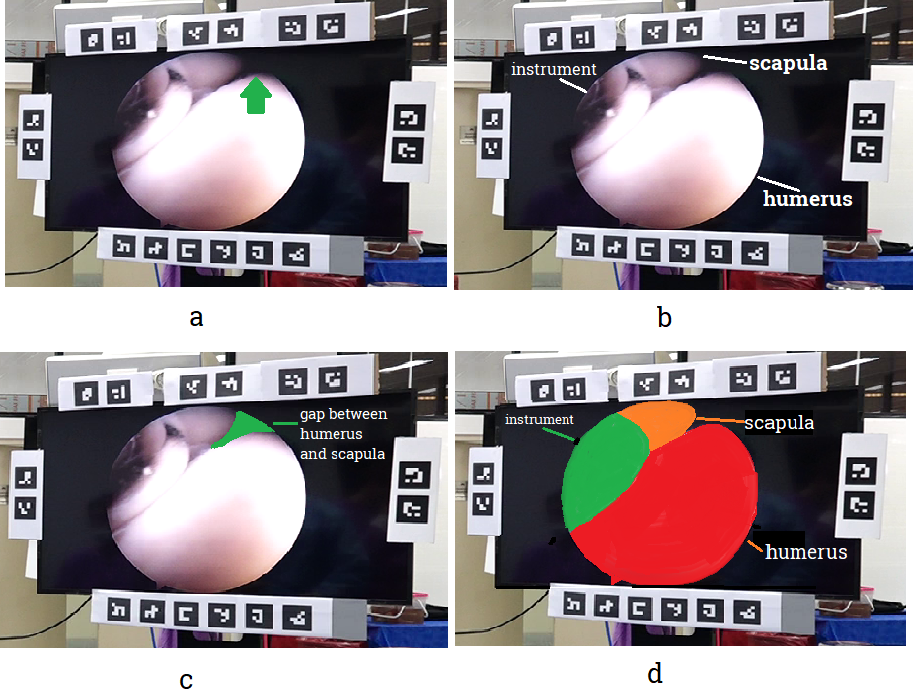}
	\caption{Assistance options for young surgeons to regain overview. }~\label{fig:surgeonSetup}
\end{figure}

\newpage

\subsection{States of confusion during arthroscopic surgery}

To enable the use of the models presented so far in the form of a diagnostic application, another feature is required. Of course, assistance options can be constantly displayed to the user during training. However, this can lead to the user being overwhelmed by the number and frequency of the assistance options, so that they may distract from the actual training. To prevent this, it makes sense to use the methods presented in Section~\ref{sec3:conf}. Because by knowing the right time for assistance options, limited to short time intervals, exact help can be displayed, so that it will only appear when it is needed.

In this context, when a certain prediction rate is reached, the detection of optimal times enables much more fine-tuned and personalized assistance. Thus, personal weaknesses can be identified online and training can be tailored specifically to the current user. However, in order to incorporate such diagnostics into an application, two key aspects must be considered. First, the prediction rate must be sufficiently high. Thus the presented model is 94\%, thus, more than adequate. And secondly, not only the right timing of the assistance but also an adequate form of it must be used. Because even if the right timing has been found, the right assistance must also be displayed to match the user's expertise class. Since different expertise classes can have different problems, these must be explicitly assigned in order to display class-compliant assistance. 
Further, in this model, the false-positive rate (detection of confusion when none is actually present) should be kept low to avoid displaying unnecessary help, but the false-positive rate does not have a large impact on the applicability of the model, since it only indicates how often the user is shown additional information that may at most distract him a little because it was unnecessary for that moment. Besides the prediction rate, a short calculation time is also an advantage. Getting assistance only after 5 seconds is not efficient, because the scene may have already changed completely in this time step. Assistance options must be provided as quickly as possible. 

Regarding the transferability of the model to other domains, there are no particular restrictions. If a think-aloud protocol is also used in that new domain, the presented methods can be used to train a model that detects and classifies states of confusion. To what extent the presented model trained on surgery data can classify new data from other domains is difficult to say. For this to work, the gaze feature expressions detected here during a state of confusion would have to be the same or at least similar to those during the confusion in the new domain. Probably, a direct transfer of the trained model without new training requires at least some similarities in the nature of the eye tracker and perhaps even the task. However, again, since there is little inter-task as well as inter-domain work, this needs to be found out in future work.

\subsection{Outlook}

One reasonable step to improve the model would first be to speed the whole pipeline up. That is, to enable the model to be used online by eye trackers with higher frequency. However, one essential step for this model is the design of adequate assistance options. For example, it does not make sense to label the arthroscopy image with the names of the bones and tissues if the surgeon is not yet familiar with them or currently learns how to navigate the arthroscope camera. Likewise, it is meaningful to specify the task beforehand and then purposefully train for one specific weakness. In football, for example, this can be done by configuring training to optimize shoulder glances. This way, the user knows what they have to pay attention to and the application can specifically display help options such as free spaces or trigger the correct direction of the gaze via sounds. There are a plethora of ways to assist young trainees, but which are reasonable, needs to be investigated in future works.

\newpage
\chapter{Ethical Considerations}
~\label{chap:ethics}

In general, when recording, recognizing, and processing personal, biometric data, which may include eye-tracking data, special care must be taken, as such data require special protection. This is true not only from a legal point of view but especially from a research ethics point of view.  All study data used in this work were collected in accordance with Article 4, No. 1. DSGVO with the consent of the subjects. Subjects have the right to view and delete their data at any time. 
Data that, in our view, have a direct personal reference, were stored anonymously and were only used for the work mentioned in this thesis in the sense of research. If data had to be published for publication, this was done in accordance with the applicable rules and laws. Likewise, where possible, all work was published under licenses prohibiting commercial use. Since in my view research is done for society. \\

In addition to the special nature of biometric data, research data on expertise recognition represent, in my view, another special feature. An identification of persons by their characteristics and the associated diagnostic data (expertise, confusion) was prevented by strict anonymization. This prevents any possible damage that could be caused by the identification of the test persons. Particularly with regard to artificial intelligence, there are often uncertainties about data protection, so I would like to make a special note here that all study data were only used anonymously so that neither outsiders nor algorithms (e.g. artificial intelligence) can establish connections to individual persons.


\newpage
\appendix

\chapter{Gaze Expertise Linkage}


\blfootnote{This chapter is based on the following publications:\\
	
	\begin{itemize}
		\item \textbf{B. W. Hosp}, F. Schultz, O. Höner, and E. Kasneci. "Soccer Goalkeeper Expertise Identification Based on Eye Movements.” In: PloS one, 16(5). 2021.  \\
		
		\item \textbf{ B.W. Hosp}, M.S. Yin, P. Haddawy, P. Sa-ngasoongsong, and E. Kasneci. "Differentiating Surgeons' Expertise Solely by Eye Movement Features”. Companion Publication of the 2021 International Conference on Multimodal Interaction (ICMI '21 Companion), October 18--22, 2021, Montréal, QC, Canada. ACM, New York, NY, USA.\\
		
		\item M. S. Yin, P. Haddawy, \textbf{B. W. Hosp}, P. Sa-ngasoongsong, T. Tanprathumwong, M. Sayo, and A. Supratak. "A Study of Expert/Novice Perception in Arthroscopic Shoulder Surgery". In Proceedings of the 4th International Conference on Medical and Health Informatics (pp. 71-77). August 2020. 
		
	\end{itemize}
}

\newpage

\section{Soccer Goalkeeper Expertise Detection Based on Eye Movements}

\subsubsection*{Abstract}
The latest research in expertise assessment of soccer players has affirmed the importance of perceptual skills (especially for decision making) by focusing either on high experimental control or on a realistic presentation. To assess 
the perceptual skills of athletes in an optimized manner, we captured 
omnidirectional in-field scenes and showed these to 12 experts (picked by DFB), 
10 regional league intermediate players and 13 novice soccer goalkeepers on virtual reality glasses. All scenes were shown from the same natural goalkeeper perspective and ended after the return pass to the goalkeeper. Based on their gaze behavior, we classified their expertise with common machine learning techniques. Our results show that eye movements contain highly informative 
features and thus enable classification of goalkeepers between three stages 
of expertise, namely elite youth player, regional league player, and novice at 
high accuracy of 78.2\%. This research underlines the importance of eye 
tracking and machine learning in perceptual expertise research and paves the 
way to perceptual-cognitive diagnosis as well as training systems. 

\subsection{Introduction}
\label{sec:intro}

Along with physical performance factors, perceptual-cognitive skills play an increasingly important role as cognitive performance factors in sports games. In perceptual research examining the underlying processes of these skills, subjects are typically placed in a situation where they have to react while their behavior is being recorded and subsequently analyzed. Such behavior can be assigned to a class, for example, to provide information about performance levels. Many studies in sports games in general, and in soccer in particular, ~\cite{berry2008contribution,catteeuw2009decision,abernethy2010revisiting,helsen1999multidimensional,abernethy1999can,farrow2002can,williams2002anticipation} have shown that athletes in a high-performance class have a more highly developed perception, leading – amongst other factors – to success in sports. However, this research is still confronted with challenges regarding experimental control and a representative presentation of the situation. Furthermore, the potential of novel technologies such as eye tracking as a means to assess the underlying perceptual-cognitive processes has not yet been fully exploited, especially with regard to the analysis of complex eye-tracking data. In this work, we research how to handle and analyze such large and complex eye-tracking data in an optimized way by applying common supervised machine learning techniques to the gaze behavior of soccer goalkeepers during a decision-making task in build-up game situations presented as 360°-videos in a consumer-grade virtual reality headset.

Latest sports-scientific expertise research shows that experts - when it comes to decision-making- have more efficient gaze behavior because they apply an advanced cue utilization to identify and interpret relevant cues ~\cite{mann2007perceptual}. This behavior enables experts to make more efficient decisions than non-experts, e.g. during game build-up by the goalkeeper. From both a scientific and practical sports perspective, of particular importance are factors that lead to successful perception, form expertise, and how these can be measured. 
To measure perception-based expertise, at first, a diagnostic system is needed for recognition of expertise, which provides well-founded information about the individual attributes of perception. These attributes are usually considered in isolation. Thus, their influence on expertise can be specifically recognized.
To allow the athletes to apply their natural gaze behavior, the experimental environment is important, but one of the main problems in perceptual-cognitive research persists in realism vs. control. In a meta-review of more than 60 studies on natural gaze behavior from the last 40 years, Kredel et al. \cite{kredel2017eye} postulate that the main challenges in perception research lie in a trade-off between experimental control and a realistic valid presentation. Diagnostic and training models are often implemented or supported by digital means.

This is nothing new, as in sports psychological research, new inventions in computer science such as presentation devices (i.e. CAVE \cite{defanti2009starcave}, virtual reality (VR) \cite{wirth2018assessment}), interface devices (i.e. virtual reality, leap motion, etc.), or biometric feature recognition devices (i.e. eye tracker \cite{nystrom2018tobii}) are used more and more often. 
As a new upcoming technology, virtual reality (VR) devices are used more frequently as stimulus presentation and interaction devices. As said, a fundamental aspect in perception research is a highly realistic presentation mode, which allows for natural gaze behavior during diagnostic. VR technology makes this possible by displaying realistic, immersive environments. However, this strength, allowing natural gaze behavior, comes less from the VR technology itself. According to Gray~\cite{gray2019virtual}, the degree to which the perceptual-cognitive requirements of the real task are replicated in such environments depends on psychological fidelity. Next to immersion and presence, Harris et al.~\cite{harris2020framework} suggest the expansion of a simulation characterization into a typology of fidelity (containing also psychological fidelity) to determine the realism of a simulation.  VR offers an immersive experience through the use of 4k 360\textdegree video, which experiences a higher level of realism than, for example, CAVE systems, by providing higher levels of psychological fidelity ~\cite{gray2019virtual,harris2020framework}. VR is therefore a popular and optimal tool for perception research. Bideau et al. \cite{bideau2010using} summarize further advantages of VR in their work. Their main contribution, however, is their immersive virtual reality that elicits expert responses similar to real-world responses.

In a narrower sense, VR is based on computer-generated imagery (CGI). One advantage of such fully CGI-based environments is the possibility of the user interacting with the environment, which presumingly increases the immersive experience. On the other hand, fully CGI-based environments contain moving avatars that are natural in appearance and hide environmental influences. This might prevent high immersion and influence the participant's gaze behavior. Therefore, we chose a realistic environment with 360\textdegree stimuli to provide a close to a natural environment that does not influence the participant's gaze behavior. As this work presents a focus on the cognitive processes of decision-making, we focus less on realistic interaction methods

Especially interesting are the developments of VR devices regarding integrated measuring devices. More and more devices have eye trackers directly integrated, which, in combination with a photo-realistic environment in VR glasses, allows for the measurement of almost optimal user gaze behavior while also showing highly realistic stimuli.  Eye trackers provide a sound foundation with a high temporal and spatial resolution to research perceptual processes. The combination of VR and high-speed eye tracking allows the collection of a massive amount and highly complex data. With the high-quality eye images and freedom of movement of a mobile eye tracker, the high speed of a remote eye tracker and the control over the stimulus in a lab setting (VR), and the naturality of in-situ stimuli by omnidirectional videos, the outcome of this combination is highly complex. Analysis of such data is a particular challenge, which emphasizes the need for new analysis methods.
As we want to infer underlying mechanisms of perceptual-cognitive expertise, tracking eye movements is our method of choice in this work. Generally, perceptual research focuses on eye tracking because, as a direct measuring method, it allows for a high degree of experimental control. Besides a realistic presentation and high degree of experimental control, VR can also be used to model the perception \cite{duchowski2000binocular} of athletes and thus creates a diagnostic system. A diagnostic system has the ability to infer the current performance status of athletes to identify performance-limiting deficits, an interesting provision of insight for the athletes and coach as well. Most importantly, such a diagnostic system forms the basis for an adaptive, personalized, and perceptual-cognitive training system to work on the reduction of these deficits.

So far, eye-tracking studies have focused on either in-situ setups with realistic presentation mode and mobile eye trackers (field camera showing the field of view of the user) or on laboratory setups with high experimental control using remote eye trackers ~\cite{uppara2018eye,grushko2014usage,bard1981considering,bahill1984can,singer1998new,vickers1992gaze}. Since mobile eye trackers are rarely faster than 100-120 Hz because saccades and smooth pursuits cannot be detected properly at such speed, investigations in an in-situ context are limited to the observation of fixations. Fixations are eye movement events during which the eye is focused on an object for a certain period of time (and thus projects the object onto the fovea of the eye) so that information about the object can be cognitively processed. The calculation of fixations with such a slow input signal leads to inaccuracies in the recognition of the start and end of the fixation. 
Only limited knowledge can be gained using such eye trackers because additional information contained in other eye events, such as saccades and smooth pursuits, cannot be computed correctly. This prevents the use of such eye trackers as robust expert measurement devices. Saccades are the jumps between the fixations that allow the eye to realign. They can be as fast as 500 \textdegree/s. Smooth pursuits are especially interesting in ball sports because they are fixations on moving objects i.e. moving players. However, especially in perception studies in soccer in VR-like environments, slow eye trackers with about 25-50 Hz are primarily used \cite{roca2020perceptual,dicks2010examination,aksum2020football,roca2018creative}. This speed limits the significance of these studies to fixation and attention distribution in areas of interest (AOI). Aksum et al. \cite{aksum2020football}, for example, used the Tobii Pro Glasses 2 with a field camera set to 25 Hz. Therefore, only fixations or low-speed information is available and no equal stimuli for comparable results between participants. In a review of 38 studies, McGuckian et al. \cite{mcguckian2018systematic} summarized the eye movement feature types used to quantify visual perception and exploration behavior of soccer players. Except for Bishop et al.  \cite{bishop2014telling}, all studies were restricted to fixations thus restricting the gainable knowledge of eye movement features. The integration of high-speed eye trackers into VR glasses combines both strengths: high experimental control of a high-speed eye tracker and a photo-realistic stereoscopic VR environment.

With more frequent use of eye trackers, and more accurate, faster, and ubiquitous devices, huge amounts of precise data from fixations, saccades, and smooth pursuits can be generated which cannot be handled in entirety utilizing previous analysis strategies. Machine learning provides the power to deal with huge amounts of data. In fact, machine learning algorithms typically improve with more data and allow - by publishing the model's parameter set - fast, precise, and objective reproducible ways to conduct data analysis. Machine learning methods have already been successfully applied in several eye-tracking studies. Expertise classification problems in particular, can be solved as shown by Castner et al. in dentistry education ~\cite{castner2020deep,castner2018scanpath} and Eivazi et al. in microsurgery ~\cite{bednarik2013computational,eivazi2012gaze,eivazi2011predicting,eivazi2017towards}.  Machine learning techniques are the current state-of-the-art for expertise identification and classification. Both supervised learning algorithms ~\cite{bednarik2013computational,castner2018scanpath} and unsupervised methods or deep neural networks ~\cite{castner2020deep} have shown their power for this kind of problem-solving. This combination of eye tracking and machine learning is especially well suited when it comes to subconscious behavior like eye movements features as these methods have the potential to greatly benefit the discovery of different latent features of gaze behavior and their importance and relation to expertise classification. 

In this work, we present a model for the recognition of soccer goalkeepers' expertise in regard to decision-making skills in build-up situations by means of machine learning algorithms relying solely on eye movements. We also present an investigation of the influences of single features on explainable differences between single classes. This pilot study is meant to be the first step towards a perceptual-cognitive diagnostic system and a perceptual-cognitive virtual reality training system, respectively.

\subsection{Methods}

The basis of this work is a pilot study on a VR system with an integrated eye tracker. This chapter describes the experimental setup, the pilot study, the eye-tracking characteristics, and the methodical procedure for the analysis with machine learning methods.

\subsubsection*{Experimental setup}

In this study, we employed an HTC Vive, a consumer-grade virtual reality (VR) headset. Gaze was recorded through the integration of the SMI high-speed eye tracker at 250 Hz. The SteamVR framework is open-source software that interfaces common real-time game engines with VR glasses to display custom virtual environments. We projected omnidirectional 4k footage on the inside of a sphere that envelopes the user's field of view, which leads to high immersion in a realistic scene.

\subsubsection*{Stimulus material}

We captured the 360\textdegree -footage by placing an Insta Pro 360 (360\textdegree camera) on the soccer field on the position of the goalkeeper. Members of a German First League's elite youth academy were playing 26 different 6 (5 field players + goalkeeper) versus 5 match scenes on one half of a soccer field. Each scene was developed with a training staff team of the German Football Association (DFB) and each decision was ranked by this team. There were 5 options (teammates) plus one "emergency" option (kick out). For choosing the option rated as the best option by the staff team, the participant earned 1 point, because this option is the best option to ensure the continuation of the game. All other options were rated with 0 points. Conceptually, all videos had the following content: The video starts with a pass by the goalkeeper to one of the teammates. The team passes the ball a few times until the goalkeeper (camera position) receives the last return pass. The video stops after this last pass and a black screen is presented. The participant now has 1.5 seconds time to report which option they've decided on and the color of the ball which was printed on the last return pass (to force all participants to recognize the last return pass realistically).

\subsubsection*{Participants}

We collected data from 12 German expert youth soccer goalkeepers (U-15 to U-21) during two youth elite goalkeeper camps. The data from 10 intermediates was captured in our laboratory and comes from regional league soccer goalkeepers (semi-professional). Data from 13 novices came from players with up to 2 years of experience with no participation in competitions and no training on a weekly basis. The experts have 8.83 hours of training each week and are 16.6 years old on average. They actively played soccer for about 9 years, which is significantly more than the novices (1.78 years), but less than the intermediates (15.5 years). This may be a result of their age difference. The intermediates are 22 years old on average but have nearly half of the training hours per week compared to the experts. Characteristics of the participants can be seen in Table~\ref{tabA:participants}.

\begin{table}[!h]

	\centering
		\def\arraystretch{1.5}
	\begin{tcolorbox}
	\begin{tabularx}{\textwidth}{lllX }
		\multicolumn{4}{c}{\textbf{Participants}} \\	\cline{1-4} \\
		\cellcolor{gray!30}Class		& \cellcolor{gray!30}Attribute 	  	  & \cellcolor{gray!30}Average 		&\cellcolor{gray!30} Std. Dev.\\	
	 
		Experts 	& Age 			      & 16.60 			& 1.54  \\
				& Active years	      & 9.16			& 5.04   \\ 
				&Training hours/week  & 8.83		    & 4.27  \\		 
	
	    Intermediates & Age			 	  & 22.00 			& 3.72   \\
				&Active years	 	  & 15.50			& 5.77  \\
			&Training hours/week  & 4.94			& 0.91  \\	
	 
		Novices 	&Age  				  & 28.64 			& 3.72   \\
				&Active years 		  & 1.78   			& 5.21 \\
				&Training hours/week  & 0.00 	    	& 0.00  \\ 
		
	\end{tabularx}
	\end{tcolorbox}
	\caption{Participants summary.}
	\label{tabA:participants}
\end{table}

\subsubsection*{Procedure}

\begin{figure}
	\centering
	\includegraphics[width=1\linewidth]{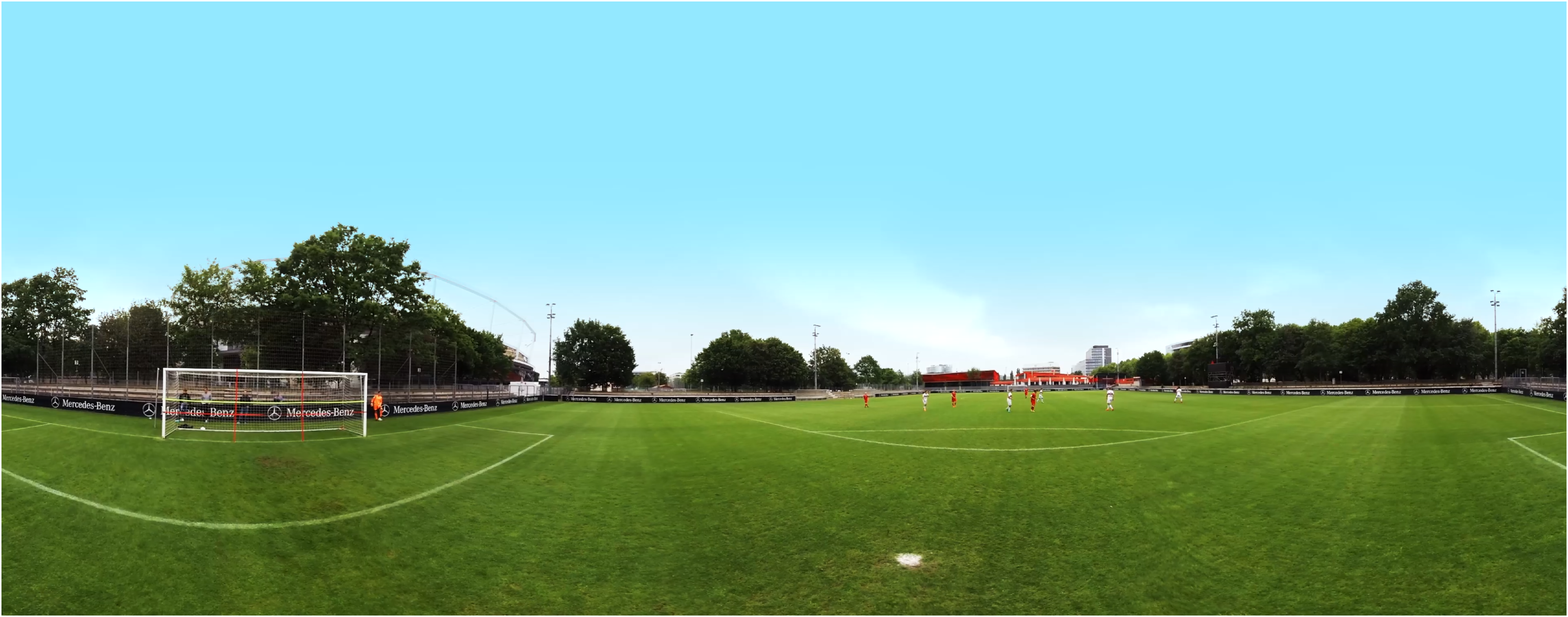}
	\caption{Example stimulus in equirectangular format. }
	\label{figA:stimulus}
\end{figure}

\begin{figure}
	\centering
	\includegraphics[width=1\linewidth]{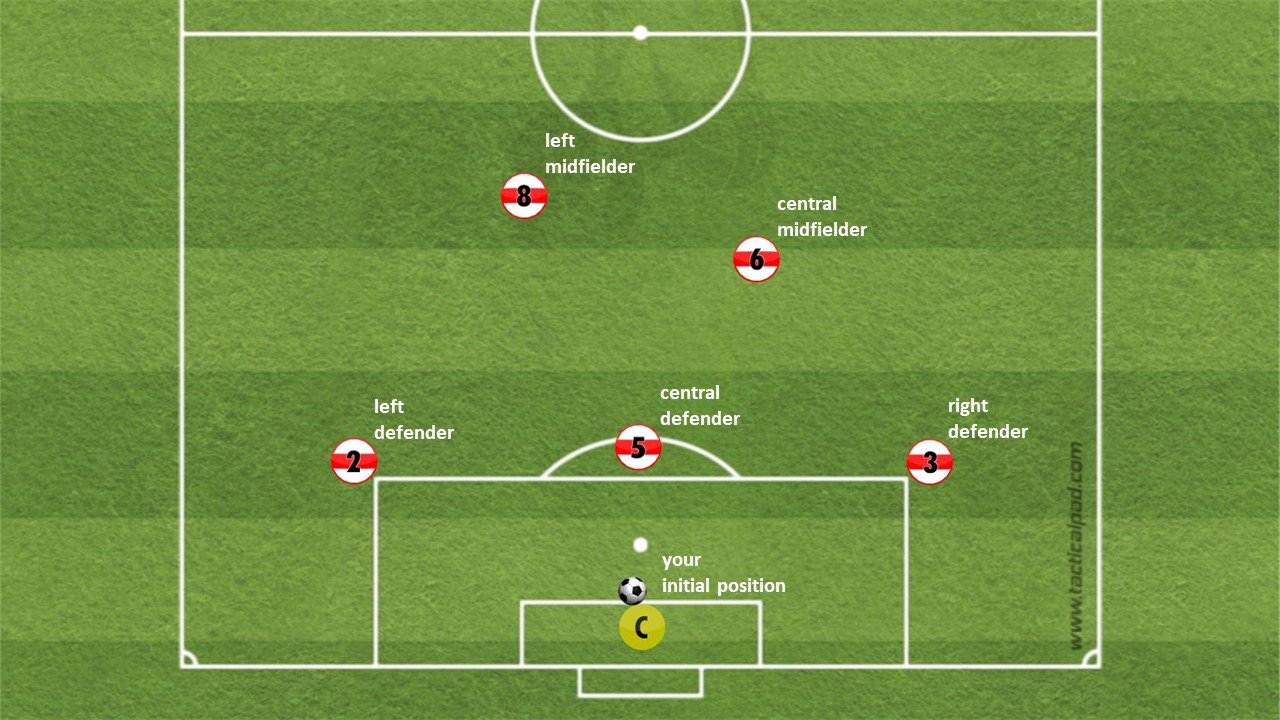}
	\caption{Schematic overview of the response options. Emergency option kick out is not shown. }
	~\label{figA:schematic}
\end{figure}

The study was confirmed by the Faculty of Economics and Social Sciences Ethics Committee of the University of Tübingen. After signing a consent form to allow the usage of their data, we familiarized the participants with the footage.

The study contained two blocks consisting of the same 26 stimuli in each (conceptually as mentioned in the stimulus material section). The stimuli in the second block were presented in a different randomized order. Each decision made on the continuation of a video has a binary rating, as only the best decision was counted as 1 (correct) while all other options were rated as 0 (incorrect). At first, 5 different sample screenshots (example view see Fig~\ref{figA:stimulus} in equirectangular form or S1 Video for a cross-section of the stimulus presentation sphere) and the corresponding sample stimuli were shown and explained to acclimate the participant to the setup. To learn the decision options, we also showed a schematic overview before every sample screenshot (see Fig.~\ref{figA:schematic}).

\subsubsection*{Eye Tracking}
~\label{sec:method}
The raw data of the SMI Eye tracker can be exported from the proprietary BeGaze software as CSV files. BeGaze already provides the calculation of different eye movement features based on the raw gaze points. As we get high-speed data from the eye tracker, we use the built-in high-speed event detection. The software first calculates the saccades based on the peak threshold, which means the minimum saccade duration (in ms) varies and is set dependent on the peak threshold default value of $40 ^{\circ}/s $. In a second step, the software calculates the fixations. Samples are considered to belong to a fixation when they are between a saccade or blink. With a minimum fixation duration of 50 ms, we reject all fixations below this threshold. As there is no generally applicable method for detection of smooth pursuits, this kind of event is included and encoded as fixations with longer duration and wider dispersion. We marked fixations with a fixation dispersion of more than 100 px as smooth pursuits. By doing this, we split fixations into normal length fixations and long fixations which we consider to be and refer to as smooth pursuits. This threshold is an empirical value based on the sizes of the players as the main stimuli in the video. The following section describes the steps that are necessary to train a model based on these eye movement features.

\subsubsection*{Feature selection}
~\label{subfeaturesSelection}

As it is not clear which subset of eye movement features explains the difference in expertise completely, we followed a brute-force method, considering all possible measures issued by the eye-tracking device and subsequently evaluating their importance. For the classification of expertise level we focus on the following features:

\begin{itemize}
	\item  event duration and frequency (fixation/saccade)
	\item fixation dispersion ( in \textdegree)
	\item smooth pursuit duration (in ms)
	\item smooth pursuit dispersion (in \textdegree)
	\item saccade amplitude (in \textdegree)
	\item average saccade acceleration (in \textdegree/$s^2$) 
	\item peak saccade acceleration (in \textdegree/$s^2$)
	\item average saccade deceleration (in \textdegree/$s^2$) 
	\item peak saccade deceleration (in \textdegree/$s^2$)
	\item average saccade velocity (in \textdegree/$s$)
	\item peak saccade velocity (in \textdegree/$s$)
	
\end{itemize}

Each participant viewed 26 stimuli twice, resulting in 52 trials per subject. First, we viewed the samples of these 52 trials and checked the confidence measures of the eye-tracking device. We removed all trials with less than 75\% tracking ratio, as gaze data below this threshold is not reliable. Due to errors in the eye-tracking device, not all participant data is available for every trial. Table \ref{table:erroneousTrials} shows an overview of the lost trials. For two participants, 11 trials had a lower tracking ratio; on participant 18, we lost 35 trials; and on participant 33, one trial was lost. This results in 1658 out of 1716 valid trials in total. 3.3\% of the trials were lost due to eye-tracking device errors.

\begin{table}[!h]
	
	\centering
	
		\def\arraystretch{1.5}
	\begin{tcolorbox}
		\begin{tabularx}{\textwidth}{XX}
			\multicolumn{2}{c}	{\textbf{Overview erroneous trials} } \\ \cline{1-2} \\
			\cellcolor{gray!30}	Participant & \cellcolor{gray!30}Number of valid trials\\ 
			
			1 & 11\\	\hline		 		
			8 & 11\\\hline
			18 & 25\\\hline
			33 & 1 \\\hline
			all others & 0\\

	\end{tabularx}
	
	\end{tcolorbox}
	\caption{ Overview of the amount of erroneous trials, based on eye-tracking device errors.}
	\label{table:erroneousTrials}

\end{table}

\subsubsection*{Data cleaning}
We checked the remaining data for the quality of saccades. This data preparation is necessary to remove erroneous and low-quality data that comes from poor detection on behalf of the eye-tracking device and does not reflect the correct gaze. Therefore, we investigated invalid samples and removed (1) all saccades with invalid starting position values, (2) all saccades with invalid intra-saccade samples, and (3) all saccades with invalid velocity, acceleration, or deceleration values. 

\begin{enumerate}
	
	\item  Invalid starting position:  0.22\% saccades started at coordinates (0;0). This is an encoding for an error of the eye-tracking device. As amplitude, acceleration, deceleration, and velocity are calculated based on the distance from the start- to the endpoint, these calculations result in physiological impossible values, e.g., over 360\textdegree saccade amplitudes.
	
	\item  Invalid intra-saccade values: Another error of the eye-tracking device stems from the way the saccade amplitude is calculated through the average velocity (Eq~\ref{eq:saccadeAmplitude}) which is based on the distance of the mean of start and endpoints on a sample-to-sample basis (see Eq ~\ref{eq:saccadeVelocity}). 3.6\% of the saccades had at least one invalid gaze sample and were removed (example see Fig \ref{fig:intra-saccadeError}).
	
	\begin{eqnarray}
		\label{eq:saccadeAmplitude}
		\oslash Velocity * EventDuration
	\end{eqnarray}
	
	\begin{eqnarray}
		\label{eq:saccadeVelocity}
		\frac{1}{n}*\sum_{1}^{n} \frac{dist(startpoint(i), endpoint(i))}{EventDuration(i)}
	\end{eqnarray}
	
	\begin{figure}[ht]
		\centering
		\includegraphics[width=0.75\columnwidth]{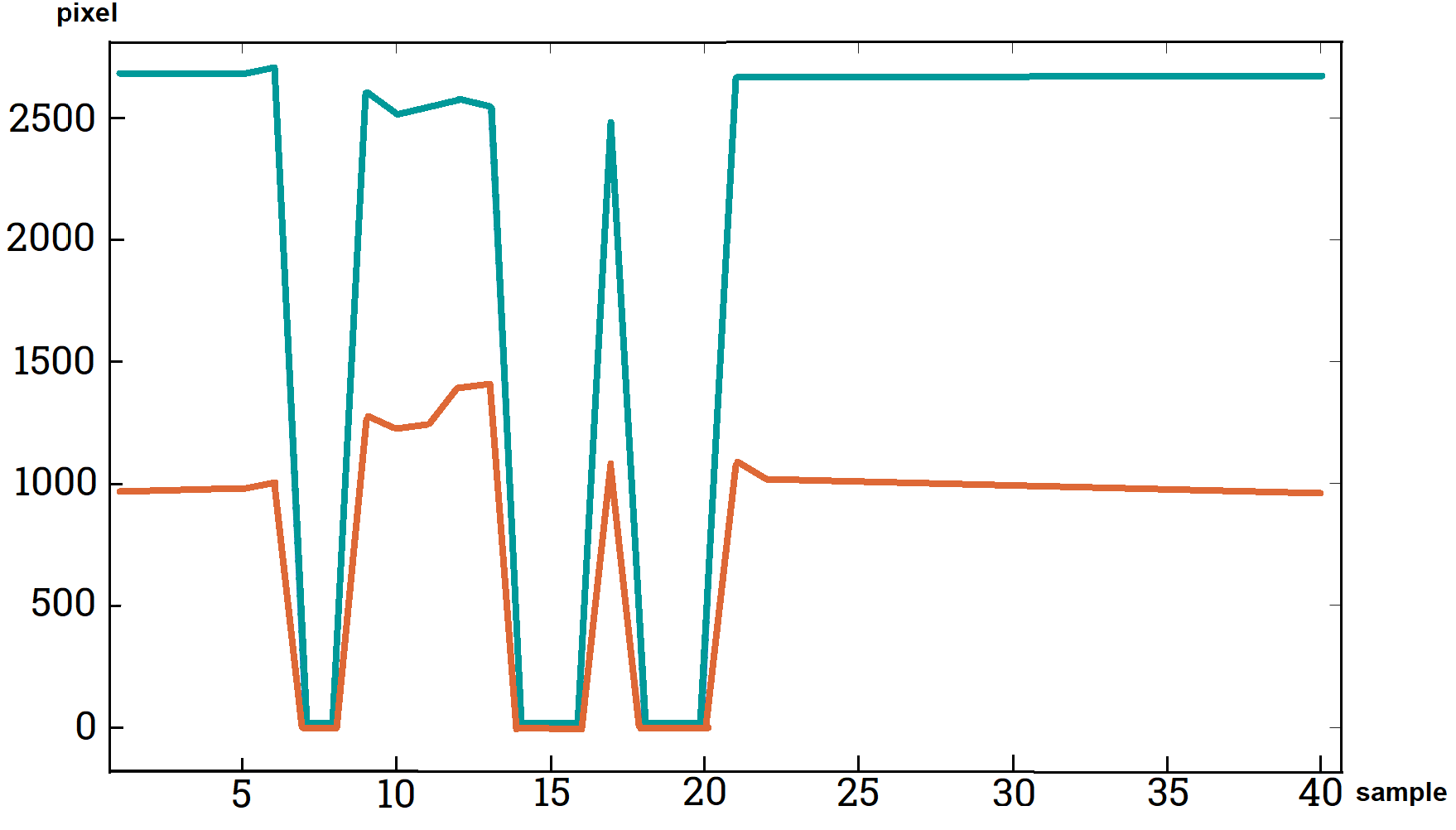}
		\caption{Example of invalid intra-saccade values. The x-axis shows the number of the gaze signal sample (40 samples, 250 Hz, 160 ms duration) and the y-axis shows the position in pixel. The blue line represents the x-signal of the gaze and the orange line the y-signal. }~\label{fig:intra-saccadeError}
	\end{figure}
	
	
	On Fig ~\ref{fig:intra-saccadeError}, the gaze signal samples 7, 8, 14-16, 18-20 (x-axis) both, the x- and y-signal (blue and red line, respectively) show zero values and thereby indicate a tracking loss. As the saccade amplitude is based on the average velocity which is calculated on a sample-to-sample Eq~\ref{eq:saccadeVelocity}, the velocity from samples 6 to 7, 8 to 9, 13 to 14, 16 to 17, 17 to 18, and 20 to 21 significantly increase the average velocity as the distances are high (on average over 2400 px for x-signal and over 1000px for y-signal, which corresponds to a turn of 225° on the x-axis and 187.5° on the y-axis in the time of 4 ms between two consecutive samples).
	
	There are two interpretations for saccadic amplitude. The first refers to the shortest distance from start to the endpoint of a saccadic movement (i.e., a straight line) and the second describes the total distance traveled along the (potentially curved \cite{holmqvist2011eye}, p.311) trajectory of the saccade. The SMI implementation follows the second definition. We could have potentially interpolated invalid intra-saccade samples instead of completely removing the complete saccade from analysis, however, this leads to uncertainties that can affect the amplitude depending on the number of invalid samples and does not necessarily represent the true curvature of the saccade. 
	
	
	\item  As the velocity increases as a function of the saccade amplitude \cite{collewijn1988binocular}, 4.8\% of the saccades were ignored because of the restriction on velocities greater than 1000\textdegree/s. Similar to extreme velocities, we removed all saccade samples that exceeded the maximum theoretical acceleration and deceleration thresholds. Saccades with longer amplitudes have higher velocity, acceleration, and deceleration, but can not exceed the physiological boundaries of 100.000 \textdegree/$s^2$ \cite{holmqvist2011eye}. 3.0\% and 4.0\%, respectively, of all saccades that exceeded this limit. As most of the invalid samples had more than one error source, we only removed 5.5 \%  of the saccades (3.5\% of all samples) in total. 

\end{enumerate}
After cleaning the data we use the remaining samples to calculate the average, maximum, minimum, and standard deviation of the features. This results in 36 individual features. We use those for classifying expertise in the following.

\begin{figure}[ht]
	\centering
	\includegraphics[width=1.0\columnwidth]{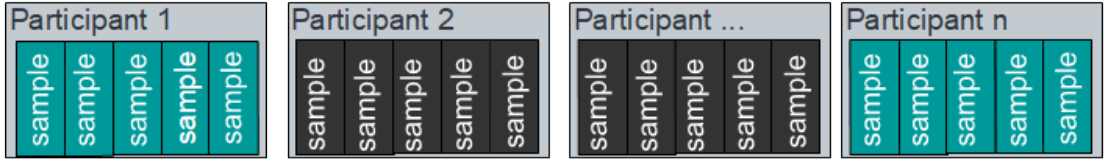}
	\caption{Example sample assignment. Top row shows a random assignment of samples, independent of the corresponding participant. Bottom row shows participant-wise sample assignment to training and evaluation set. }~\label{fig:sampleAssignment}
\end{figure}

\subsubsection*{Machine learning model}
In the following, we refer to \textit{expert samples} as trials completed by an elite youth player of a DFB goalkeeper camp, \textit{intermediate samples} as those of regional league players, and \textit{novice samples} as those of amateur players. We built a support vector machine model (SVM) and validated our model in two steps: cross-validation and leave-out validation. We trained and evaluated our model in 150 runs with both validations. For each run, we trained a model (and validated with cross-validation) with samples of 8 experts, 8 intermediates, and 8 novices samples, and used the samples of two participants from each group of the remaining participants to predict their classes (leave-out validation). The experts, as well as the intermediates and the novice samples in the validation set, were picked randomly for each run.\\

\subsubsection*{Sample assignment}

We found that the way in which the data set samples are split into training and evaluation sets is very important and a participant-wise manner should be applied. By randomly picking samples independent of the corresponding participant, participant samples usually end up being distributed on the training and the evaluation set (illustrated in Fig  ~\ref{fig:sampleAssignment}). This leads to an unexpected learning behavior that does not necessarily classify expertise directly, but, rather, matches the origin of a sample to a specific participant thereby indirectly identifying that participant's level of expertise. This means that a model would work perfectly for known participants, but is unlikely to work for unseen data. Multiple studies show that human gaze behavior follows idiosyncratic patterns. Holmqvist et al. \cite{holmqvist2011eye} show that a significant number of eye-tracking measures underlay the participants' idiosyncrasy, which also means that the inter-participant differences are much higher than intra-participant differences. A classifier learns a biometric, person-specific measure instead of an expertise representation.

\subsubsection*{Model building}
\begin{figure}[ht]
	\centering
	\includegraphics[width=0.9\columnwidth]{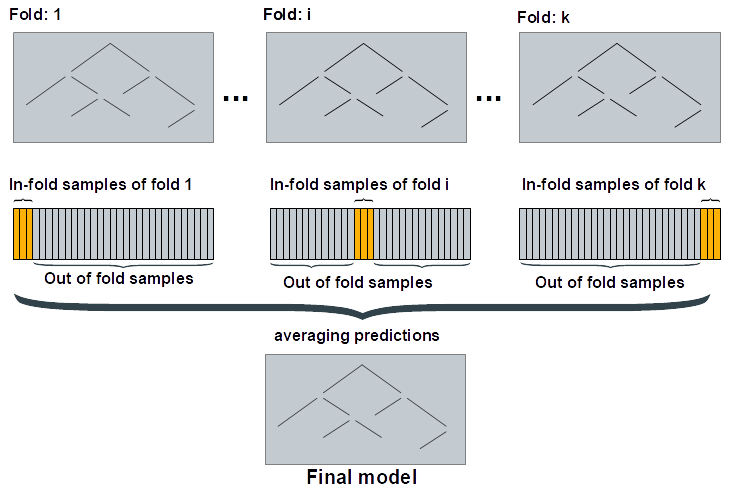}
	\caption{Illustration of the k cross-validation procedure. Each of the k models has a different out-of-fold and in-fold data set. We build the final model on the average of all predictions from all k models. }~\label{fig:trees}
	
\end{figure}

To find a model robust to high data variations, we applied cross-validation during training. The final model is based on the average of k=50 models, with k = number of folds in the cross-validation. For each model $m_i$, with $i \in \{1, \dots ,k\}$, we use all out-of fold data of the i-th fold to train and evaluate $m_i$ with the in-fold data of the i-th fold (this procedure is illustrated in Fig~\ref{fig:trees}). The final model is evaluated with a leave-out validation. The cross-validation step during training is independent of the leave-out validation with totally new data (never seen by the model). Information from cross-validation is used during the building and optimizing of the model and leave-out validation solely provides information about the prediction accuracy of the model when using completely new data. 

With a total of 810 valid samples, equally distributed on expert, intermediate, and novice samples, we built a subset of 552 samples for training the model and a subset of 258 samples for evaluation. As each sample represents one trial, our approach here is to predict whether a trial belongs to an expert, intermediate, or novice class. We tested assumptions in different approaches.\\

\subsubsection*{Classifiability}

Firstly, we used all 46 features to check the classifiability of this kind of data. The first approach contains all features from section \textit{Feature selection} ~\ref{subfeaturesSelection} with their derivations, (namely average, maximum, minimum, and standard deviation) to build an SVM  model (Tables ~\ref{table:novicesAll}, ~\ref{table:intermediatesAll} and ~\ref{table:expertsAll} show all features with their derivations, split by class). When the binary case (expert vs. intermediates) results point out classifiability, the ternary case (expert vs. intermediate vs. novice) should be investigated.

\begin{table}[!ht]

	\centering
	\def\arraystretch{1.5}
	\begin{tcolorbox}
	\begin{tabularx}{\textwidth}{lcccc}
		\multicolumn{5}{c}{	\textbf{ Novices }}\\  \cline{1-5} \\
			        		& average & std. dev. 	& minimum & maximum 	\\   
     	
     	\multicolumn{5}{l} {\cellcolor{gray!30}\textbf{Fixation}}\\   
     	frequency (Hz)      & 0.21  & - 			& -		 & -   \\
     	duration (ms)       & 214.01 & 31.92 		& 190.49 & 239.30   \\ 
     	dispersion (pixels) & 72.09  &	25.68		& 24.67  & \cellcolor{teal} \color{white} 110.52 \\

     	\multicolumn{5}{l}{	\cellcolor{gray!30}\textbf{ Saccade }}\\ 			
     	frequency (Hz) 			& \cellcolor{teal} \color{white}0.07		& -								& - 							& - \\ 
     	duration (ms) 			& \cellcolor{teal} \color{white}71.68    	&  \cellcolor{orange!70}38.86	& \cellcolor{orange!70} 26.514  &\cellcolor{teal} \color{white} 175.46\\ 
     	amplitude (\textdegree)	&\cellcolor{teal} \color{white}9.29 		& \cellcolor{teal} \color{white}9.41 &0.57 		&\cellcolor{teal} \color{white}51.40 \\

     	\multicolumn{5}{l} {\cellcolor{gray!30}	\textbf{Saccade mean acceleration}}\\  			
     	mean (\textdegree$/s^2$) & 4263.38 	& \cellcolor{teal} \color{white}2482.01 & \cellcolor{teal} \color{white}366.66 & 13984.56 \\ 
     	peak (\textdegree$/s^2$) & 9322.48& 5777.27							 &231.83 								 & 28355.22 \\

     	\multicolumn{5}{l}{\cellcolor{gray!30}\textbf{Saccade deceleration}} \\  							
     	peak ($°/s^2$)&-6848.10&\cellcolor{orange!70}4166.26&-35563.64&\cellcolor{teal}\color{white}-411.76\\
     	
     	\multicolumn{5}{l} {\cellcolor{gray!30}	\textbf{Saccade velocity }} \\   							
     	mean (\textdegree$/s$) & 105.46 & \cellcolor{teal} \color{white}65.02 & \cellcolor{teal} \color{white}20.28 &\cellcolor{teal} \color{white}298.13\\ 
     	peak (\textdegree$/s$) & 215.24 & \cellcolor{orange!70}129.29 		& \cellcolor{teal} \color{white}40.31 & 766.15\\

     	\multicolumn{5}{l} {\cellcolor{gray!30}	\textbf{Smooth pursuit }}\\ 								
     	duration (ms)		&302.63 						& 278.11								& 75.62 						&1026.32\\ 	
     	dispersion (pixels)	& \cellcolor{orange!70}622.80	&\cellcolor{teal} \color{white}201.26 	&\cellcolor{orange!70}185.43 	& \cellcolor{orange!70}1085.90\\ 
     	
	\end{tabularx}
	\end{tcolorbox}

	\caption{All 46 features with their derivations. Novice class. Green cells show features with significant differences between classes. Orange cells stand for the most frequent feature.}
	\label{table:novicesAll}
	
\end{table}

\begin{table}[!ht]

	\centering
	\def\arraystretch{1.5}
	\begin{tcolorbox}
	\begin{tabularx}{\textwidth}{lcccc}
		
		\multicolumn{5}{c}{	\textbf{ Intermediates }}\\   \cline{1-5} \\
							 & average & std. dev. & minimum & maximum  \\ 
							 
		\multicolumn{5}{l} {\cellcolor{gray!30}\textbf{Fixation}}\\  
		frequency (Hz)      & 0.25&	-	&-	&- \\ 
		duration (ms)       &255.22 & 53.37& 215.83& 299.62 \\ 
		dispersion (pixels) &73.17 &  26.54  & 23.07 & \cellcolor{teal} \color{white}114.76 \\ 
				
		\multicolumn{5}{l}{	\cellcolor{gray!30}\textbf{ Saccade }}\\  			
		frequency (Hz) &\cellcolor{teal} \color{white} 0.08&	-&	-& -\\ 
		duration (ms) &\cellcolor{teal} \color{white} 84.34 & \cellcolor{orange!70} 59.72 & \cellcolor{orange!70}26.12  & \cellcolor{teal} \color{white}246.12\\ 
		amplitude (\textdegree)&\cellcolor{teal} \color{white}9.88 &\cellcolor{teal} \color{white} 10.674 & 0.57 &\cellcolor{teal} \color{white}54.83\\ 
		
		\multicolumn{5}{l} {\cellcolor{gray!30}	\textbf{Saccade mean acceleration}}\\  				
		mean (\textdegree$/s^2$) & 4123.97 & \cellcolor{teal} \color{white}2685.99 &\cellcolor{teal} \color{white} 315.34 & 15472.88 \\ 
		peak (\textdegree$/s^2$) & 8920.17 & 5989.25 & 216.72 & 28266.00 \\ 
		
		\multicolumn{5}{l} {\cellcolor{gray!30}	\textbf{Saccade deceleration}}  \\  								
		peak (\textdegree$/s^2$)&-6948.49 & \cellcolor{orange!70} 4770.06 &-36334.13& \cellcolor{teal} \color{white}-231.35 \\
		 
		\multicolumn{5}{l} {\cellcolor{gray!30}	\textbf{Saccade velocity }} \\  								
		mean (\textdegree$/s$) & 104.19 & \cellcolor{teal} \color{white}	66.68 & \cellcolor{teal} \color{white}21.52 & \cellcolor{teal} \color{white}331.11 \\ 
		peak (\textdegree$/s$) & 213.83 &\cellcolor{orange!70} 136.52 & \cellcolor{teal} \color{white}40.10 & \cellcolor{gray!40}764.02 \\ 
		
		\multicolumn{5}{l} {\cellcolor{gray!30}	\textbf{Smooth pursuit }}\\  								
		duration (ms)& 291.09 & 278.71 & 73.83 &977.12 \\ 	
		dispersion (pixels)&  \cellcolor{orange!70}425.08 &  \cellcolor{teal} \color{white}124.85 &  
		\cellcolor{orange!70}168.32 &  \cellcolor{orange!70}694.37 \\ 
			
	\end{tabularx}
	\end{tcolorbox}	
	\caption{All 46 features with their derivations. Intermediate class. We consider samples as belonging to a smooth pursuit when the dispersion of the samples is greater than 100 px. As the size of the players in the stimulus varies around 90 pixel + a buffer. }
	\label{table:intermediatesAll}
\end{table}

\begin{table}[!ht]
	\centering
	\def\arraystretch{1.5}
	\begin{tcolorbox}
	\begin{tabularx}{\textwidth}{lcccc}
		
		\multicolumn{5}{c} {	\textbf{Experts}}\\ \cline{1-5} \\ 
		Features & average & std. dev. & minimum & maximum  \\ 
		\multicolumn{5}{l}{\cellcolor{gray!30}	\textbf{Fixation }}\\  
		frequency (Hz)& 0.24 &	-& 	- & - \\ 
		duration (ms)& 241.50 &  58.62 & 198.13 &  291.72 \\ 
		dispersion (pixels)&  72.83 &	25.989&21.73 & \cellcolor{teal} \color{white}114.54\\ 		
		\multicolumn{5}{l}{\cellcolor{gray!30}	\textbf{ Saccade} }\\  				
		frequency (Hz)& \cellcolor{teal} \color{white}0.00 & - &- &- \\ 
		duration (ms) &\cellcolor{teal} \color{white}65.47 &  \cellcolor{orange!70}35.54 &\cellcolor{orange!70} 25.01 &\cellcolor{teal} \color{white} 163.41 \\ 
		amplitude (\textdegree)& \cellcolor{teal} \color{white} 8.93 &\cellcolor{teal} \color{white}9.430 &  0.56& \cellcolor{teal} \color{white}52.02\\ 
		\multicolumn{5}{l}{\cellcolor{gray!30}	\textbf{Saccade mean acceleration}}\\  				
		mean (\textdegree$/s^2$) &4769.65 & \cellcolor{teal} \color{white}3064.34 & \cellcolor{teal} \color{white}390.09 &  18965.94 \\ 
		peak (\textdegree$/s^2$) & 10026.45 &  7094.930 & 175.24 & 39445.12\\ 
		\multicolumn{5}{l} {\cellcolor{gray!30}	\textbf{Saccade deceleration }} \\  								
		peak (\textdegree$/s^2$)& -7912.19 & \cellcolor{orange!70} 5492.28 &-43479.91 & \cellcolor{teal} \color{white}-362.39\\ 
		\multicolumn{5}{l} {\cellcolor{gray!30}	\textbf{Saccade velocity  }}\\  								
		mean (\textdegree$/s$) & 110.67 &\cellcolor{teal} \color{white} 72.73 & \cellcolor{teal} \color{white}21.18 & \cellcolor{teal} \color{white}375.36 \\ 
		peak (\textdegree$/s$) & 238.37 & \cellcolor{orange!70} 157.74 &\cellcolor{teal} \color{white} 40.26 & 935.51 \\ 
		\multicolumn{5}{l} {\cellcolor{gray!30}	\textbf{Smooth pursuit}}\\  								
		duration (ms) & 276.78 & 265.67 & 74.40 & 953.66 \\ 	
		dispersion (pixels)&   \cellcolor{orange!70}399.93 &\cellcolor{teal} \color{white} 112.41 &  
		\cellcolor{orange!70}336.01 & \cellcolor{orange!70} 505.03 \\ 
			
	\end{tabularx}
	\end{tcolorbox}
	\caption{All 46 features with their derivations. Expert class.}
	\label{table:expertsAll}
\end{table}

\subsubsection*{Significant features}
\label{subsubsec:wilcoxonFeatures}
Secondly, we had a look at the features themselves and checked for differences between the single features according to their class and as well as checking for the significance level of feature differences under 0.11\%. We built a model based on the features that have a significance level under 0.11\% (Tables \ref{table:novicesAll}, \ref{table:intermediatesAll} and \ref{table:expertsAll} all white cells, gray cells mean there is no significant difference between the groups).

\subsubsection*{Most frequent features}
\label{subsubsec:mostfrequentfeatures}
In a third approach, we reduced the number of features by running the prediction on all 46 features 150 times. By taking the most frequent features in the model, we search for a subset of features that prevent the model from overfitting and allow for interpretable results representing the differences between expertise classes with a minimum amount of features. These most frequent features are imperative for the model to distinguish the classes. During training, the model indicates which features are the most important for prediction in each run. The resulting features with the highest frequency (and therefore highest importance for the model) in our test can be seen in Tables \ref{table:novicesAll}, \ref{table:intermediatesAll} and \ref{table:expertsAll}, in orange.

\subsection{Results}

\label{sec:results}

We first report the results of an intra-expert classification test to see 
whether inter-experts differences are smaller than inter-class differences. 
Then, since we first need to know whether there are differences between experts 
and novices, the classifiablity test (binary classification) provides a deeper 
analysis of the model trained with all features for distinguishing experts and 
novices. The remaining chapter describes two ternary models which are based on 
a subset of features obtained through 1) their significance level and 2) their 
frequency in the all feature model.

\subsubsection*{Intra-expert classification}
To strengthen the implicit assumption of this paper that it is possible to distinguish between novices, intermediates, and experts based on their gaze behavior, we evaluated our expert data separately by flipping a subset of experts with intermediates. After 100 iterations in which half of the experts were randomly labeled as intermediates, the average classification accuracy was below chance-level, meaning the model can not differentiate between experts and flipped experts properly. This strengthens our assumption that inter-expert differences are smaller than inter-group differences between experts, intermediates, and novices.

\subsubsection*{Binary classification}
\begin{figure}[h]
	\centering
	\includegraphics[width=0.5\linewidth]{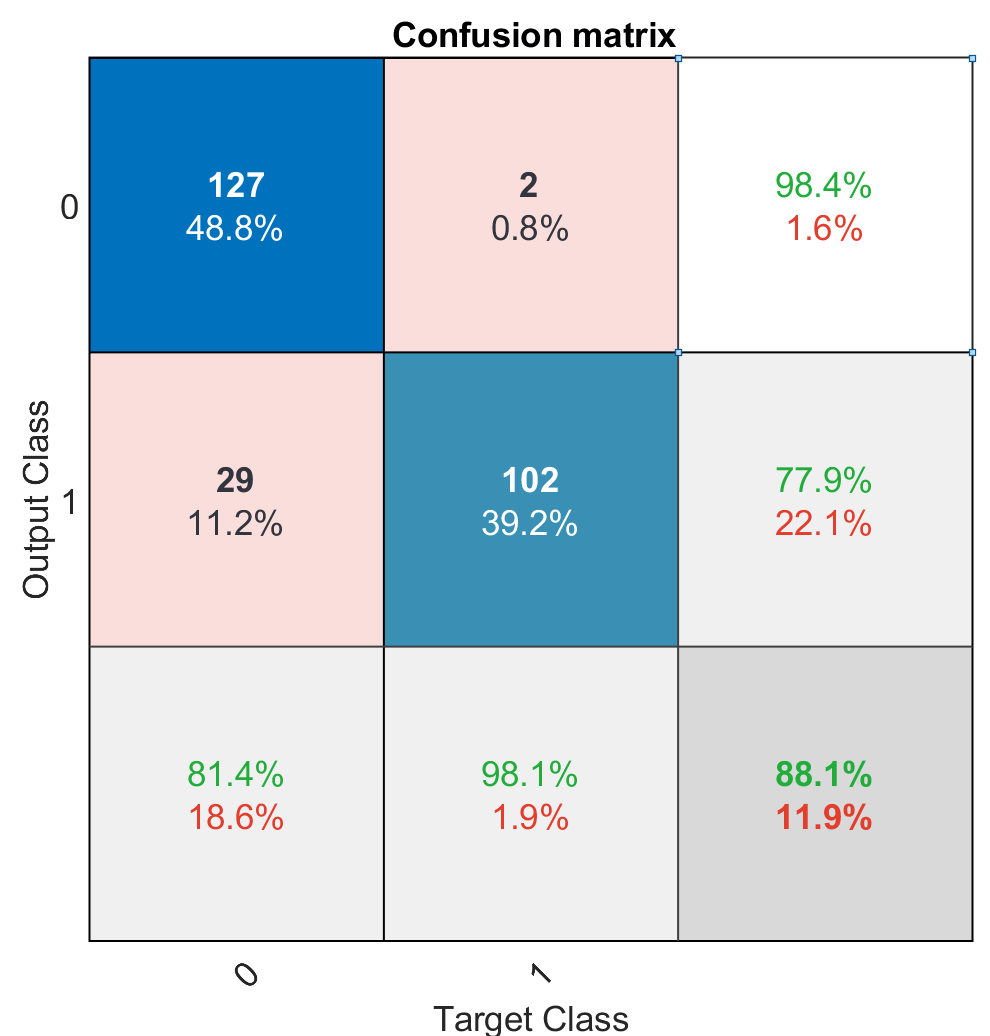}
	\caption{Binary confusion matrix about predictions on 100 randomized runs. 
	}~\label{fig:confusionMFF}	
\end{figure}

The classifiability test shows promising results. This binary model is able to distinguish between experts and intermediates with an accuracy of 88.1\%. The model has a false negative rate of 1.6\% and a false positive rate of 18.6\%. This means the binary model predicted two out of 260 samples falsely as class zero and 29 samples that are class zero as class one. As the false-negative rate is pretty low, the resulting miss rate is low (11.9\%) as well. The confusion matrix (Fig.~\ref{fig:confusionMFF}) shows the overall metrics. The binary model is better in predicting class zero samples (intermediates) than class one samples (experts). The overall accuracy of 88.1\% is sufficient to investigate a ternary classification. In the following, we show deeper insights on the ternary approaches by looking at accuracy, miss rate, and recall of the ternary models and compare those values between the All-feature model (ALL), most frequent features model (MFF), and significant features model (SF). This is to see if there is a better-performing model with fewer features.

\subsection*{Accuracy}
~\label{subsubsec:classifiability}

\begin{figure*}[ht]
	\centering
	\includegraphics[width=1\linewidth]{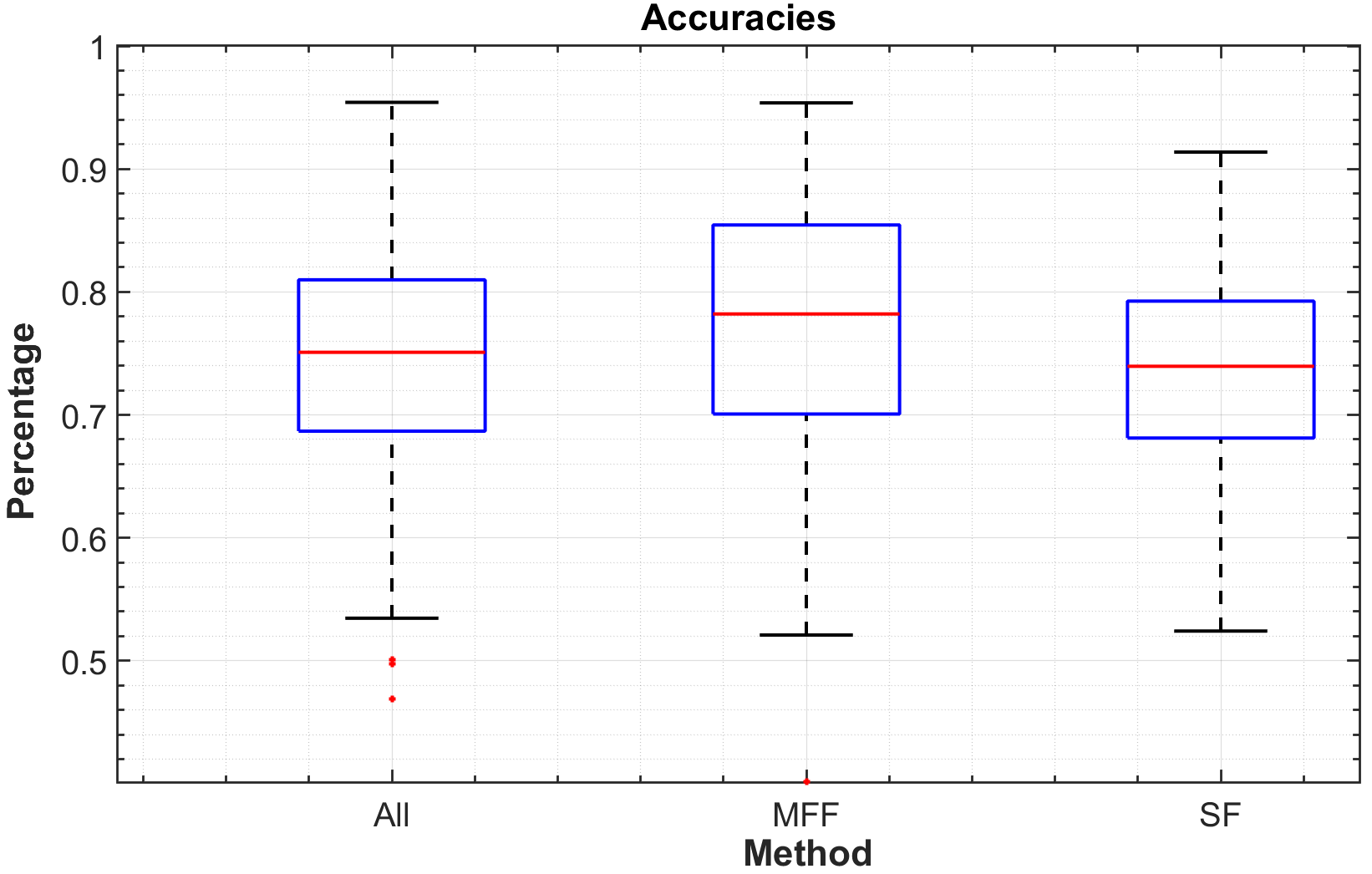}
	\caption{Box plot showing the accuracy values of the ternary methods. All 
		three models have median accuracy values $\sim 75-80\%$}
	~\label{fig:accuracies}

\end{figure*}

The differences in accuracy between the three approaches are barely visible 
when looking at the median (ALL: 75.08\%, MFF: 78.20\%, SF: 73.95\%), but even 
greater when comparing the 75th percentile (ALL: 80.989\%, MFF: 85.44\%, SF: 
79.25\%, see Fig.~\ref{fig:accuracies}). All models show a wider range of 
accuracy values which means these models might over fit more on some runs and 
under fit on others. The lower adjacent of all models is higher than the chance 
level (ALL: 53.46\%, MFF: 52.93\% and SF: 52.41\%), which means all models 
perform better as guessing. The chance level for 3 classes is 33.33\%. A system 
that would only guess the correct class would usually end up with an accuracy 
of about 33.33\%. Although not in each run, on average all models show a much 
better performance. Even the worst classification is over 20\% higher than the 
chance level. Successful performance for classification expertise in machine 
learning models is usually when their average accuracy is between 70\% and 
80\%. A statement about the performance of a model with lower than 70\% 
accuracy depends on the task and how much data is available. Sometimes there 
are only a few people in the world who can be considered experts. As the 
accuracy is a rough performance metric that only provides information about the 
number of correct predictions (true positives and true negatives), we offer a 
more detailed look into the performance of the methods by comparing the miss 
rates of the single approaches.

\subsection*{Miss rate}

The miss rate is a metric that measures the rate of wrongly classified samples belonging to class x but predicted to belong to class y. The models are better at predicting the membership of samples belonging to expert and intermediate classes than the novice class. This results in miss rates that are only a little lower than the chance level when looking at the median miss rates (All: 28.12\%, MFF: 23.81\% and SF: 26.80\%, see Fig.~\ref{fig:missRates}). The upper adjacent shows a high range of miss rates reaching even values of over 43.19\% for the SF-model. The MFF-model has the lowest median miss rate of all three methods with a miss rate of 41.96\%.

\subsection*{Recall}

\begin{figure}[ht]
		\centering
	\includegraphics[width=1\linewidth]{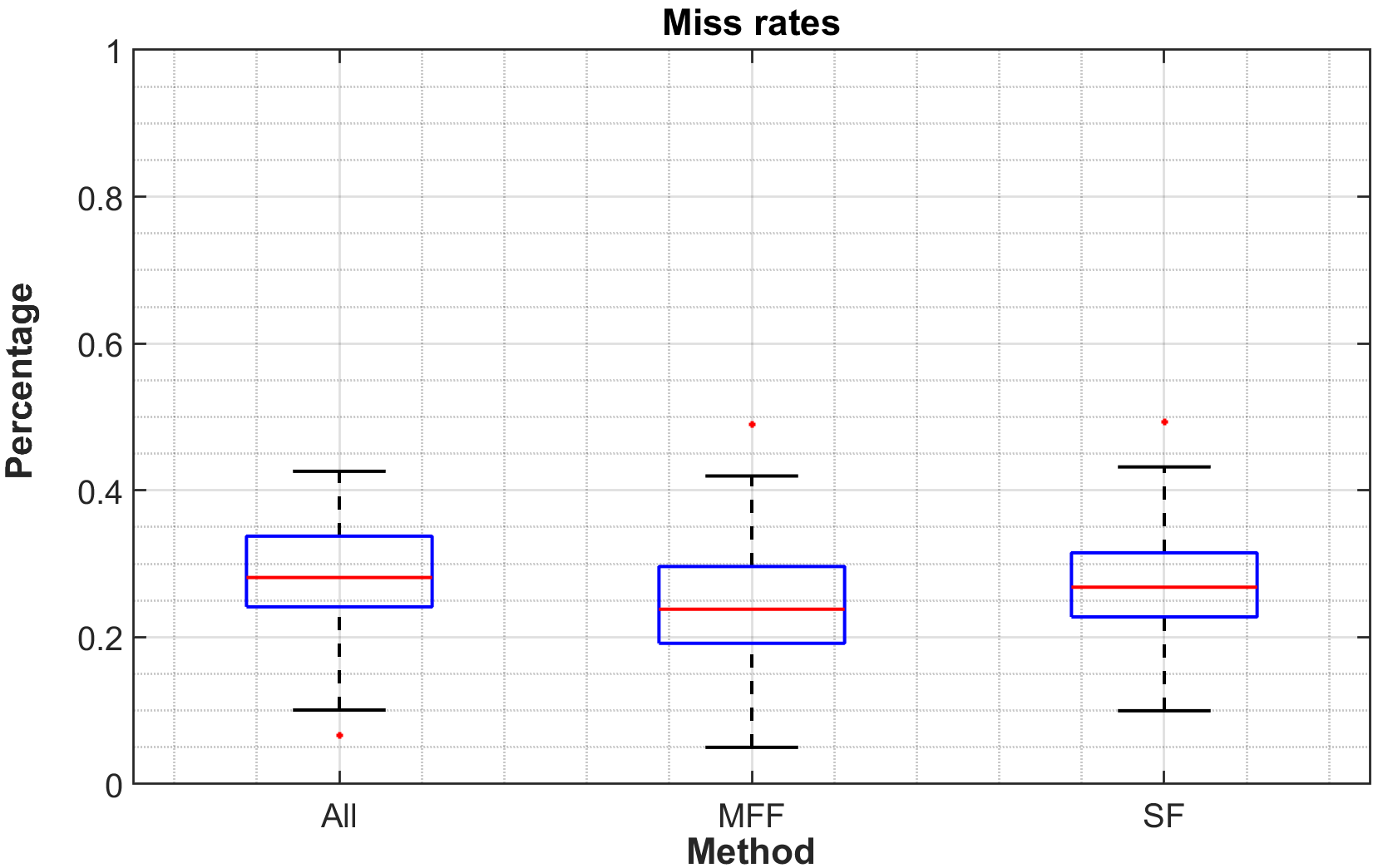}
	\caption{Miss rates of ternary methods. }~\label{fig:missRates}	
	
	\centering
	\includegraphics[width=1\linewidth]{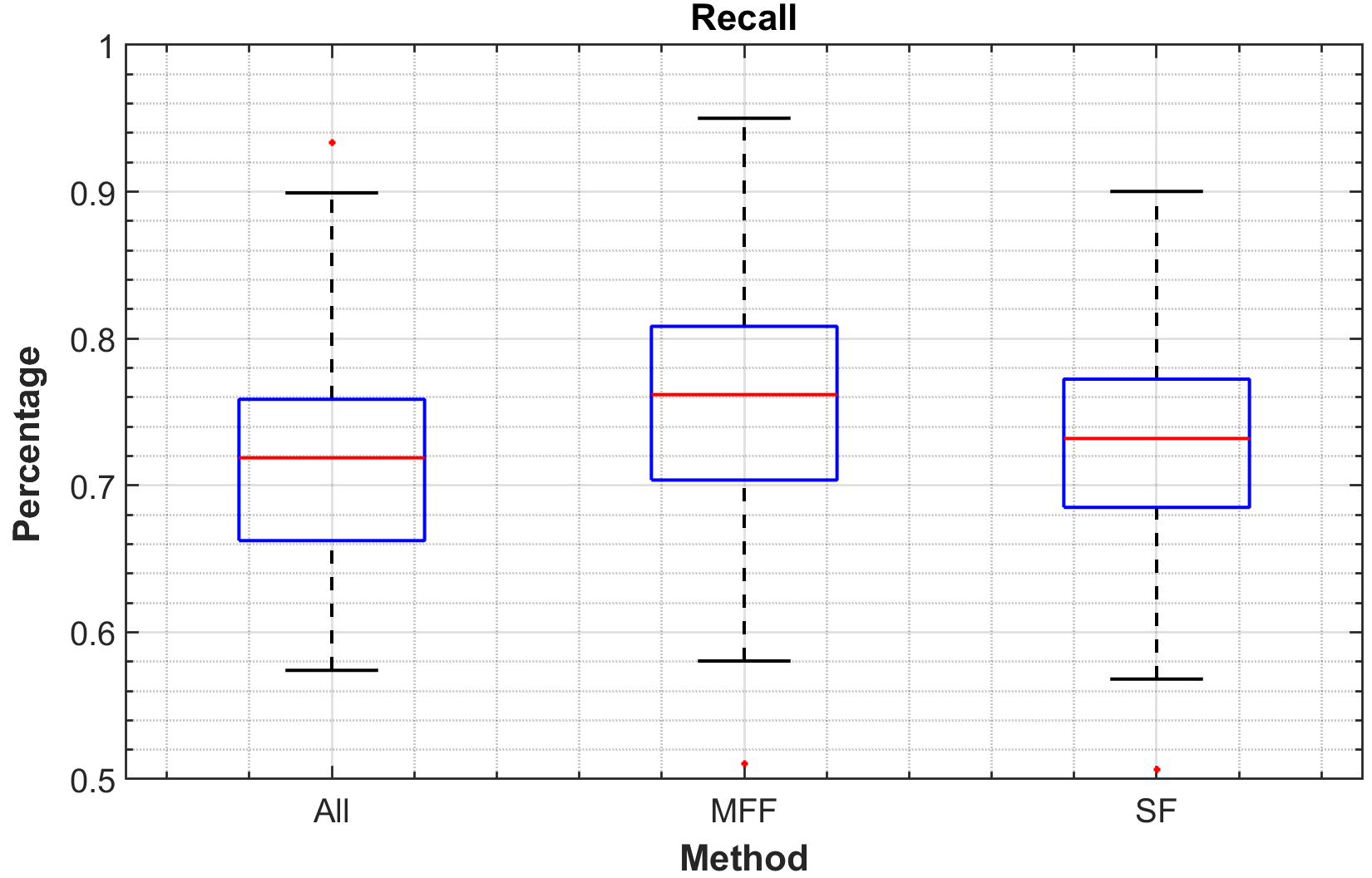}
	\caption{Recall values of ternary methods. }~\label{fig:recalls}
\end{figure}

Recall provides information about the rate of predicted samples belonging to class x in relation to the number of samples that really belong to class x. All three models have a median recall of over 70\% (as can be seen in Fig.~\ref{fig:recalls}). In the ternary case, the chance level is at 33.33\% which means all models have a recall over two times higher than the chance level as the lower adjacent of all three models is higher than 33.33\%. The MFF-model median is the highest at 76.18\% followed by the SF-model at 73.19\% and the ALL-model at 71.87\%. Again the MFF-model has the best performance values of all three methods.

\subsection*{Most frequent features}

The most frequent features in 100 runs are summarized in Table \ref{table:mff1}. Only the minimum of the saccade duration has $p > 0.011$. This means the differences are not statistically significant. All other features show significant differences, signifying that a Mann-Whitney-U-test discards the null hypothesis that there are no differences with $p < 0.011$ for each of the features.

\begin{landscape}
	\centering
		\def\arraystretch{1.5}
	\begin{table}[p]
	\begin{tcolorbox}[colback=PP-blue!60,{colupper=white}]
			\begin{tabularx}{\textwidth}{lcccccc}

				\multicolumn{7}{c}{\textbf{ Most frequent 
				features} }\\  \cline{1-7} \\
				
			\cellcolor{PP-blue!90}	& \cellcolor{PP-blue!90}derivation &\cellcolor{PP-blue!90} novice &\cellcolor{PP-blue!90} intermediate &\cellcolor{PP-blue!90} expert & \cellcolor{PP-blue!90} p-value & 
				\cellcolor{PP-blue!90}significant?   \\

				\rowcolor{PP-blue!50}saccade duration (ms) & std. dev. & 38.869 & 59.726 & 35.548 & 3.33*e-08 & 1\\ 	
				\rowcolor{PP-blue!50}saccade duration (ms) & minimum & 26.514 &  26.127 & 25.019 & 	0.242  & 0\\ 
				\rowcolor{PP-blue!50}peak saccade deceleration (\textdegree$/s^2$) & std. dev.  & 4166.262 & 4770.063 & 5492.287 & 2.49*e-18 & 1\\ 
				\rowcolor{PP-blue!50} peak saccade velocity (\textdegree$/s$) & std. dev.  & 129.294& 	136.529 & 157.740 &  6.19*e-07 & 1\\ 
				\rowcolor{PP-blue!50} smooth pursuit disp. (pixels) & average  & 622.805 & 	425.089 & 399.939 & 9.66*e-82 & 1\\ 			
				\rowcolor{PP-blue!50}smooth pursuit disp. (pixels) & minimum  & 185.437& 168.320 & 336.016 & 5.44*e-12 & 1\\ 			
				\rowcolor{PP-blue!50}smooth pursuit disp. (pixels) & maximum  & 1085.903& 	694.370 & 505.031 & 1.52*e-81 & 1\\ 			
							\end{tabularx}
			\end{tcolorbox}
			\caption{\textbf{ All most frequent features.}} 
			\label{table:mff1}	
			
	\end{table}
	
\end{landscape}

%
%
%
%
%
%
	

\subsection{Discussion}
~\label{sec:disc}


In this work, we have presented a diagnostic model to classify the eye movement features of soccer goalkeepers into expert, intermediate, and novice classes. We further investigated how well the features provided by the diagnostic model led to explainable behavior.
Our model has shown that eye movement features are highly informative and well suited to distinguish different expertise classes. Based on a support vector machine as a simple machine learning model, we were able to classify three different expertise groups at an average accuracy of 78.2\% (compared to the baseline of 33.3\% in a three-class classification problem), thus a quality result for current machine learning techniques. As the performance values differ, the real-world application has to be further evaluated with larger subject groups. A closer look at the classification results reveals that our model can distinguish correctly between experts and intermediates. This is due to the fact that experts and intermediates have already been tested in the sense that they play in higher leagues and have already proven their ability. Thus, there is ground truth for these classes.  A limitation of the classification model is currently the novice group. Since our novice group consists of participants with no regular training or involvement in competitions, novices can be equally talented players regarding their gaze behavior who have simply not yet proved their ability in a competition. This assumption is especially evident in the false-negative rate of 1.6\% and the false positive rate of 18.6\% from the binary model, respectively. This means that 18.6\% of novice samples are classified as intermediate samples, but only 1.6\% of the intermediate samples are classified as a novice. As is usual in expertise research, a proportion of low performers (novices) can also be found in higher classes. Our models confirm that the correct classification of novices is considerably more difficult than other classes since there is, to date, no objective ground truth.
Despite this limitation, our model achieved a very good average accuracy of 78.2\%. Most likely, a model with more subjects and finer graduation of the novices would offer a much better result. Machine learning models are data-driven and therefore can learn more from more data. However, the number of elite youth goalkeepers in Germany who can provide samples for the expert class is highly restricted. Out of 56 in total, we collected data from 12 for our study. An additional step would be to define a more robust ground truth for participants classified as novices. As it is more important that the model does not downgrade participants with higher expertise to a lower class, it can still be used as a diagnostic model. As aforementioned, the false positive rate only shows, that some novices with limited experience can perform better than others and therefore be classified into a higher class. This is correct because their gaze behavior is closer to intermediates than it is to typical novices.


By examining the individual eye movement features in more detail we have shown that, on the one hand, a subset of features is sufficient to create a solid classification and, on the other hand, that the differences in eye movement behavior between the individual groups are difficult to interpret. 
We only investigated the most frequent features since these features built the best-performing model. The differences are noticeable, but hard to interpret as there is no simple characteristic behind these features. 

There are indications that 1) experts (std. dev. 35.54 ms) as well as novices (std. dev. 
38.86 ms) have a more homogeneous saccade behavior compared to intermediates (std. 
dev. 59.72 ms). The lengths of the saccades differ less. However, it would be a fallacy to 
attribute the same viewing behavior to novices and experts due to the similar 
standard deviation and minimum duration of the saccades (novice: 26 ms, intermediate: 
25 ms, expert: 25 ms). It is clear that both groups have similarly long saccades, but the novices have similarly long saccades and the experts similarly short saccades. 
Conversely, this means that the experts might have longer fixations than the novices and intermediates. These findings are in line with Mann et al. [20] who show that experts are over-represented in fewer, but longer fixations. Their visual strategy is often based on longer fixations to avoid saccadic suppression (which might lead to information loss). In our statistics, fixation durations did not exhibit to have significant differences between the three groups. This is in line with the findings of 
Klostermann et al. \cite{klostermann2020fewer}. It also might be based on the split of the fixation values in short fixations and smooth pursuits. The source of these differences may also be the age difference between the single groups (see Table ~\ref{tabA:participants}). With the current data, this is not rigorously answerable.

Further differences between the groups can be found in the maximum peak deceleration of the saccades. There is a continuous increase in the maximum 
deceleration speed of the novices' saccades (4166.262 \textdegree$/s^2$) to intermediates(4770.063 
\textdegree$/s^2$) to experts (5492.287\textdegree$/s^2$), which is in line with the findings of Zwierko et al. \cite{zwierko2019oculomotor} who found that the deceleration behavior can be inferred from different expertise classes. 

One observation made by the experimenter during the study was that novices often follow the ball with their gaze for a long time. This behavior is less evident among experts. 
They tend to only look at the ball when it has just been passed or when they themselves are not in play. At these times, the ball can not change its path. This observation is supported by the values of the smooth pursuit dispersion. With 505.031 pixels maximum and 336 pixels minimum, experts have a very narrow window of smooth pursuit lengths. Basically, the maximum smooth pursuit of the experts (505.03 pixels) is less than half as long as the novices (1085.90 pixels), and the minimum smooth pursuits 
(expert: 399 pixels, intermediate 425 pixels, novices 622 pixels) is still 1/3 shorter than
the novices. The intermediates are placed in the middle between the two groups. Again,
the values are continuously decreasing. Based on the continuity of the average smooth pursuits that correlate negatively with the classes, as well as the maximum and standard deviation, it can be concluded that experts tend to make smooth pursuits of a 
more regular length. One explanation for this could be that, in addition to the opponents and players, the ball, as an almost continuously moving object, attracts a 
high level of attention. In order to maintain a clear overview in the decision-making 
process, soccer players are taught the following behavior: Shortly before the ball arrives at the pass goal, you look at it. This is done until the ball is passed away. Since the path of the ball can only be changed by a player who is in possession of the ball and not in the middle of a pass, it is only necessary to follow the path of the ball at the beginning and end of the pass. In the meantime, players should scan the environment for changes to keep track of options in the field. This leads to short smooth pursuits around the ball before the end and at the beginning of each pass so that experts can appreciate the ball and follow the ball with similarly long smooth pursuits. On the other hand, as aforementioned before, novices often follow the ball's path almost continuously or, at least, very often. The characteristics of the smooth pursuit support this theory. 
The characteristics of smooth pursuits differ significantly from one another in the 
three groups with an average, minimum, and maximum significant p-value of less than 
$1*10^-12$. The novices with 622.81 pixels make, on average, much longer smooth pursuits than the intermediates (525.09 pixels) and significantly more than the experts (399.93 pixels). 
With 185.44 pixels, the shortest smooth pursuits of the novices are smaller than those of the intermediates (168.32 pixels) and the experts with 336.01 pixels. The maximum values show a uniform behavior. With 1085.9 pixels, the novices have the highest 
maximum values after the intermediates with 694.37 pixels and the experts with 505.03 
pixels. 

Although the standard deviation of the lengths of the smooth pursuits does not belong to the MF features, clear differences can be seen here as well. The dispersions of 
the smooth pursuits with 201.27 pixels scatter far more among the novices than among 
the intermediates (124.85 pixels) and experts (112.41 pixels). These findings lead us to believe that a stimuli oriented investigation on gaze distribution for expertise 
recognition might reveal even more pronounced differences, i.e correlation between ball 
movement and smooth pursuits.

\subsection{Conclusion \& Implications}

After the ternary classification of expertise, the next step should be the evaluation of a more robust classification model. As machine learning techniques are data-driven, adding more subjects to each group should, presumably,  provide better results. As soon as a robust model is built, a finer-grained gradation should be considered to achieve a more sensible model that allows for the classification of participants in more classes by predicting their class in a more nuanced fashion. In our further work, we plan to expand our data set to more subjects in the current groups, add more nuanced classes and add a physical response mode to infer speed and correctness in a standardized, controllable and objective manner, thus increasing the immersion. however, a fully interactive mode will only be possible when CGI can provide high enough quality and cost-efficient environments. Another step is to focus on the research of person-specific, gaze-based expertise weakness detection. As soon as a robust model is achieved, another point is to integrate the model into an online diagnostic system. To use the model online, the gaze signal can be directly drawn online at 250 Hz from the eye tracker by using the provided API of the vendor. Using a multi-threaded system, the data preparation and feature calculation can be done directly online in parallel to data collection. Only the higher level features (e.g. std. deviations) need to be computed when the trial ends and fed as a feature vector to the already trained model in order to estimate the class of the current trial. As predicting is completed by solving a function, the prediction result is supposed to be available a few moments after the trial ends. This is necessary as the prediction is the input for the adaption of the training. This work will be implemented in an online system for real-time gaze-based expertise detection in virtual reality systems with an automatic input for the presentation device to ensure dynamic manipulation of a scene's difficulty. With a prototype running in VR, we are planning to expand the system to be used in-situ with augmented reality glasses (AR). This may further pronounce the differences and lead to even better classifications. A more sensible model would allow, by mapping expertise on a larger number of classes, the dynamic manipulation of the difficulty level of a training system exercise or game level in virtual environments. Next to a training system for athletes and other professional groups, the difficulty level in a VR game can be dynamically adjusted based on the gaze behavior of the user. We are, however, aware that the small sample size restricts potential conclusions that can be drawn and may lead to contentious results. Another limitation of this work is the restriction presented by head movement unrelated eye movement features and the absence of a detailed smooth pursuit detection algorithm, which might be important. Therefore, in our future work, we will implement an appropriate event calculation method i.e. based on the work of Agtzidis et al. \cite{agtzidis2019ground}.
This work, however,  strengthens the assumption that there are differences between the gaze behavior of experts, intermediates, and novices, and that these differences can be obtained through the methods discussed.
Using machine learning techniques on eye-tracking data captured in a photo-realistic environment on virtual reality glasses can be the first step towards a virtual reality training system (VRTS). Objective expertise identification and classification leads to adaptive and personalized designs of such systems as it allows for a definition of certain states in a training system. A VRTS that can be used at home and, based on its objective and algorithmic kind, allows for self-training at home.  The choice of difficulty can be adapted based on the expertise of the user. For higher-skilled users, the level of difficulty can be raised by pointing out fewer cues or showing more crowded, faster/more dynamic scenes to increase the pressure placed on decisions. With enough data, it is also possible to adapt the training level based on personal deficiencies discovered during expertise identification in a diagnostic system. This can result in a system that knows a user's personal and perceptual weak spots to provide personalized cognitive training (e.g. different kinds of assistance like marking options, timing head movements, showing visual and auditory cues). Such a system is also potentially applicable in AR as the findings on the photo-realistic VR setup can be used in AR settings (i.e. in-situ). For uses such as AR-trainings - that can enhance physical training - the fundamental findings must be based on real gaze signals.
As a second step, training systems can be developed based on the diagnostic findings. As, in addition to physical training, perceptual-cognitive training forms are increasingly being researched \cite{appelbaum2018sports,wilkins2020early,burris2020visual,klemish2018visual}.

\newpage

\section{Differentiating Surgeons' Expertise Solely by Eye Movement Features}

\subsection*{Abstract}

Medical schools are increasingly seeking to use objective measures to assess surgical skills. This extends even to perceptual skills, which are particularly important in minimally invasive surgery.  Eye tracking provides a promising approach to obtaining such objective metrics of visual perception.  In this work, we report on results of a cadaveric study of visual perception during shoulder arthroscopy. We present a model for classifying surgeons into three levels of expertise using only eye movements. The model achieves a classification accuracy of 84.44\% using only a small set of selected features. We also examine and characterize the changes in visual perception metrics between the different levels of expertise, forming a basis for development of a system for objective assessment.

\subsection{Introduction}

Arthroscopy is a popular minimally invasive surgical procedure that improves patient outcomes while at the same time conserving hospital resources.  According to Monson et al. \cite{monson1993advanced}, patients experience less pain, have fewer complications and recover faster than with traditional open surgery. However, a surgeon needs advanced technical skills for this type of operation \cite{hermens2013eye}.  Arthroscopy involves inserting instruments and a scope into the joint (e.g. shoulder or knee) through small incisions.  A key capability in performing arthroscopic surgery is the ability to use the scope to navigate through complex anatomy of the joint for inspection, diagnosis, and to locate the surgical site.  The scope can rotate in multiple dimensions and casts its image on a screen placed next to the patient, which surgeons largely rely upon during surgery. Navigation is challenging due to complex anatomy, limited field of view, projection of the 3D space onto the 2D monitor, and the rotation of the monitor from the instrument plane.  

Due to these technical challenges, there is growing interest within the medical community to optimize training, including having objective measures of performance for tasks like navigation. Since navigation is a psychomotor task in which visual perception plays a crucial role, it is natural to look to eye tracking for such a measure.  Indeed, the role of eye movements is increasingly being investigated in surgery \cite{hermens2013eye}. In particular, the role of eye movements is increasingly being investigated (for an overview see \cite{hermens2013eye}). To determine whether eye tracking can serve as a basis for an objective measure in arthroscopy, first it must be determined whether, and to what extent, differences in surgeons' expertise are reflected by their eye movements. The findings from this study are significant for the design of adequate training and evaluation scenarios for perceptual-cognitive diagnostic and training systems.
In this work, we consider the perception of surgeons using eye movement patterns from three expertise levels in a human cadaveric study of diagnostic arthroscopy of the shoulder. We selected this task since it focuses on navigation skill in which perception plays a major role.  We use stimulus-independent eye movement patterns to develop a model to classify the subjects into the three levels of expertise.  Using only a small number of selected features, our model achieves a classification accuracy of over 84\%. We further investigate differences in eye movement patterns among the three classes in order to understand how these patterns evolve with increasing levels of expertise. We hope that such an understanding can assist in developing specialized training to provide the appropriate support to surgeons at different expertise levels.

\subsection{Related Work}

In eye tracking studies, using artificial forms of presentation like virtual reality (VR)~\cite{law2004eye,zheng2011surgeon} or images \cite{sodergren2010hidden,eivazi2012gaze} could omit important perceptual details requiring the participants to fill in through inference which often subsequently leads to the higher levels of frustration \cite{zheng2011surgeon}.
To provide a presentation mode that is as natural as possible, we use so called soft cadavers that provide natural tactile sensation while maintaining the naturalness of the scene. Although remote eye trackers are commonly used in lab studies ~\cite{law2004eye}, as soon as the participant changes to another direction (e.g. down at the cadaver), they can no longer capture the gaze signal. 
To allow the participant to use normal gaze behavior and move freely without data lost, we use a head-mounted eye tracker in combination with a 4k-screen. This setup supports natural gaze behavior as well as high control of the stimulus allowing us to capture highly detailed information of the tissue on a screen with high resolution and gaze signals on the cadaver, both with the same field camera.
Eye tracking studies in surgery are differed in how they evaluated the gaze signal. The gaze signal on the stimulus was considered, i.e. target gaze behavior, switching behavior (alternating gaze between target and instrument), or following behavior (eye following the instrument) \cite{law2004eye}. Other studies focused on quiet eye periods \cite{wilson2011perceptual}. However, there are also studies that have gained insights at the feature level. For example, Kocak et al.\cite{kocak2005eye} used stimulus-independent eye features in their binary classification and found significantly lower saccade rates, as well as significantly higher peak velocities for experts, which was confirmed by other studies \cite{hermens2013eye}. Tien et al. \cite{tien2011quantifying} found a higher fixation rate in experts. 
Eivazi et al. \cite{eivazi2012gaze} show differences in time to first fixation and mean fixation duration. However, theses differences were not confirmed by Sondergren et al. \cite{sodergren2010hidden}, as in both studies fixation durations are analyzed differently and the choice of regions of interest plays an important role. These results show that eye movements can be used to assess the surgical expertise and to define differences between groups. 

Many studies have focused on the detection of differences in expertise between experts and novices \cite{wilson2011perceptual,zheng2011surgeon} 
and only few studies have focused on the development of eye movements. Studies focusing on development have used mostly simulators \cite{kocak2005eye} or images \cite{sodergren2010hidden}. Hidden Markov models (HMM) used in the latter study reveal differences in eye movement patterns between high and low performers. So far, several algorithms have been introduced to eye tracking including supervised methods like support vector machines ~\cite{hosp2021soccer,castner2018scanpath} 
and neural networks \cite{castner2020deep}. Ahmidi et al. \cite{ahmidi2010surgical} mixed instrument movements with eye movement data and achieved a binary classification accuracy of 82.5\% for skill level classification.
All these studies show that eye movement data can be used to differentiate between experts and novices and that it is not necessary to determine exactly where the surgeons were looking to measure their skill accurately.

\subsection{Participants and Methods}

\begin{figure}[!th]
	\centering
	\includegraphics[width=1\textwidth]{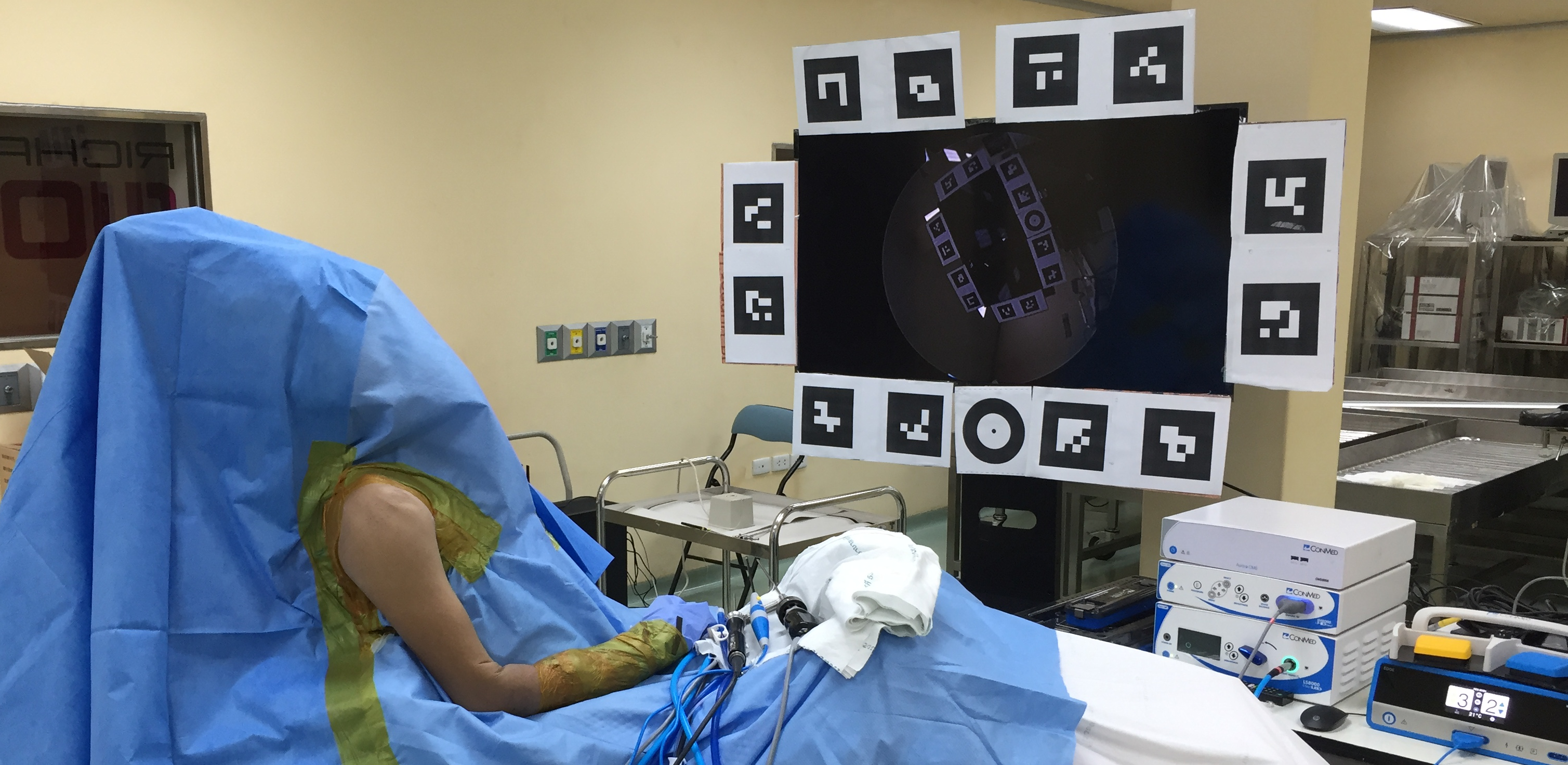}
	\caption{Experimental setup showing cadaver, arthroscopic equipment, and 4k monitor with ARUCO markers.}
	\label{fig:setup1}
\end{figure}
\subsubsection*{Procedure}

This work makes use of the eye tracking data set from the work of Yin et al. \cite{yin2020study}. Their data set contains eye movement data for three classes of surgeons: 3rd year residents (R3), 4th year residents (R4), and fellows. Each class consists of five (n=5) participants, equally. We even use the data of the two participants that were left out in their study because of a gaze signal offset. Since we only use relative features, we can use the data of these participants too. In their study, participants were placed in front of the cadaver and four feet away from the 4k, 52-inch screen where the output of the arthroscope was displayed (Figure ~\ref{fig:setup1}). Each participant was familiarized with the setup and asked to navigate and diagnose 12 anatomical landmarks in the shoulder, while wearing a Tobii Pro Glasses 2 eye tacker. The gaze was recorded with Tobii software.

\subsubsection*{Data preparation}

The Tobii Glasses 2 were set to a frame rate of 100 Hz, thus a gaze sample is available every 10 ms and saved with a timestamp, x-, and y-coordinates. The samples are used to calculate fixations and saccades metrics using the Tobii Fixation Filter, with a sliding window averaging method and the feature classification algorithm.
These samples are used to calculate metrics like fixations and saccades. To calculate these metrics, we used the Tobii Fixation Filter, using a sliding window averaging method. The feature calculation is based on the classification algorithm 
of Olson \cite{olsson2007real} with a default velocity threshold of 0.7 pixels/ms. The raw eye tracking data, as well as the fixations and saccade metrics, are exported from the Tobii Studio software. From the fixations, velocity of the saccades, saccade duration, values of the gyroscope (yaw, pitch and roll) as well as the amplitude of saccades, we use the person-specific average, minimum, maximum and standard deviation as features. While Tobii provides metrics about the first saccade and first fixation too, we did not include them.  Since our participants were familiarized with the glasses for different lengths of time when the trials started, we end up with chaotic first saccades, which have no informative character. 

As our aim is to infer which features contribute to expertise differences, we first used all the exported features from the Tobii Studio Software and added common metrics to them. Subsequently, we evaluated their frequency in the model building process and rated the most frequent used features to build a model with this subset of features for expertise acquisition. To incorporate uncertainties, we trained the model 150 times and calculated the most frequently used features by taking the features with the maximum number of occurrences in the training process.
We added certain typical eye movement features which we calculated by ourselves. The fixations were split into small fixations and smooth pursuit fixations. As the Tobii Software does not provide calculations of smooth pursuits, which are assumed to help differentiating different expertise classes, the smooth pursuit events were encoded in the fixations. We therefore treated fixations with a dispersion over 30 pixels as smooth pursuit fixations. This threshold was empirically defined during data analysis. \newpage
The set of features was:

\begin{itemize}
	\item Saccade duration (average, min, max. std. dev.)
	\item Fixation duration (average, min, max. std. dev.)
	\item Smooth pursuit dispersion (average, min, max, std. dev.)
	\item Fixation frequency
	\item Saccade frequency
	\item Pupil diameter (average, min, max. std. dev.)
	\item Gyroscope X,Y,Z (average, min, max. std. dev.)
\end{itemize}

We decided to include the gyroscope values because they could provide information about head movement between screen and cadaver may be revealed. 
The integration of pupil diameter features is based on the assumption that experts may have less fluctuating pupil diameter since their mental effort is considered to be smaller. Vice versa, the pupil diameter of intermediates and novices may reveal expertise differences by such effects.

\subsubsection*{Machine learning model}

We used all 38 features to build a support vector machine (SVM) model in 150 independent runs. On each of the 150 runs we keep out one participant (leave-one-out validation). This participant is our test set and has never been seen by the model (of the current run) before. Therefore, in each run we take all data of the remaining 14 participants to train the model and test it with the unseen data of the test set participant. While the training algorithm iterates over the same procedure it changes the participant for the test set (sequentially iterating over the participant numbers from 1 to 15) 150 times. Thus, each participant is used as test set 10-times in total. By having 10 runs for each participant, we are taking statistical fluctuations into account. To ensure independence between runs, we train a new model on every run and report the accumulated accuracy values of the 150 runs. Thus, in each run, the model is trained with 14 participants and tested with the test set data of one participant, which is unseen by the current model. A strict separation of data in a participant-wise manner is very important, as mixing up samples of one person into training and testing data would allow the model to remember person-specific (idiosyncratic) features and restrict a real expertise learning process. 

On each run, the data of the  14 participants of the training set is split into 10-folds. This is called a 10-fold cross-validation. The cross-validation is important to protect the model against over fitting. In each fold, $\ceil{\frac{1}{10}}$ of the 14 participants that belong to training set is used to validate the model that is trained with $\floor{\frac{9}{10}}$ of 14 participants. Which participant belongs to training or validation set, is decided randomly. However, the split is always done participant-wise to prevent an idiosyncratic learning behavior of the model. 

In a first model, we use all 38 features to check the classifiability of the data set and afterwards reduce the amount by taking the four most frequent features of 150 runs. The most frequent features are features that have the highest importance values for a single model prediction. In each run we built a queue of all 38 features sorted by importance for the current model. Subsequently, we computed their overall frequency over all models.

\subsection{Results}
\begin{table}[b!]
	\centering
			\def\arraystretch{1.5}	
	\begin{tcolorbox}
	\begin{tabularx}{\textwidth}{Xll}
		\multicolumn{3}{c}{\textbf{ Most frequent and 
		important features}}\\ \cline{1-3} \\
	\cellcolor{gray!30}	&\cellcolor{gray!30}Feature & 
	\cellcolor{gray!30}derivation\\
				 		
		1. &Peak velocity of saccades & standard deviation\\	
		2. &Amplitude of saccades & minimum \\
		3. &Total amplitude of saccades & sum\\
		4. &Saccade duration & standard deviation\\
		
	\end{tabularx}
	\end{tcolorbox}
	\caption{ The most important and frequent features on 150 runs.}
	\label{tbl:mff1}
\end{table}

Our first classification model shows promising results with an average accuracy of 60\%.
As a system that would simply guess the class, would only reach a chance-level of 33.33\%, the all feature model can already be considered as well-performing. But as we want to specify the results to allow a precise statement about a high performing classification with the least amount of features, we continued by collecting all features and their importance values on 150 runs of the all feature model and took the most frequent features (MFF) as a new set. With this subset of four features, shown in Table~\ref{tbl:mff1}, earlier counteracting features may be avoided and a precise statement about the differences of the groups can be stated. 
The final SVM model with the four MFF uses a linear kernel and a box constraint of 11.0174. We adopted one-vs-all approach for multi-class classification with the kernel scale remains 1. Before training, we standardized the data. Training took about 56.03 sec.

\subsubsection*{Performance metrics}
\begin{figure}[ht]
	\centering
	\includegraphics[width=0.5\linewidth]{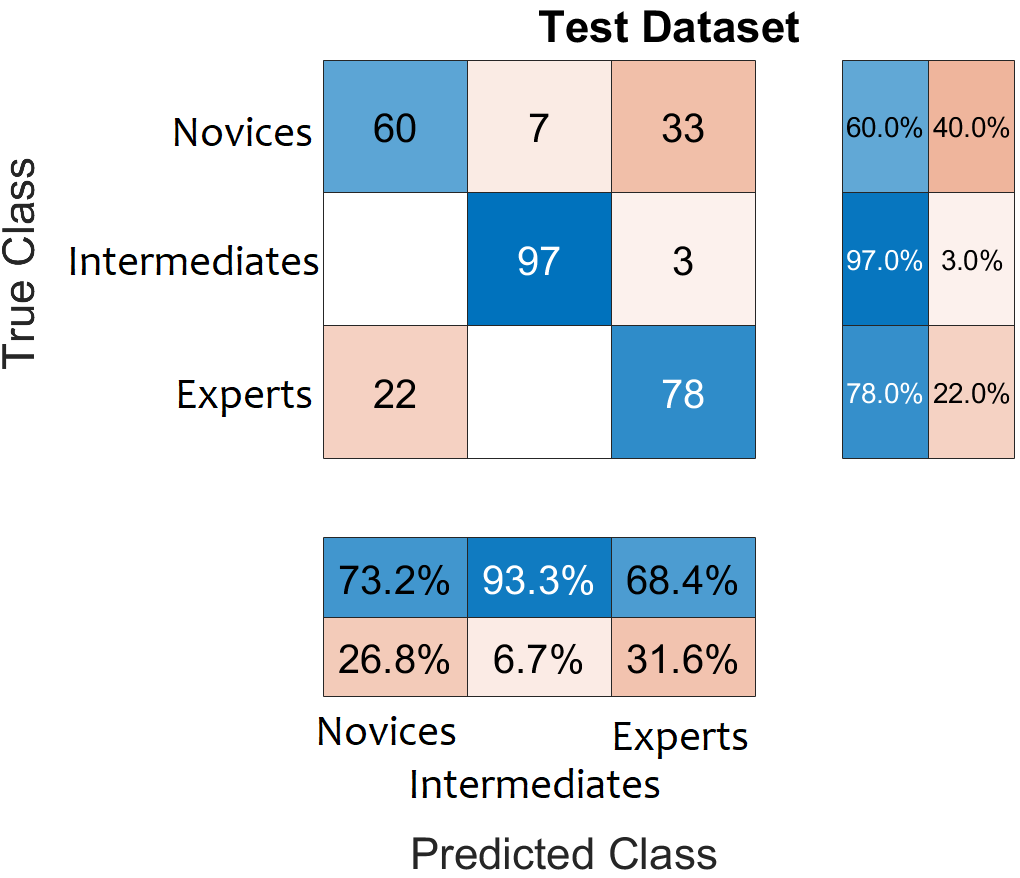}
	\caption{Performance values on 100 runs. }
	~\label{fig:all1}
\end{figure}

With an accuracy of 84.44\% the model improved over 20 percentage points, compared to the all feature model. Figure~\ref{fig:all1} shows the confusion matrix after 100 runs. 7 samples of the novice class were classified as intermediate and 33 as expert. This results in a class accuracy of 60\%. The classification of the intermediates peaked at 97\%, as only 3 samples were classified as experts and non as novices. This is especially interesting since the intermediates are in between the other classes and are therefore more likely to spread to both sides. The expert samples were with 78 samples correctly and 22 samples as novice samples, the second best classified class. The average recall is with 95.43\% extremely high which is confirmed by the average miss rates. Only 4.57\% of samples were misclassified. For the SVM model we achieve an area under the curve (AUC) of 0.91.

\subsubsection*{Feature evolution}

As we consider there is a cognitive process going on, forming the optimal gaze behavior from novice to expert, we have a look at the evolution of the gaze features between the classes to describe such process as good as possible. 
To analyze these evolutionary steps, we have a look at the single feature characteristics separately. We do that with the four MFF from Table~\ref{tbl:mff1}. Table~\ref{tbl:evolution1} contains the characteristics of the four MFF features.

\begin{table}[!h]
	\centering	
		\def\arraystretch{1.5}
	\begin{tcolorbox}	
		\begin{tabularx}{\textwidth}{lccc}
		\multicolumn{4}{c}{\textbf{Average Feature Evolution}} \\\cline{1-4} \\
		 \cellcolor{gray!30} &\cellcolor{gray!30} Fellow & 
		 \cellcolor{gray!30}R4 & 
		 \cellcolor{gray!30} R3\\ 
				 		
		Saccade peak velocity (STD) & 93.26 \textdegree/s & 121.72 \textdegree 
		/ s & 117.45 \textdegree/s \\ 
		Saccade amplitude (min) & 0.86 \textdegree & 0.40 \textdegree & 0.64 
		\textdegree \\ 
		Total saccade amplitude & 481.32 \textdegree & 1120.74 \textdegree & 
		1956.21 \textdegree \\  
		Saccade duration (std. dev.) & 18.96 ms \textdegree & 16.58 ms 
		\textdegree & 23.54 ms \textdegree \\ 
				
	\end{tabularx}
	\end{tcolorbox}
	~\caption{Average feature evolution between classes.}
	~\label{tbl:evolution1}
\end{table}

The table shows that experts have a smaller standard deviation of the peak velocity of the saccades (93.26 \textdegree/s). This feature is hard to interpret, but one assumption may be that experts have a more uniform distribution of saccade velocities. This means they do more saccades at the same speed, in a structured and planned way, compared to intermediates and novices. Interestingly, intermediates as the middle class between expert and novice show a much more diverse saccade peak velocity behavior (121.72 \textdegree/s). Novices are in the middle between experts and intermediates. A higher value for the standard deviation of the saccade peak velocities could be an indicator for a more chaotic gaze behavior, but it is hard to draw a conclusion about such a feature. When having a look at the minimum saccade amplitudes, we can see the same differences. The experts have on average a larger minimal saccade of length 0.86 \textdegree, compared to the intermediates with 0.40\textdegree and the novices with 0.64\textdegree. Again, we can see that the novices are in between the experts and intermediates. Only the total amplitude of all saccades shows a uniform evolution. The experts do a total of 481.32\textdegree of saccade length, where intermediates do more than twice the experts (1120.74\textdegree) and novices (1956.21\textdegree) even more than five time the experts and nearly double the intermediates. Another interesting feature evolution can be seen in the standard deviation of the saccade durations. This feature is also hard to interpret, but one possibility could be that experts with 18.96 ms and intermediates with 16.58 ms have slightly more order in their saccades than novices. Though the differences are very small and should be confirmed with more data.

\subsection{Discussion}

In this work we developed a model with supervised machine learning techniques that is able to distinguish three levels of expertise solely on the basis of eye movements during an arthroscopic surgery of the shoulder. With an accuracy of 82.33\% the model can be considered as performing well on this 3-class problem. Thus, it can be stated that expertise differences between three different groups of expertise are reflected by their eye movements. 
To further understand the differences between the three levels of expertise, we had a look at the four most frequent features of the model and analyzed the evolution of the characteristics between the groups. Except for the total amount of saccade amplitudes, the remaining three of the four most frequent features show a uniform evolution. First, novices tend to have a more chaotic gaze behavior and distribute their gaze over a larger portion of the scene by making many different saccades with different speed. They also tend to look more at the outside than the center. The evolution to intermediates shows an atypical behavior, as they tend to still gaze over a larger area of the scene than the experts, but do smaller saccades with a still diverse velocity. This might indicate, that they try to focus on more specific visual clues and start to concentrate on the center of the scene. In the next evolution step, the saccade velocities shrink significantly, which signifies a more planned scanning behavior, with somewhat longer saccades, concentrated more on specific areas. To summarize our findings, one can state that the evolution of novices to intermediates first tends to lead to a partly more chaotic gaze behavior, then turning to be more precise. With the investigations on the evolutionary steps, we can also define class dependent weak-spots in perception for each class. An evolution between the single classes is clearly recognizable. Thus, opening the way to a class-specific training system that is optimized for different steps in perceptional evolution. We also showed that for a high accuracy classification there are not many features needed. A subset of four features describing the gaze behavior is already enough to distinguish different classes. Luckily, the four features are easily calculated, which would allow the usage of the classification as an online classification system. Though, the classification would need to be done segment-wise after a certain period of time. 

Further steps are to add more participants to each class, and refine the number of classes. This would allow a much finer classification and therefore a better understanding of the differences between the levels of expertise. A finer classification is important to robust assumptions made by the model about gaze behavior and optimize the recognition of class-specific weak-spots to be used in a training system.

\newpage

\section{A Study of Expert/Novice Perception in Arthroscopic Shoulder Surgery}

\subsection*{Abstract}

Arthroscopic shoulder surgery is an advanced orthopedic surgical procedure, which is particularly challenging due to the complex anatomy of the shoulder, and tight spaces for navigation, which also limits the view from the arthroscope. In carrying out arthroscopy, the ability to quickly and effectively navigate through the joint to reach a desired location is essential. Novices often experience confusion in trying to triangulate the information from arthroscopy output with the background knowledge of anatomy while orienting and navigating the instruments. In this paper, we report on the results of the first cadaveric eye-tracking study of arthroscopic surgery in which we investigate differences in perception between experts and novices. Novices’ perception is analyzed with cognitive load analysis throughout the procedure and specifically, during the portions of the procedure in which subjects are observed to be confused. In investigating such portions, the gaze data analysis is supplemented with head rotations and acceleration information from gyroscope and accelerometer sensors from the eye tracker. We also use the gathered eye tracking metrics to construct a model to classify subjects into expert/novice.  We find statistically significant relations between head movement as well as pupil diameter and periods of confusion.  We identify a subset of the metrics that we use to build a simple classifier that is able to distinguish between novices and experts with accuracy of 84\%.   

\subsection{Introduction}

Arthroscopic shoulder surgery is an advanced orthopedic surgical procedure, which is particularly challenging due to the complex anatomy of the shoulder, and tight spaces for navigation, which also limits the view from the arthroscope. It is used to treat a number of disorders such as repair of torn tendons and rectifying chronic dislocation, as well as for diagnosis.  In all of these procedures, the ability to quickly and effectively navigate through the joint to reach the desired location is essential. An important aspect of navigation is the ability to quickly recognize anatomical landmarks and to focus attention on the appropriate region of the arthroscope image. For assessment and training it is important to have an objective assessment of such perceptual and attentional aspects and to detect portions of the procedure where students may become confused. 
In this paper, we report on the results of the first cadaver-based study to analyze and compare expert and novice eye movement patterns in performing arthroscopic surgery. We study the diagnostic arthroscopic shoulder surgery task since it involves navigating to various parts of the shoulder and inspecting them and thus allows us to focus purely on navigation skills.  The existing studies on comparing eye movement patterns between experienced surgeons and novices have predominantly used VR training simulators ~\cite{tien2010measuring,sodergren2010evaluation,wilson2010psychomotor,tien2011quantifying,zheng2011surgeon,atkins2012saccadic}, still images of the surgery  ~\cite{sodergren2011orientation,eivazi2012gaze} or physical box trainers ~\cite{kocak2005eye}.  We use so-called soft cadavers, which are specially prepared so as to retain the natural tissue properties. This means that our study is able to capture important aspects of the surgery such as tactile feedback and surgical setup not captured by simulations. Our work is also the first to study arthroscopic shoulder surgery.  Previous eye-tracking studies of surgery have concentrated predominantly on laparoscopic surgery which usually involves anatomy of the abdomen. In contrast, the diagnosis of the shoulder requires the surgeon to navigate the arthroscope through bones and muscles inside the rounded shoulder joint.
Experts can usually smoothly maneuver the arthroscope instruments with the automaticity developed through experience. In contrast, novices often experience confusion in trying to locate the anatomical landmarks from the magnified view of the operating site on the arthroscope output. Previous studies in the area of human-computer interfaces and intelligent tutoring have found pupil size and head movement to be associated with periods of confusion ~\cite{ehlers2018view,mccuaig2010detecting}. We sought to determine whether these metrics can also be used to detect confusion during shoulder arthroscopy and found positive relationships between both and novice states of confusion. Ours is the first study to attempt to use objective metrics to detect confusion during surgery.
An effective assessment instrument should be able to distinguish between performance of subjects with varying levels of experience and expertise.  We thus analyze the differences in gaze metrics between experts and two groups of novices of varying experience.  We identify a small subset of the metrics with good discriminatory power and use them to build a simple classifier that is able to distinguish between novices and experts with high accuracy. This leads us to conclude that there are significant differences in perceptual parameters between novices and experts in arthroscopic surgery that could be used for objective assessment as well as tutoring

\subsection{Related Work}

Arthroscopic skills are difficult to acquire because they require use of multiple tools, using both hands while viewing the surgical site on a two-dimensional display, with constant vigilance to the operating environment ~\cite{tauro2017arthroscopic}. Arthroscopic surgery is taught as a core component in a majority of orthopedic residency programs. Cadavers are often the first choice of surgeons for practice because they provide a real anatomical experience ~\cite{karahan2015effective}.  Other methods that have been tested with varying success in orthopedic teaching include interactive computer simulation~\cite{tofte2017knee}, physical simulation environments~\cite{tuijthof2010first} and virtual reality simulators~\cite{pedowitz2002evaluation,gomoll2008individual}. Approaches in assessing arthroscopic surgical skills include Global Rating Scales~\cite{hoyle2012validation}, motion analysis~\cite{howells2008motion}, virtual reality simulators~\cite{pedowitz2002evaluation,gomoll2008individual},  and simple bench model arthroscopic simulators~\cite{goyal2016arthroscopic}.

Eye tracking studies comparing experts and novices have been carried out in a number of surgical domains.  Tien et al.~\cite{tien2015differences} compared the gaze behaviors of experts and junior surgeons during key stages of a live open inguinal hernia repair. They found that experts have a higher fixation frequency and concluded that it could be due to lower mental demand resulting from automaticity developed through practice. Similar findings are reported by Erridge et al.~\cite{erridge2018comparison} during live laparoscopic gastric bypass surgery. Novices were found to pay less attention to the operative site but more to the sterile field. A number of studies of eye movement patterns of experts and novices ~\cite{wilson2010psychomotor,law2004eye,hermens2013eye} found that experts tend to fixate on the target more often than the instruments. Meanwhile, Law et al.~\cite{law2004eye} reported that novices either alternate their gaze between the target and instruments, focus on objects in between the target and the instruments, or follow the instrument on its way to the target. A study by Hermens et al.~\cite{hermens2013eye} also found differences in eye movement statistics between experts and novices. The experts in their study reportedly had lower saccadic rates and higher peak velocity, independent of where these eye movements were aimed. Similarly, in a study of global eye movement parameters of expert and non-expert participants, Kocak et al.~\cite{kocak2005eye} found that experts had significantly lower saccade rates and higher peak velocity than non-experts.  
Beyond analysis of eye movement metrics, a number of studies have used the metrics to build models to classify subjects into expert and novice. Eye metrics and tool motion data have been considered as features in assessing the skill of a surgeon while performing functional endoscopic sinus surgery~\cite{ahmidi2010surgical}. Hidden Markov models were built for seven different surgeries in two levels of expertise using the eye-gaze locations and the surgical tools motions. The findings revealed that eye-gaze data contains the skill-related structures, and combining it with the surgical tool motion data improves the classifier performance. Richstone et al.~\cite{richstone2010eye} used eye movement metrics to develop models to classify surgeons into experts and non-experts.  In a simulated surgery they achieved 91.9\% and 92.9\% accuracy with the linear discriminant analysis and neural network analysis, respectively and 81.0\% and 90.7\% accuracy in a live operating room setting. Eivazi et al.~\cite{eivazi2017towards} used a random forest classifier to classify micro-surgeons in the cutting and suturing tasks and achieved a 70\% recognition rate for the detection of expert and novice groups. Rose and Pedowitz~\cite{rose2015fundamental} investigate the assessment of basic arthroscopy skills using virtual reality modules developed through task deconstruction. Participants with the most arthroscopic experience performed better and were more consistent than novices on all 3 virtual reality modules. Greater arthroscopic experience correlates with more symmetry of ambidextrous performance.
While no work has investigated detection of confusion during surgery, detection of cognitive affective states such as confusion and boredom has been studied in the field of Intelligent Tutoring Systems. Pachman and colleagues~\cite{pachman2016eye} used eye tracking for early detection of confusion in a digital learning environment. In their study, the participants were asked to solve problems while their eye trajectories were recorded and this data was triangulated with self-ratings of confusion and cued retrospective verbal reports. Delucia and colleagues~\cite{delucia2014eye} sought to determine whether eye movements reflect confusion while users completed tasks with two simulated devices.  They measured confusion using a subjective Likert measure in which subjects were asked to rate their agreement with the statement “I was confused” and were not able to find consistent common correlation patterns between the variables for both devices, but they found that higher confusion ratings were positively correlated with the total fixation time on the whole screen, mean fixation duration and task completion time.  Lallé and colleagues~\cite{lalle2016predicting} included pupil diameter and head distance to the target as the predictors of the user’s confusion. They studied various combinations of gaze, pupil diameter, head distance and mouse events as predictors. The authors concluded that features of pupil size are strong predictors of confusion, which is consistent with the fact that pupil size is correlated with cognitive load, which plausibly correlates with confusion.

\subsection{Participants, Materials and Methods}

\begin{figure}[ht!]
	\centering
	\includegraphics[width=0.7\linewidth]{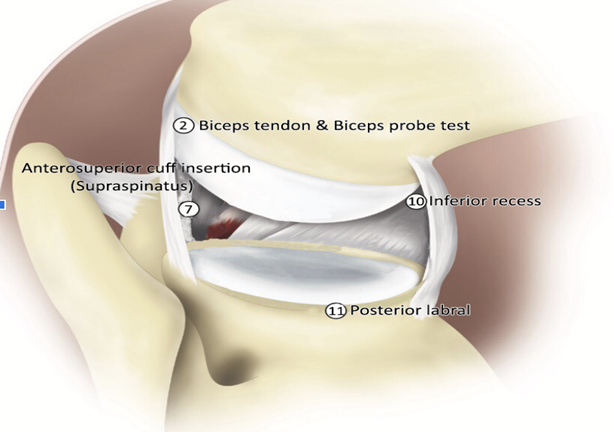}
	\caption{Portion of the shoulder anatomy with Landmarks 2, 7, 10, and 11 }
	~\label{fig:1shoulder}
\end{figure}

After obtaining approval from the Mahidol University Institutional Review Board, a total of thirteen participants (4 Females) were recruited.  They consisted of four fellows (two to ten years of experience) from the Department of Orthopaedics, Faculty of Medicine Ramathibodi Hospital, Mahidol University, and nine residents from the Orthopaedic Surgery Residency Program there.  Five of the residents were in the third year and four in the fourth year.  The residents were at an early stage of orthopedic training and were without prior arthroscopy experience. All the participants had normal or corrected-to-normal vision.
Eye gaze data was recorded using the Tobii Pro eye tracker (Tobii Glasses 2.0, Tobii Sweden), which was calibrated by looking at a marker placed near the arthroscopic output screen.  The cadaver (Male, 52 years old) was set up in the beach-chair position. An expert surgeon prepared the arthroscope setup (ConMed Linvatec) and inserted the primary portals into the shoulder prior to the procedure. The arthroscope camera output was displayed on a 52-inch screen which was placed four feet away from the participant. ARUCO markers were also placed around the screen in order to identify the screen in a later stage. Each participant was first acquainted with the cadaver setup, the diagnostic shoulder arthroscopy steps, and the evaluation study protocol. Each participant was asked to navigate and diagnose twelve anatomical landmarks within the shoulder in sequence (Table~\ref{tbl:1landmarks}).  The portion of the shoulder anatomy from viewing with the scope in the posterior portal and four visible landmarks 2, 7, 10 and 11 are shown in Figure~\ref{fig:1shoulder}. Among them, some are easy to navigate to and diagnose while some are more difficult. The landmarks which are categorized by the expert as hard to diagnose are highlighted and explanations are provided in Table~\ref{tbl:1landmarks}.
For each landmark, the expert provided explicit verbal instructions with the name of the landmark (e.g. “Start Biceps tendon”) to navigate to and upon arrival at the landmark, the expert called out its name (e.g. “reached Biceps tendon”). The start and end times for each landmark navigation task were recorded as part of the data stream. Throughout the procedure, a think-aloud protocol was used and the participants were asked to describe their immediate objective, actions and any points at which they became confused (when they could not find the landmark or they did not recognize the part of the anatomy they were in). 
In addition to the self-reported confusion, a member of the investigation team also monitored the participants and recorded portions of the performance as confusion in situations when a participant paused or made non-goal directed movements for a period of time which was followed by the attending surgeon’s assisting intervention. The study spanned two days, with the left shoulder of the cadaver used on the first day for six participants, and the right shoulder used on the second day for eight participants.

\begin{table}[ht!]

	\centering	
		\def\arraystretch{1.5}
		\begin{tcolorbox}
\begin{tblr}[\textwidth]{cX}

	 		\multicolumn{2}{c}{\textbf{Anatomical Landmarks}} \\
	 		\cline{1-2}\\
		1. & Rotator interval \\ 
		2. & \SetCell{l, gray!25} Biceps tendon \& Biceps probe test: easy to find long head biceps (LHB) but difficult for use probe to handle LHB (need another hand to control the probe)\\  
		3. & Biceps anchor\\  
		4. & Labral superior to anterior\\  
		5. & IGHL\\  
		6. & Subscapularis tendon and insertion\\  
		7. & Anterosuperior cuff insertion (Supraspinatus)\\  
		8. & \SetCell{l, gray!25} Posterosuperior cuff insertion (Infraspinatus): difficult to move from supraspinatus to infraspinatus (need to control the camera backward along the tendon).\\
		9. & Bare area\\  
		10. & \SetCell{l, gray!25} Inferior recess: difficult move from the posterior chamber downward direction to the inferior chamber\\  
		11. & \SetCell{l, gray!25} Posterior labral: difficult to slide the camera from inferior chamber to posterior than to superior chamber (the camera could easily back out from the trocar due to the limited space)\\ 
		12. & Back to Rotator interval\\

	\end{tblr}\end{tcolorbox}
	~\caption{Twelve anatomical landmarks to diagnose (The landmarks which are categorized by the expert as hard to diagnose are highlighted.)}
	~\label{tbl:1landmarks}

\end{table}

\begin{figure}[h!]
	\centering
	\includegraphics[width=0.5\linewidth]{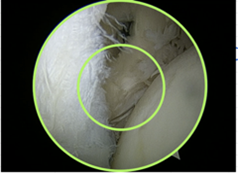}
	\caption{Inner and outer circles on the scope output. }
	~\label{fig:2innerOuter}
\end{figure}

\begin{figure}[h!]
	\centering
	\includegraphics[width=0.5\linewidth]{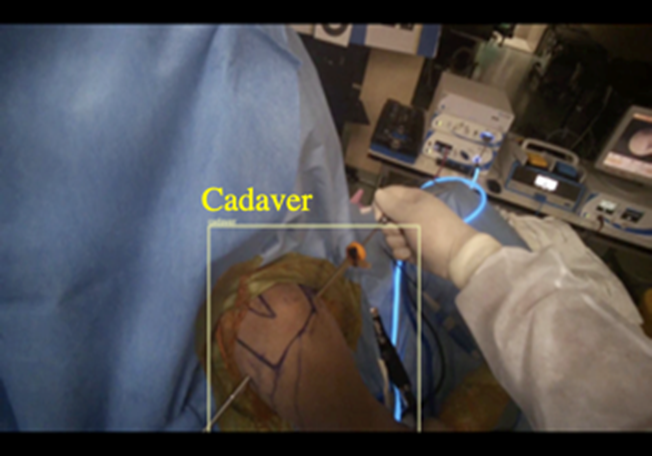}
	\caption{Detected cadaver shoulder on the video. }
	~\label{fig:3DetCadaver}
\end{figure}

\subsection*{Data Preparation}

From a preliminary study, we found that while surgeons diagnose a landmark, they tend to look at the center of the scope image and tend to look at the area near the circumference of the scope image in the direction of the next landmark to visit before moving the scope. We, therefore, define four areas of interest (AOIs): the center area of the scope image (the inner circle) (Figure~\ref{fig:2innerOuter}), the outer area of the scope image (outer circle) (Figure~\ref{fig:2innerOuter}), the arthroscope output screen (outside of the scope image), and the shoulder area on the cadaver (Figure~\ref{fig:3DetCadaver}). 
Eye-tracking metrics considered in this study are the rate and duration of fixations/saccades, the time to first fixation and the duration of the first fixation were calculated with the Tobii-I-VT Attention Filter using default parameters. Fixation is the visual gaze on a single location and saccades are the rapid movements of the eyes that abruptly change the point of fixation. The field view videos of the eye tracker were processed to demark the AOI’s.  The arthroscope output screen was detected using ARUCO markers and the scope view on the screen was detected using a simple circle detection method (cv2.circle()). The cadaver area in the video frames was detected by using the YOLOV3 CNN object detection model~\cite{redmon2018yolov3}  trained using transfer learning.  The cadaver shoulder in the video frames was labeled using the video labeler app from Matlab (R2019b). We used the video frames from three participants for the left shoulder and from two participants for the right shoulder area.

\subsection{Analysis and Discussion}

The two most commonly studied features of eye movement are fixations and saccades.  Fixations are visual gazes on a single location whereas saccades are rapid eye movements between fixations. Among the large number of possible eye tracking metrics, those commonly used in medical studies are fixation rate (number of fixations per second), saccade rate (number of saccades per second), fixation duration (length of each fixation), saccade duration, average time to first fixation, and duration of first fixation ~\cite{kocak2005eye,tien2015differences,richstone2010eye,merali2017eye,atkins2013surgeons}. We thus chose these metrics for the current study.  Along with the eye metrics, we used the completion time as an objective measure of skill. We categorized participants into three groups: four experts as E, four third-year residents RY3, and five fourth-year residents as RY4.

\subsubsection*{Gaze Data Analysis}

As shown in Table ~\ref{tbl:2eyegaze}, the average fixation rate of experts is higher than novices, but the expert’s average fixation duration is the lowest among all the groups. The average saccade rate and duration (ms) of experts is higher than the RY4 group. The expert’s average time to the first fixation is the lowest among the three groups, the average fixation duration is less than that of RY4.

\begin{table}[ht!]

	\centering	
		\def\arraystretch{1.5}
	\begin{tcolorbox}	
	\begin{tabularx}{\textwidth}{lXXX}
		\multicolumn{4}{c}{\textbf{Eye Gaze Metrics}} \\\cline{1-4}\\
			\cellcolor{gray!30}	&\cellcolor{gray!30}\textbf{Expert}	& \cellcolor{gray!30}\textbf{RY3} &\cellcolor{gray!30}	\textbf{RY4} \\ 
		Avg. fixation rate &	3.01&	1.62&	1.93 \\ 
		Avg. saccade rate		&	0.71&	0.39&	0.82\\ 
		Avg. fixation duration (ms)&	411.24&	490.37&	466.33\\ 
		Avg. saccade duration (ms)&	29.90	&35.77&	28.24\\ 
		Avg. time to first fixation (ms)&	50.00&	155.00&	450.00\\ 
		Avg first fixation duration (ms)&	1,039.50&	499.80&	1,269.25\\
		
	\end{tabularx}
	\end{tcolorbox}
	~\caption{Eye gaze metrics.}
	~\label{tbl:2eyegaze}
	
\end{table}

Overall, experts have higher fixation rates compared to the novices and the majority of their fixations fell on the scope image.  To investigate the fixation patterns of the expert and novice in the inner and outer circles AOIs of the scope, we considered 80\% of the process of navigating from one landmark to another into finding the general area of the landmark and another 20\% as zeroing in on the landmark. We found that during the 80\% portion experts and novices both tended to fixate more on the outer circle in a ratio of roughly 2:1. During the 20\% portion the experts fixated on the inner circle with a ratio of 2:1 while the novices continued to fixate on the outer circle with roughly the same ratio as before.  This shows that the experts adjust their focus of attention to suit the portion of the navigation task, while the novices keep their focus primarily in only one area. This could be explained by the fact that an expert would be expected to know that they are getting close to a landmark whereas a novice might not.  

\subsubsection*{Confusion}

With a handful of reference anatomical regions within the joint, novices often miss the target landmark to diagnose during the procedure. Failure to recognize landmarks may result in disorientation and confusion as a student seeks to navigate through the shoulder joint.    Since previous studies in user interfaces and intelligent tutoring had identified significant relationships between user confusion and metrics of pupil diameter and head movement, we sought to determine whether such relationships exist in this surgical domain as well.   
As head movement metrics, we used the gyroscope and accelerometer data available from the Tobii eye tracker. Six novice participants (3 RY3, 3 RY4) reported a total of 14 confusion points while navigating and diagnosing at landmarks 1, 3, 6, 7, 8, and 12. The number of confusion points per landmark ranged from one to five with the highest frequency of three times reported at the landmarks 1, 6 and 8. 

The follow-up interviews with the experts revealed that novices might get confused in landmark 1 due to a lack of recall of the background knowledge. At landmark 1, instead of looking for the void triangular space of the rotator interval between the subscapularis and glenoid and supraspinatus, the novices tended to look at the nearby structure.  While in landmark 6, the novices need to locate the insertion of supraspinatus on the humerus. In the experts’ opinion, the novices mostly focus on the tendon part, while all experts specifically focus on the tendon insertion point. This may be related to the level of knowledge of the pathological area on this tendon. The infraspinatus at landmark 8 is a tendon posterior to supraspinatus tendon. These tendons are blended together and have the same texture. Therefore, the location of infraspinatus can be identified only by understanding the exact location of infraspinatus (posterior half of these blended tendons). 

In terms of the time taken to complete the task, the experts completed the task with the least amount of time to diagnose at each landmark and had the least variation in task times. We observe that some landmarks require more time to navigate to and diagnose, particularly landmark 2 and 6 which are categorized as hard to diagnose. On average, the six novices who became confused took 1.5 times and 2 times longer than other novices in hard and easy landmarks, respectively.

The Percentage Change in Pupil Diameter (PCPD) is an objective measure of cognitive skills. Kruger et al.~\cite{kruger2013measuring} studied PCPD as a measure of cognitive load and compared it with different cognitive load metrics including EEG, heart rate and blink rate when students were watching a recorded academic lecture, with and without subtitles. They found that higher cognitive loads were associated with higher PCPD values. We expect that the subject’s cognitive load will increase while navigating the arthroscope in the landmarks where confusion was recorded. To determine that, we need a period of low cognitive load as a baseline.  We used the period from the end of the previous landmark until the beginning of the current (confused) landmark as the baseline period since during that period the subject just is not actively navigating through the joint.  The PCPD value was computed by subtracting the average diameter from the (confusion) landmark from the baseline diameter and divided it by the baseline diameter. From the six participants who became confused, the PCPD ranged from a minimum of 0.91\% (left eye) and 0.97\% (right eye) to a maximum of 1.22\% (left eye) and 1.12\% (right eye).  On average, during the periods of confusion the pupil diameter changed by 1.02\% in the left eye and 1.03\% in the right eye relative to the baseline.  The minimum values came from two novices at five different landmarks; all others had positive change in PCPD.  

We investigated the head movement of the novice participants during the landmarks with confusion using the information from the gyroscope and accelerometer sensors of the eye tracker. Confusion was not reported in landmark 2 (L2: Biceps tendon \& Biceps probe test) for any of the novices and hence it was considered as the baseline.  We compared the head rotation and acceleration information between novices with and without reported confusion by computing the differences between the minimum and maximum values in x-, y- and z-axes. T he differences are compared with the baselines using a paired t-test for each participant with confusion reported. The differences are significant in all three axes for head movements from the accelerometer as well as in y- and z-axes from the gyroscope sensors (p-value = 0.05). As shown in Table~\ref{tbl:3gyroAcc}, the average differences between the two groups are substantial in the x-axis for head rotations and the z-axis for acceleration.

\begin{figure}[h!]
	\centering
	\includegraphics[width=1\linewidth]{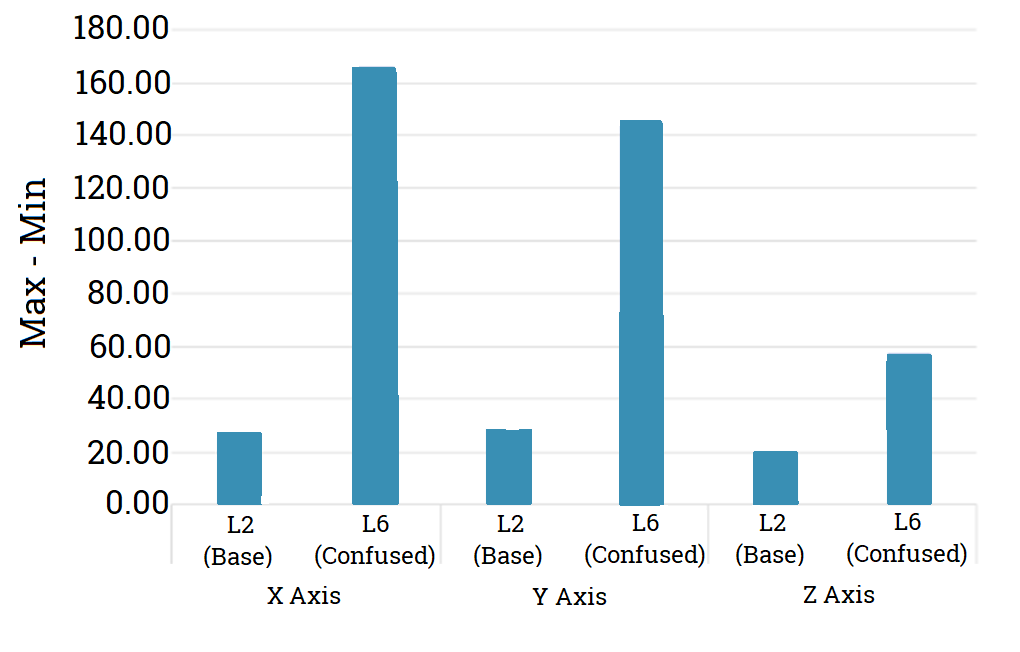}
	\caption{(a) The differences between minimum and maximum in three axes at the baseline landmark (L2) and the landmark with confusion (L6) from the gyroscope sensor.}
	~\label{fig:4_a}
\end{figure}

\begin{figure}[h!]
	\centering
	\includegraphics[width=1\linewidth]{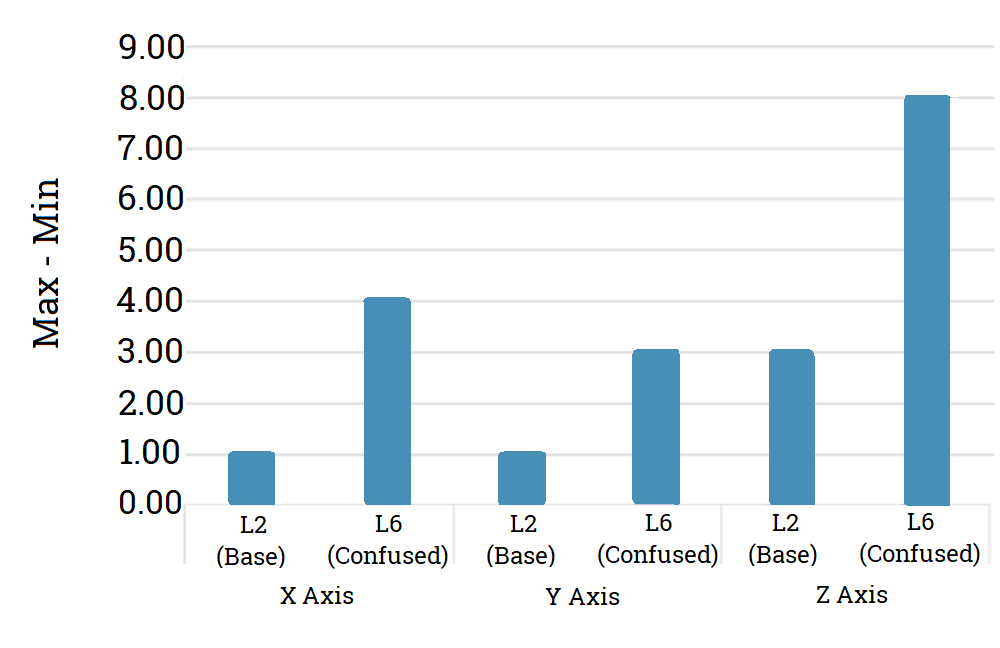}
	\caption{(b) The differences between minimum and maximum in three axes at the baseline landmark (L2) and the landmark with confusion (L6) from the accelerometer sensor}
	~\label{fig:4_b}
\end{figure}

\begin{table}[ht]

\centering
		\def\arraystretch{1.5}
	\begin{tcolorbox}
	\begin{tabularx}{\textwidth}{lXlll}
		\multicolumn{5}{c}{\textbf{Gyroscope and Accelerometer}} \\\cline{1-5}\\

		\cellcolor{gray!30}\textbf{Sensor}	 	 &    		\cellcolor{gray!30}                		  &			\cellcolor{gray!30}\textbf{x – axis}        &		\cellcolor{gray!30}\textbf{y– axis}	     & 		\cellcolor{gray!30}\textbf{z – axis} \\ 
				
		Gyroscope  	 		 &	Novices with confusion    & 106.67   				 &	52.50       		 & 28.17\\ 
		        			 &  Novices without confusion & 22.93	 				 & 30.12	    		 & 13.08\\ 
		Accelerometer		 &	Novices with confusion	  &	2.00	 				 & 2.50	     		  	 & 4.67\\ 
				     		 &	Novices without confusion &	1.37	  			     & 0.99	    			 & 1.83 \\		
	\end{tabularx}
	\end{tcolorbox}
		~\caption{Comparison in average differences in minimum and maximum values in three axes between novices with confusion reported, novices without confusion }
	~\label{tbl:3gyroAcc}
	
\end{table}

Figure~\ref{fig:4_a} the rotation from the gyroscope sensor and~\ref{fig:4_b} shows the acceleration from the accelerometer along the x, y, z axes of a novice participant (RY4). As shown in the figures, this particular novice rotates the head along the x-axis and moves along the z-axis while navigating the arthroscope to the landmark 6 and performing the insertion (Subscapularis tendon and insertion).

\subsection*{Classification}

In order to evaluate whether the eye-tracking metrics can be used to assess level of expertise in arthroscopic shoulder surgery, we sought to build models to classify participants as novice or expert.  Due to the small size of the data set, we used leave-one (participant)-out to validate the classifiers. We applied Synthetic Minority Over-sampling Technique (SMOTE) repeatedly to the remaining twelve participants' gaze features. In each iteration, we randomly selected three novices and four experts, and generated one instance of novice with SMOTE and added it back to the novice data pool. The process was repeated until we reached a total of 100 novices. In the same manner, we generated expert data instances until we achieved 100, resulting in a balanced data set with 200 instances. Features considered for the classification model included twelve gaze metrics extracted from the eye data including fixation and saccade rates for the whole procedure and three AOIs, average fixation and saccade rates, time to first fixation, and duration of first fixation. We selected the best five features using the information gain ratio (Table~\ref{tbl:4gainRatio}). 
With the logistic regression model, we achieved a classification accuracy of 84\%.  The logistic regression model misclassified an expert and an R3 novice who have similar fixation rates (gaze points/sec) and an R3 novice with similar time to first fixation with an expert. The results show that in the domain of arthroscopic shoulder surgery, although the differences in eye-movement data are multidimensional, the two groups of participants can be classified with high accuracy by a simple model.

\begin{table}[ht!]
	\centering	
		\def\arraystretch{1.5}
	\begin{tcolorbox}
		\begin{tabularx}{\textwidth}{llX }
			\multicolumn{3}{c}{\textbf{Selected Features}} \\\cline{1-3}\\
			
		\cellcolor{gray!30}\textbf{Feature}	 			& \cellcolor{gray!30}\textbf{Gain Ratio} & \cellcolor{gray!30}\textbf{Min, Max, Mean} \\ 
		
		Time 1st fixation (ms)		& 0.482			  &  Expert: 25.0,75.0, 69.2  		\newline	 Novice: 75.0, 1075.0, 181.2 \\ 
		Fixation Rate					& 0.418			  &  Expert: 2.0, 3.6, 3.3 	\newline	Novice: 1.2, 2.6, 1.6  \\
        Fixation Rate AOI In	    	& 0.381			  &	 Expert: 9.6, 15.0, 12.0  	\newline	 Novice: 5.9, 13.3, 10.0 \\ 
        Avg Fixation Dur. (ms)	& 0.358			  &  Expert: 221.7, 705.9, 302.4 	\newline Novice: 354.2, 640.8, 489.4 \\ 
		Avg Saccade Dur. (ms)	& 0.306			  &  Expert: 26.0, 35.0, 28.4 		\newline Novice: 25.4, 42.9, 34.2 \\

		\end{tabularx}
	\end{tcolorbox}
	~\caption{Selected features with the information gain ratio. }
	~\label{tbl:4gainRatio}
	
\end{table}

\subsection{Conclusion}

The required skill set for arthroscopy is complex, due to an indirect view of the surgical site through the arthroscope, limited tactile feedback, and complex hand-eye-coordination. The operative time, probe path length, and number of movements are commonly utilized as surrogate markers for assessing skills. While previous studies have centered around the dexterous aspects of motor skills, we investigate cognitive aspects by studying the differences in perception between participants of differing experience~\cite{pedowitz2002evaluation}. During the arthroscopic surgery, surgeons rely primarily on visual information. Perception and attention are two separate but related processes. Initially attention occurs, and perception follows.  
This study has shown that there are significant differences between expert and novice focus of attention during the arthroscopic navigation task both overall and during particular portions of navigation. We investigated a number of other questions such as the relationship between user confusion and metrics of pupil diameter and head movement, as well as whether the eye-tracking metrics can be used to classify the experts and novices. In contrast to the existing studies, the gaze measures in our study are collected with the cadaver specimens which provide the most realistic experience.  We have demonstrated the potential of eye-tracking to provide reliable tools for automatic performance assessment in arthroscopic shoulder surgery. This leads us to the conclusion that gaze data carries important information about the skills of arthroscopic surgeons which could contribute to automated objective assessment. The future steps of this research include the development of an intelligent training system in the virtual reality environment that dynamically detects novice confusion and classifies surgeon’s performance based on eye-movement data

\subsection*{Acknowledgement}

This work was partially supported through a fellowship from the Hanse-Wissenschafts- kolleg Institute for Advanced Study (HWK), Delmenhorst, Germany to Su Yin for collaborative work with the University of Bremen, and through a study group grant from HWK to Haddawy.  It was also partially supported through a grant from the Mahidol University Office of International Relations to the MIRU joint unit.

\chapter{Cross-Domain Generalization}


\blfootnote{This chapter is based on the following publications:\\
	
	\begin{itemize}
		\item \textbf{B.W. Hosp}, F. Schultz, O. H{\"o}ner, and E. Kasneci. "In Search of A Superior Gaze Behavior: Cross-Domain Shared Expertise-Related Gaze Features." \\
		
		\item 	\textbf{B. W. Hosp}, F. Schultz, E. Kasneci, and O. H{\"o}ner. “Expertise classification of soccer goalkeepers in highly dynamic decision tasks: A deep learning approach for temporal and spatial feature recognition of fixation image patch sequences,” Frontiers in Sports and Active Living, vol. 3, p. 183, 2021.

	\end{itemize}
}

\newpage


\section[Cross-Domain Shared Expertise-Related Gaze Features]{In Search of A Superior Gaze Behavior:Cross-Domain Shared Expertise-Related Gaze Features.}

\subsubsection*{Abstract}

When we talk about perceptional expertise, we usually talk about it in certain limits like a domain or a task. However, so far, there has been no proof found that states expertise is restricted to such limits. Perceptional expertise might also have some kind of domain- or task-independent source,  which is shared by experts from different domains. Such perceptional expertise is considered to prove that it is possible to generalize gaze behavior and describe it as a domain-independent skill. Seeing generalized, cross-domain perceptual expertise definition as a far-reaching aim, a first step is to find commonalities and differences between experts from different fields. Therefore, we are investigating a minimal set of features from one domain to build a machine learning model and predict the expertise of samples from two other domains. The diversity of the performance values might indicate that not domain but task similarity or other boundary conditions are more important for generalization.

\subsection{Introduction}

On the one hand, it is assumed that experts develop their optimal methods of perception by solving highly similar tasks for many years and
optimizing their perception in the process. Thus, expertise forms over years of experience and practice. On the other hand, however, it is assumed that there are certain commonalities in the gaze behavior of experts. In addition to
these commonalities, the differences between levels of expertise are also of particular
interest for research \cite{ericsson1991toward,mann2007perceptual,bertrand2009effects,williams1998visual,vaeyens2007effects,williams1994visual,roca2011identifying,roca2013perceptual}. This interest originates from the ability to
derive findings of perception at different developmental stages, but also from the
ability to develop the diagnostics as a foundation for possible support options, based on
findings of perception research. Different expertise classes show different similarities
in perception so that a beginner needs completely different assistance than an advanced
user \cite{hosp2021differentiating}. In recent years, perception of experts has been investigated in various fields and tasks \cite{castner2020gaze,tanaka1997expertise,murphy2021esport,hosp2021soccer,hosp2021differentiating}. In sports psychology, expertise has been linked to more efficient gaze behavior in decision-making tasks \cite{mann2007perceptual,bertrand2009effects,williams1998visual,vaeyens2007effects,williams1994visual,roca2011identifying,roca2013perceptual}. However, aspects of perception, that allowed separation of expertise classes, were often found, but could not lead to consistent results. Thus, findings are often dependent on domain and task type. While a look at the current research situation shows a mass of expertise research studies, only little inter-domain or inter-task work is done, so most work is somehow limited to a task or domain. Gegenfurtner et al. \cite{gegenfurtner2013transfer} show that it is possible to transfer expertise from familiar tasks to semi-familiar tasks, but not to unfamiliar tasks. Likewise, they took the same subjects for both tasks, which introduces a high risk of enabling recognition of subject-specific characteristics instead of expertise. Thus, while differences have often been found, only little is known about inter-task or at least inter-domain expertise that is transferable or generalizable. The problem of a missing generalizable feature set that works
for more than one task or domain, has yet not been confirmed. So far, no dedicated set of traits has been found that is better suited to recognize expertise than others. Therefore, previous study results could hardly or not at all be transferred to other studies and were always limited to one field, task or at least data set \cite{klostermann2020fewer}. However, since it is expected that experts in the same task exhibit certain commonalities in gaze behavior, in a subsequent step, experts could also exhibit certain commonalities regardless of the task or even domain. To prove this hypothesis, studies are needed that evaluate the gaze behavior with the same methods, on different tasks, or in different domains. The overall question is
whether expertise-related features derived from visual behavior are consistent across
domains and whether experts from different domains share (at least some) visual strategy features.
A superior set of perceptual properties would lead to a complete overturning of our
understanding of expertise. Such perceptional expertise is considered to prove that it is possible to generalize gaze behavior and describe it as a domain-independent skill. Seeing generalized, cross-domain perceptual expertise as a far-reaching aim, a first step is to find commonalities and differences between experts from different fields.  Therefore, we are investigating a set of features, shared by three different domains. We use a minimal feature set to infer expertise classes by training a machine learning model with data of one domain and testing the model with unknown data from the two other domains, by predicting the expertise classes of the new data. 

Especially, when looking at highly dynamic tasks from a more generous perspective, one can see that it typically consists of fast movements. Decisions need to be made in little time and have usually a high impact on the continuation of a task. To capture underlying cognitive processes with eye tracking, we use features from even volatile and fast movements recordings of the eyes.

\subsection{Methods}

\label{sec:methods}

In the first step, we collected all the gaze data from three distinct studies that we conducted. 
Each of the data sets contains samples of subjects that were previously assigned (based on their skill) to one of the following classes: expert, intermediate, or novice. As we use supervised learning algorithms, we need an external classification to label some of the samples we collected as belonging to the correct class. We do this for a certain amount of samples but equally distributed on each of the present classes. This bunch of data is called the training data set. With such data, we train our algorithm to recognize the connection between a training sample (defined representation of the gaze behavior in form of samples for each video, operation, image, etc.) and its correct class (novice, intermediate, and expert). By training the algorithm, we want to use a representation of the gaze data that optimally describes the gaze behavior of a sample. The better the representation can describe the commonalities of samples from the same class and differences of the samples from different classes, the better our algorithm can be taught how to predict the class of new, unknown samples. In the next step, some samples, that have not been labeled yet are fed to the model. The model has no idea about the class membership of this bunch of samples, which is called the testing data set. By feeding the model this unknown data, we can estimate how well the model behaves when we collect more data and predict their class membership. It is therefore a quality measure. We applied this method to all three of the following studies. Thus, we build a model, that 1) can recognize different skill levels of subjects based on their gaze behavior, 2) prove how well it behaves on unknown data, 3) allows the whole process to be reproducible and objective and 4) if possible, provides insights about the perceptional development between the expertise classes in a cross-domain manner.

\subsubsection{Study A}

Data set A contains the samples of 33 soccer goalkeepers and 28 soccer field players from two studies in virtual reality on decision-making. For these, we took an HTC Vive with an integrated SMI eye tracker, which is capable of recording the eyes with 250 Hz. 
We defined typical in-game scenarios of soccer. Resulting in unique videos, in which youth players of the VfB Stuttgart played the defined scenes on the training space of the youth performance center of the VfB Stuttgart. While they were acting the scene, a  360° camera was placed at the position of the subject to capture the realistic field of view. The task of the subjects was to decide how to continue the game after the last return pass to the position of the subject on the field, as the screen went black after the last pass. We collected data of n=12 experts (expert youth soccer goalkeepers from U-15 to U-21) during two youth elite goalkeeper camps of the DFB. Data of the n=10 intermediate players were recorded in our lab. The intermediates are goalkeepers from the regional league in Germany (semi-professional). The novices (n=13) had no experience in competitions and no training on a weekly basis, but up to 2 years of experience. The study was confirmed by the Ethics Committee of the Faculty of Economics and Social Sciences of the University of Tübingen. 

The subjects of the field player study (n=14) are all from the VfB Stuttgart youth elite program. Therefore, they are all considered to be experts. They all play higher than the regional league and have a lot of experience in competitions.

\subsubsection{Study B}

The second study was made in arthroscopic surgery, where, usually everything takes place in front of the surgeon. As such, a field of view camera provides a much better resolution of a smaller area, which is technically better suited. Especially in arthroscopic surgery, the main focus of the surgeon lays on the patient and the scope output which is usually a big screen in front of the surgeon where the arthroscopic camera sends its video feed. With such a camera, more details can be captured. 
We asked surgeons to navigate an arthroscope through a portal on a soft cadavers' shoulder to the operating site where the tendon of the shoulder needs to be repaired. The surgeons were standing in front of the soft-cadaver and 4 feet further away we placed a 4k, 52-inch screen which showed the output of the arthroscope. We gave surgeons a head-mounted eye tracker ( Tobii Glasses 2) during that arthroscopic surgery, namely a shoulder tendon repair operation. The study was approved by the Mahidol University Institutional Review Board. 
We captured the data of n=15 subjects in an operating room of the Ramathibodi Hospital of the Mahidol University in Bangkok, Thailand. The expert group (n=5) are fellow surgeons from the Orthopaedics Faculty of Medicine of the Mahidol University, who have 4-10 years of experience in arthroscopic surgery. A second group, we now and later call intermediates, consists of n=5 surgeons being in their fourth year of the Orthopaedic Residency program. In the last group, we call novices, we collected data of n=5 surgeons being in their third year of the Orthopaedic Residency program. The intermediates as well as the novices had no experience in arthroscopic surgery before. The difference between intermediates and novices is mainly based on the one year of medical education between them.

\subsubsection{Study C}

The data of study C is coming from a more static task. In study C we collected data of 58 dentists during OPT analysis. N=17 subjects are novices. On recording day, they have been in their 6th semester of dental studies before their first course in OPT reading. The intermediates (n= 14) have been in their 10th semester. Thus, they have more experience than the novices and already visited two courses in OPT reading. The dental experts are dental physicians who have already practiced for several years in their field. All subjects from this study are students from the University of Tübingen and/or are working at the University Hospital Tübingen.
The task for the dentists was to mark anomalies in multiple radiographs. Thus, the task was quite static. Therefore, we limited the time for each radiography to be marked by a dentist. This leads to a more dynamic gaze behavior. The data of the dentists were captured with an SMI RED 250 remote eye-tracker which was attached to a common laptop. The whole procedure of marking anomalies has been performed on such laptops.

\subsubsection{Procedure}

The experts in all data sets are classified based on either years of experience in the task or being picked by talent scouts. The novices are defined as beginners of the field or having no experience in the task. The intermediates are loosely defined as in between, with more experience than novices but way less than experts. As all data sets were captured with a different eye-tracking device, we first looked at all the features from all data sets and defined a subset of features that are shared by all of the data sets. In the next step, we split the data of experts, intermediates, and novices in each data set. We defined a balanced training set of a randomly picked data set and trained a bagged tree model. With this first model, we used an MRMR technique for feature selection. Subsequently, we ranked the features by their importance for the model during cross-validation. With the new subset of features, we now used data from the two remaining data sets to test the model on other domains. In the following, we will talk about our observations.

\subsection{Results}
~\label{sec:results}

After the first feature ranking, we end up with a sub-set of features that has the highest impact on accuracy. We, therefore, pick them as candidates for a subset of features that are shared by all data sets.
The most promising subset of features was the following:

\begin{itemize}
	
	\item maximum saccade peak velocity
	\item maximum fixation dispersion
	\item standard 	deviation of saccade peak velocity
	\item maximum saccade amplitude
	\item minimum smooth pursuit dispersion
\end{itemize}


\begin{figure*}[ht]
	\centering
	\includegraphics[width=0.5\columnwidth]{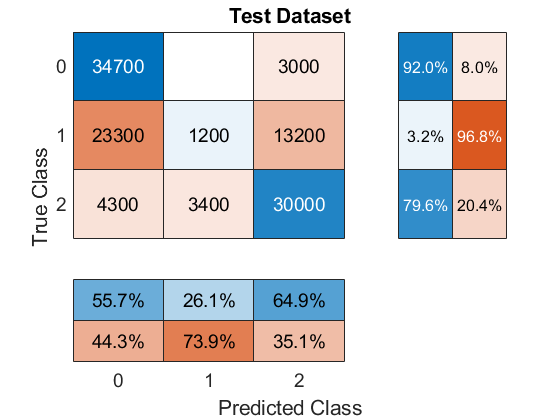}
	\caption{Confusion matrix showing predictions with data of new data set. \\ Class 0 = novices, class 1 = intermediates, and class 2 = experts. }~\label{figureC:conf}
\end{figure*}

With these features, we were able to achieve an accuracy performance of 58\%. This sounds quite low, but we need to remember that we are looking at a three-class problem. Thus, the chance level of picking the right class is 33.33\%. With 58\% we are slightly worse than the doubled chance level. An accuracy of over 66\% would lead to the fact, that single samples might be classified incorrectly, but the majority is classified correctly. Therefore, also the majority of a subject's samples are classified correctly and subsequently the subject in total, too. Looking at the two data sets that were classified, the dentist data set (study C), had a total classification accuracy of only 29\%. The intermediates were classified with 7.7\%, the experts with 35\%, and the novices with 45 \%. Therefore, the dentist data set is slightly worse than the chance level and thus, we might have not the most optimal features for that data set. Another reason for this classification might also be the different task of static diagnostics. On the soccer data set (study A), which task was much more similar to the surgeons, we reached an accuracy for the novices of 83.4\%, 0.8\% for the intermediates, and 98.4\% for the experts. Again, because of the miss classifications of the intermediates, the average accuracy is at 60\%. With the mentioned features we were able to train a model with one data set and classify the two other data sets with an accuracy of 58\%. On a deeper look at the single classes, we can see that the novices (92.0\%, see Figure~\ref{figureC:conf}) were nearly optimally detected, the intermediates with 3.2 \% not at all, and the experts still with an accuracy of over 79.6\%. From 100 runs 34,700 samples were correctly classified as novice and 3,000 falsely as an expert. This is not a problem, as we know in expertise research there are subjects acting better than their initial classification. More problematic is the amount of samples that belong to the expert class but is classified as novice or intermediate. In 100 runs 30,000 samples were classified correctly as expert samples. 4,300 samples incorrectly as a novice, and 3,400 samples as intermediate.  At first, these results look complex to understand, but a closer look at how the intermediates are defined reveals the ambivalence of these results and a weak point in the classification. This will be discussed in the discussion part of this paper.

\subsubsection{Shared, latent expertise features}

Having a deeper look at the features and their characteristics, we can see three important correlations. As we normalized the data based on their data set, the values can be positive as well as negative. For comparison, this is important, as the correlations are only visible there. The surgeons' experts e.g. have a maximum saccade peak velocity of -221.560 °$/s$, followed by the intermediates with -0.7267 °$/s$ and the novices with 222.287°$/s$. Comparing the values with those of the soccer players, we see that the experts also have a highly negative value of -593.31 °$/s$ followed by a high value of 234.56 °$/s$ and an even higher value of 1211.377 °$/s$ for the novices. Soccer players have more or less the same trend between the expertise classes. In the data set of the dentists, we cannot see this trend. Only experts and novices show similar values, thus, intermediates will be miss classified as novices (their values correspond much closer to the novices). For the dentists, we found a correlation between the trends of the standard deviation of the peak velocity. The dentists, as well as the surgeons, follow the same trend (experts: ca. -15°$/s$, intermediates: ca. 6.5 °$/s$, and novices: ca. 10°$/s$). Here the data of the soccer players do not fit at all. A feature whose values correlate with both other data sets' experts and novices, is the minimum smooth pursuit dispersion. The values for the expert groups are slightly positive (0.019 to 2.25 pixels), while the values of the novices are slightly negative (-4.8 to -0.15). Only, again, the soccer players' intermediates correlate with the surgeons by being negatively close to zero.


\subsection{Discussion}
\label{sec:discGeneral}

In this paper, we defined a feature set that is thought to explain expertise in multiple data sets from different domains. We trained a machine learning model with one data set and predicted the classes of two other data sets. With an average accuracy of 58\%, the total performance is quite sobering, but on a more detailed look, the performance value can be understood quite easily. The differences in the domains seem to be not important, as it is possible to detect novices in dentistry with 45\% accuracy. Thus, there might be some kind of general gaze behavior explainable throughout the two domains. The detection of the expert dentists is 34\% much lower and close to chance-level. Thus, there are differences between the two data sets that hinder a generalization. As the data set of soccer players show high-performance values, we cannot say that different domains need different gaze behaviors, as our model is trained with surgeons' gaze behavior and could predict novices and experts of soccer pretty well. A much more important difference than domain, are the boundary conditions that we need to take into account. A possibly important difference between the dentist and the other two data sets was their low dynamics. The dentists were observing images on a laptop screen while the surgeons as well as the soccer players were allowed to move their heads completely free. Likewise, surgeons and soccer players needed to gain an overview, navigate, and decide how to continue, while dentists only marked anomalies on a fixed image. Thus, we support the findings of Gegenfurtner et al. \cite{gegenfurtner2013transfer}, that it is possible to transfer knowledge about gaze behavior of one task to gaze behavior of a familiar-tasks, but not unfamiliar-task. To allow a statement in the direction of cross-domain expertise-related gaze features, there need to be more investigations with data sets from different domains and/or different tasks, but similar hardware setups (same eye tracker, same speed, etc), but with this current work we can state that to generalize expertise-related gaze behavior between different domains, the task seems to be much more important than the domain itself.

\newpage

\section{Expertise Classification of Soccer Goalkeepers in Highly-Dynamic Decision-Tasks:  A Deep-Learning Approach for Temporal and Spatial Feature Recognition of Fixation Image Patch Sequences}
\sectionmark{Expertise Classification of Soccer Goalkeepers: A Deep Learning Approach}

\subsubsection*{Abstract}

The focus of expertise research moves constantly forward and includes cognitive factors like visual information perception and processing. In highly dynamic tasks, such as decision-making in sports, these factors become more important in order to build a foundation for diagnostic systems and adaptive learning environments. Although most recent research focuses on behavioral features, the underlying cognitive mechanisms have been poorly understood, mainly due to a lack of adequate methods for the analysis of complex eye-tracking data that goes beyond aggregated fixations and saccades. There are no generally applicable statements about specific perceptual features that explain expertise. However, these mechanisms are an important part of expertise, especially in decision-making in sports games as highly trained perceptual-cognitive abilities can provide athletes with some advantage. We developed a deep learning approach that independently finds latent perceptual features in fixation image patches. It then derives expertise based solely on these fixation patches which encompass the gaze behavior of athletes in an elaborately implemented virtual reality setup. We present a CNN-BiLSTM-based model for expertise assessment in goalkeeper-specific decision tasks on initiating passes in build-up situations. The empirical validation demonstrated that our model has the ability to find valuable latent features that detect the expertise level of 33 athletes (novice, advanced, expert) with 73.11\% accuracy. Our model is a first step in the direction of generalizable expertise recognition based on eye movements.

\subsection{Introduction}

In general, expertise research spans many different areas. Expertise research based on behavioral data has found its way especially into several fields, i.e. dentistry \cite{castner2020deep}, surgery \cite{eivazi2011predicting,kubler2015automated,hosp2021surgeons}, and sports  \cite{hosp2020eye,kredel2017eye,snegireva2018eye,moran2018implications,discombe2015eye,fegatelli2016use}. In all of these areas, the assessment of user expertise is a fundamental task. By estimating the expertise of a user as accurately as possible, adaptive systems can be built to model different, distinct expertise classes and potentially create tasks specifically adapted to the expertise class. 
For diagnostics within the framework of sports science expertise research, groups of different performance levels are examined using the ‘expert-novice paradigm’ \cite{chi1981categorization}. According to Tenenbaum et al. \cite{tenenbaum2000anticipation}, this is the most efficient way to study the development of cognitive and motor skills. Based on this paradigm, Ericsson et al. \cite{ericsson1991toward} developed the frequently used framework of the 'Expert Performance Approach'. 
This approach assumes that a subjects’ behavior in a laboratory task is closest to their behavior on the pitch if the laboratory setting is as realistic as possible. It is therefore required to establish the highest possible ecological validity of laboratory tests, taking into account the internal validity \cite{kredel2017eye}. According to this assumption, within the Expert Performance Approach, sports-specific scenes are often selected as stimuli for diagnostic \cite{romeas20163d}. However, in previous studies, the video stimuli were mostly presented on large screens or PC monitors and often from a third-person perspective (for review, see e.g. \cite{mann2007perceptual,murr2020decision}). This classical laboratory setting results in a low external validity \cite{marasso2014developmental} (for an overview see \cite{travassos2013expertise}).
The trade-off of these validities plays an important role. Especially in highly dynamic environments, it is difficult to obtain robust and natural data. Robust data is obtained in highly controlled environments while natural data is obtained in natural environments. Therefore, these two aspects are opposites and relative to discussions about the tension between the internal and ecological validity of scientific studies. This is especially true in fields such as sports where besides, tactical and physical components, highly refined perceptual-cognitive abilities are key to success \cite{berry2008contribution,catteeuw2009decision,abernethy2010revisiting,helsen1999multidimensional}. Due to the fact that in high-level sports the physical strain of the athletes is significant due to intensive training schedules, enhancing cognitive factors like decision-making without additional physical training is gaining in importance \cite{appelbaum2016international}. For this reason, research efforts to identify the major cognitive factors leading to differences in performance, especially in regard to decision-making in the sports game, have increased in recent years. One aim of these efforts is the development of valid diagnostics that can, for example, identify the gaze behavior of experts engaged in successful decision-making. Accordingly, by teaching this gaze behavior it may be possible to design training programs that lead to improved decision making. 

Due to ongoing technological development in the field of virtual reality (VR), it is now possible to present 360°stimuli from a first-person perspective in head-mounted displays (HMD). This increases the feeling of ‘presence’ for participants, which is defined as the psychological experience of ‘being there’ \cite{cummings2016immersive}. An increased feeling of presence should lead to more valid results as compared to presentations on screens \cite{bird2019ready,slater2018immersion}. In addition to the valid stimulation and recording of behavior, an analysis of the underlying mechanisms of expertise is necessary to formulate explanatory approaches for identified performance differences.
In recent years, cognitive processes (e.g., decision-making under pressure or anticipation of the continuation of a scene) in sports games have been studied. Thereby, new developments in image processing, measurement methods, machine learning, and eye tracking may be used to control the stimuli or utilized as non-invasive methods that do not influence the natural behavior of the athlete. The developments in eye tracking have shown that these methods of measurement hardly disturb natural behavior, but, instead, become increasingly accurate and informative because cognitive processes like perception are very simple, non-invasive, and meaningful to track.

In sports science, the non-invasive method of eye tracking is considered a common and objective research method for the analysis of visual attention and the intake of visual information (for an overview see \cite{hagemann2006training}). Here it is also assumed that the measurement of athlete gaze behavior in real sports situations generates the highest ecological validity.
Mobile eye trackers have disadvantages (e.g. inaccurate measurements due to slippage, low frequencies), that can be circumvented by eye trackers integrated into the HMD. Due to the 360° videos that can be presented there, gaze behavior can be recorded at high frequency (up to 250 Hz) in ecologically valid environments with high experimental control.

The type of analysis also plays an important role because up until this point eye-tracking data has mainly been evaluated manually, visually or with statistical methods \cite{blascheck2017visualization}. 
A newer and popular technique to classify expertise is to train a model by a brute-force approach of all possible features available from the data. Hosp et al. \cite{hosp2020eye} use this technique to investigate the expertise of soccer goalkeepers by recording their gaze during the game build-up. In their approach to expertise recognition, they take all possible features provided by the eye-tracking vendor and add derived statistical features on top. They find a support vector model (SVM) with high accuracy. However, this feature crafting is highly time-consuming and does not necessarily provide the most suited features. There is no real evidence that certain features or feature combinations highlight expertise. Fixations, saccades, and their frequencies and lengths are often used, but can not lead to a full understanding of expertise as Klostermann and Moeinirad \cite{klostermann2020fewer} revealed. They conclude that single features describing gaze behavior are only conditionally suitable to classify expertise differences or, at the very least, have yet to be found. Rather, expertise comes from the optimized perception of helpful gaze locations and the sequence of these locations also called scan path. To explore the gaze locations and their temporal succession, our approach is to let artificial intelligence (AI) describe the features around these gaze locations (albeit very abstract). In doing so, the AI itself decides which shapes, colors, corners, and edges in the fixation locations are considered important for distinguishing expertise. This does not lead to new insights about the features of gaze behavior in athletes. However, the sequence of fixation locations from the stimulus can be used first to automatically recognize expertise and differences in the scan path and second, given sufficient data, to generate an optimal scan path. Ultimately, this scan path can help one understand important expertise-related fixation locations and their sequence in the gaze signal. Furthermore, with an optimal scan path, one can infer the importance of opponents, teammates (or at least parts of such), or the ball for the decision-making process. By looking at the fixation patches and running an object or person detection, a successful orientation of the scene can be achieved. This leads to a large amount of data which is advantageous for machine learning as machine learning algorithms show their strengths in regression and the classification of large amounts of data. Even in supervised machine learning algorithms we often face the problem of choosing optimal features because there is no indication as to which set of features can best show the expertise of a class. 

Next to supervised learning algorithms, where features need to be selected first, other approaches work in an end-to-end learning fashion where features do not need to be identified beforehand. The most important representatives in this field are the convolutional neural networks (CNNs) and recurrent neural networks (RNN), i.e. bidirectional long short-term memory networks (BiLSTM). CNNs are well used in a range of applications like semantic segmentation and object recognition and can learn to distinguish relevant patterns and shapes or to derive abstract objects. Next to CNNs, RNNs and particularly long short-term memory networks (LSTMs) \cite{hochreiter1997long}, which can find temporal relationships \cite{tian2019multimodal,liu2018deep}, are also widely used. LSTMs optimize RNNs by minimizing the impact of vanishing and exploding gradients. By using a special function block, LSTMs implement a long short-term memory, which pushes the performance of neural networks. These function blocks allow for the remembering of long-time dependencies and previous information. The network learns which information from the past is important for the current output and which can be forgotten (by a forget gate). As the gaze signal is continuous, LSTMs are predesignated to be used in the analysis of temporal patterns in the gaze signal. Currently, both kinds of machine learning techniques are well used for expertise identification in different domains, e.g. in dentistry education ~\cite{castner2020deep,castner2018scanpath} or microsurgery ~\cite{bednarik2013computational,eivazi2012gaze,eivazi2011predicting,eivazi2017towards}.  Neural networks ~\cite{castner2020deep} and supervised learning algorithms ~\cite{bednarik2013computational,castner2018scanpath,hosp2021surgeons,hosp2020eye} have both shown their power in objective expertise identification based on gaze behavior. They found major differences in gaze behavior and could link these differences to different expertise classes. This means both machine learning techniques provide suitable methods to deal with large amounts of data and analysis in a fast, objective, and reproducible way.

In this work, we introduce gazePatchNet which combines the strengths of CNNs to detect latent spatial feature relationships, and BiLSTMs to detect temporal feature relationships in fixation patches. To evaluate gazePatchNet, we conducted a study where we showed participants 360° stimuli of defined soccer game situations from the natural perspective of a goalkeeper on a consumer-grade HTC Vive HMD. The gaze was recorded by the integrated SensoMotoric Instruments (SMI) eye tracker with a frequency of 250 Hz. Each stimulus shows a build-up scene and ends after a pass to the user. We used our model to classify the expertise of our participants into three classes, namely, novice, advanced, and expert.
This model is meant to serve as a step in the direction of a perceptual-cognitive training system. If our model is robust enough, the discovered knowledge can be used to identify optimal synthetic scan paths that can then be used to train the gaze behavior of athletes. The underlying hypothesis is that an improved gaze strategy leads to a more reliable recognition of cues and better decision-making based on these cues. 

\subsection{Methods}

\subsubsection*{Stimulus}

\begin{figure*}[ht]
	\centering
	\includegraphics[width=1\columnwidth]{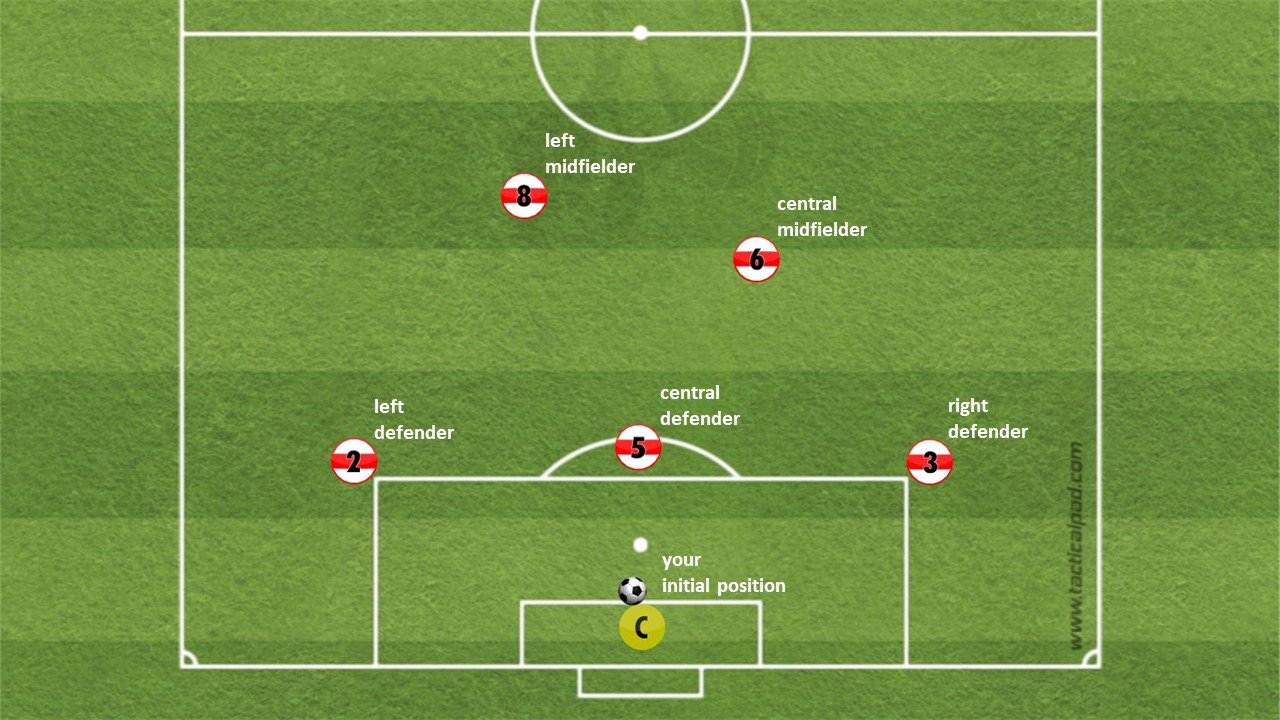}
	\caption{Schematic overview of the response options. The sixth option (kick out) is missing. }~\label{fig:schematic}
\end{figure*}

To show the stimulus video material in virtual reality, we used the SteamVR framework prefab in Unity. SteamVR is an open-source framework that allows common real-time game engines, like Unity, to interface with HMDs. Instead of an artificial recreation of the environment (simulation) within the game engine, we projected realistic footage of 4k omnidirectional videos we captured onto the inside of a sphere around the participant (3840x1920 pixel). This allowed us to display a natural stimulus with high immersion in a realistically mimicked scene. The 360\textdegree-footage was captured in cooperation with the German Football Association (DFB) at the training space of the elite youth academy of a German first league club. To capture the footage, we placed a 360° camera at the position of the goalkeeper while 5 teammates and 5 opponents were physically replaying the defined scenarios on the training space. The camera captured the scene with 30 FPS. Each scene was developed based on common scenarios during a match, each with unique movements. After a video finished, participants had to choose one of six options (five teammates to pass the ball or kick out) to continue the game. In each video, there is one optimal option. This option is counted as one. All five other options are counted as zero. This leads to binary answers in each video. All stimuli were captured on-field and acted out by youth elite players. The plausibility of the scenes, movements, and rating of the decision options were evaluated by an expert trainer team of the DFB. Only stimuli with a single good decision option were included in the experiment. An overview of the options can be seen in Figure \ref{fig:schematic} (except "kick out").

\begin{figure*}[ht]
	\centering
	\includegraphics[width=1\linewidth]{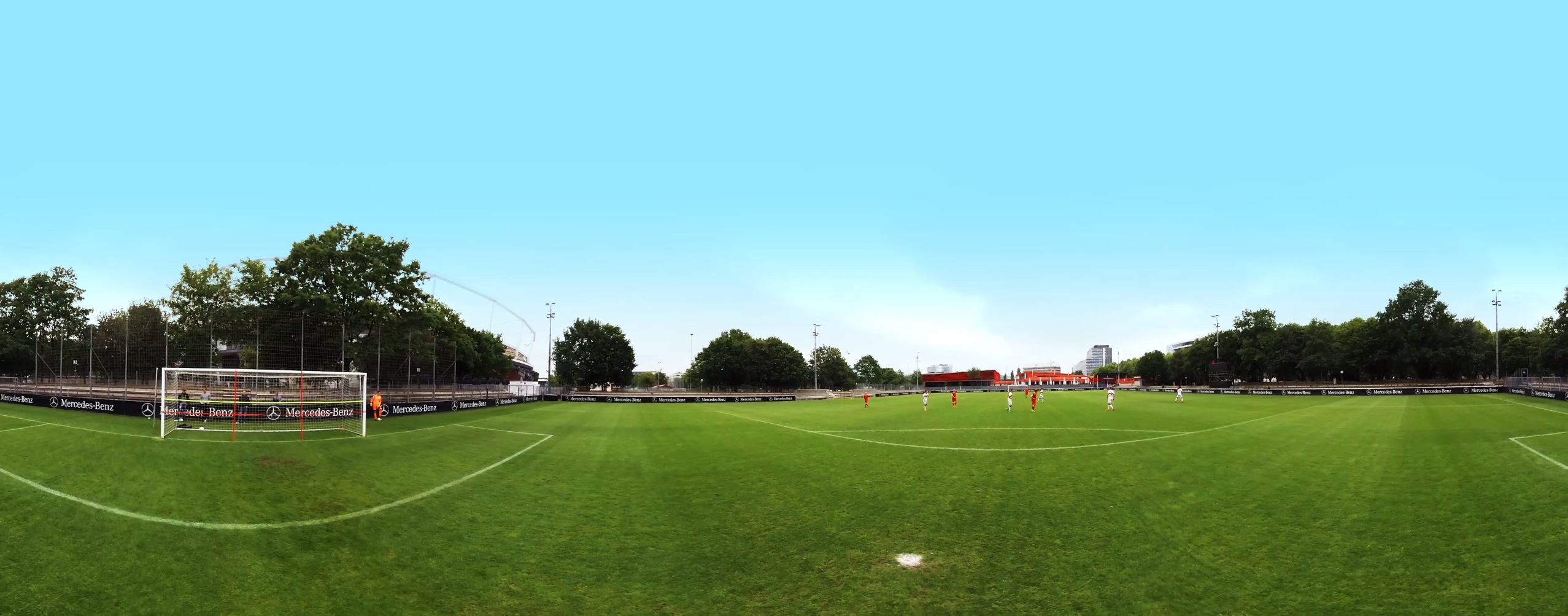}
	\caption{Example stimulus in equirectangular format. }~\label{fig:stimulus}
\end{figure*}

\subsubsection*{Data collection}

Participants' responses were relayed verbally and finally rated as either right or wrong with only one right decision available per video. The correct decision is a pass to the only teammate who is not covered by an opponent. In total, each participant saw 52 trials, consisting of 26 videos with unique movements, repeated in a different order. Each video trial of each participant counted as one sample.

\subsubsection*{Participants}

\begin{table}[!h]
	\renewcommand*{\arraystretch}{1.5}
	\begin{tcolorbox}

	\begin{tabularx}{\textwidth}{lXXc}
		\multicolumn{4}{c}{\textbf{Participants}} \\\cline{1-4} \\
		\cellcolor{gray!30}{Class} & \cellcolor{gray!30}{Age (Mean/SD)} & \cellcolor{gray!30} {Training hours/week (Mean/SD)} & \cellcolor{gray!30}{Active years (Mean/SD) }\\	 		 	
		Novice (n=13)& 28.64 / 3.72 & 0.00 / 0.00 & 1.78 / 5.21  \\
		Advanced (n=8)& 22.00 / 3.72 & 4.94 / 0.91 & 15.50 / 5.77  \\
		Expert (n=12)& 16.60 / 1.54 & 8.83 / 4.27 & 9.16 / 5.04  \\

	\end{tabularx}
	\end{tcolorbox}
	\caption{Participants summary.}	~\label{tab:participants}
\end{table}

Characteristics of all participants can be seen in Table~\ref{tab:participants}.
Data of n=12 experts were collected during a DFB goalkeeper camp, where the DFB gathers the top German elite soccer goalkeepers (U-15 to U-21) for specialized training. These experts are among the top 15 youth elite goalkeepers in Germany. These are the only expert youth players available in Germany. The data of the n=8 advanced and data of n=13 novice athletes were collected in the university's lab. The advanced players belong to different soccer teams playing in the southern regional league (semi-professional, 4th level) in Germany. The novices have very little to no experience in amateur leagues up to district league with no participation in competitions and no training on a weekly basis.

\subsubsection*{Procedure}

\begin{figure*}[h]
	\centering
	\includegraphics[width=1\linewidth]{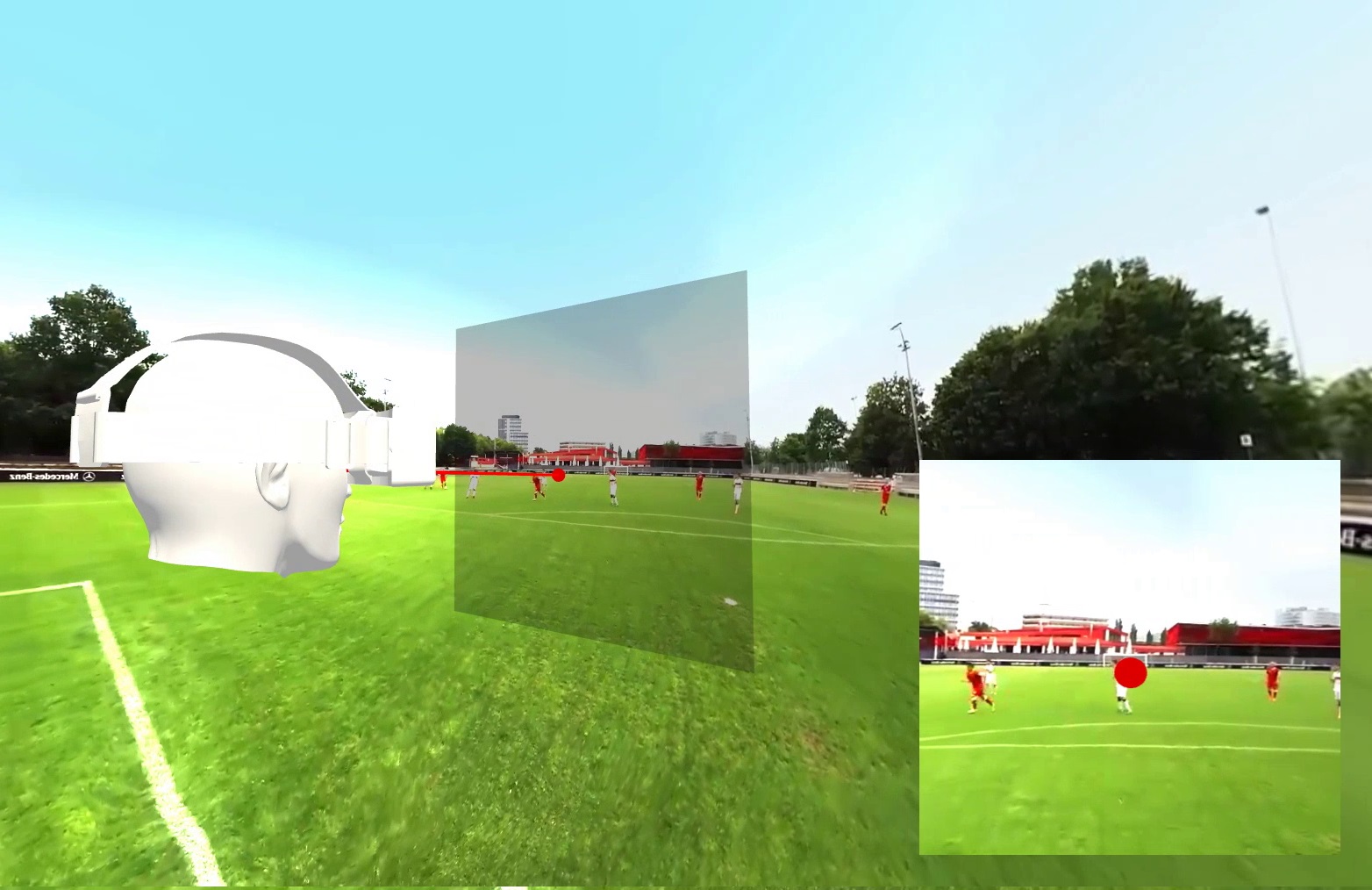}
	\caption{Screenshot of an animation showing our system setup. The red line and dot are the gaze signal. The gray rectangle is the field of view of the user inside the HMD. The content of the field of view is shown in the small rectangle on the bottom right side. The users are able to freely move their head/perspective in the scene. }~\label{fig:perspective}
\end{figure*}

The study was confirmed by the Faculty of Economics and Social Sciences Ethics Committee of the University of Tübingen. After completing a consent form, we started familiarizing the participants with the stimulus presentation and response mode. During the familiarization phase, we showed a sample 360° screenshot of a video on the HMD to allow free exploration of the scene followed by a schematic overview of the field (see Figure~\ref{fig:schematic}). After that, the video scene (see Figure~\ref{fig:stimulus} for an example) was played and stopped (black screen) after the last return pass to the position of the participant. In each scene, we manipulated the color of the ball with a colored dot during the last return pass. This was done in order to force the participants' gaze on the ball during this important phase. As soon as the screen went black, the participant had to report the color of the ball as well as their decision for an option to continue the game. The decision selection is identical to the initial schematic overview of the field (Figure~\ref{fig:schematic}) plus an emergency option to "kick out". This procedure was repeated 5 times.

After this learning phase, we started the first block of 26 trials. The second block contained the same 26 videos, but in a different order. Between the blocks, participants could take off the HMD for a break. Figure~\ref{fig:stimulus} shows a screenshot during data collection and Figure~\ref{fig:perspective} shows a simulation to visualize the setup of the eye-tracking and VR setting.

\subsubsection*{Image patch extraction}

\begin{figure*}[ht]
	
	\centering
	\includegraphics[width=\columnwidth]{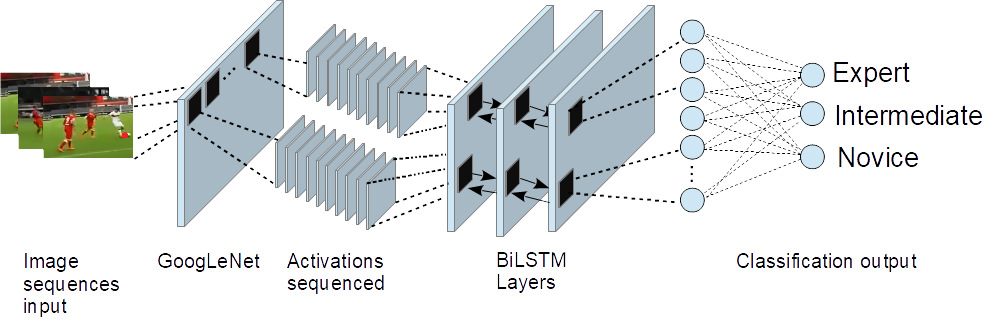}
	\caption{gazePatchNet: Our CNN-BiLSTM-based model architecture for expertise classification } ~\label{figD:model}
\end{figure*}

\label{sec:method}
As introduced above, our method (coined gazePatchNet) includes 1) finding latent features in the image patches around the participants' fixations and 2) classifying the scan path as a sequence of the consecutive fixation image patches. The whole process is illustrated in Figure~\ref{figD:model}. As not all data collection went smoothly because of slippage of the head-set (too loose) or bad calibration results, we reviewed the gaze signal quality of all samples. We only considered a sample valid if the tracking ratio was higher than 75\%. We assigned either class 0 (for novice samples), class 1 (for advanced samples), or class 2 (for expert samples) to each sample. After removing invalid data points, we collected all gaze signal samples for each fixation (timestamp,x,y) and saved them with the corresponding omnidirectional video file. The fixations were calculated with the vendor's velocity threshold-based event detection filter (I-VT) ~\cite{salvucci2000identifying} algorithm using a threshold of 50\textdegree/s. We calculated the temporal as well as the spatial center of the fixation based on the averaged gaze signal samples of the fixation. Afterward, we looped over the video file frames to find the corresponding frame by timestamp and cut out an image patch around the fixation on the frame. The size of the patch fits the input size of the input layer of the GoogLeNet CNN (224x224x3 pixels), which we used to extract features later. As soon as we had all the fixation image patches of one trial, we created sequences that fit our BiLSTM. These sequences were essentially fixation image patches in order of their occurrence in the stimulus video. 

\begin{figure}[h]
	
	\centering
	\includegraphics[width=1\textwidth]{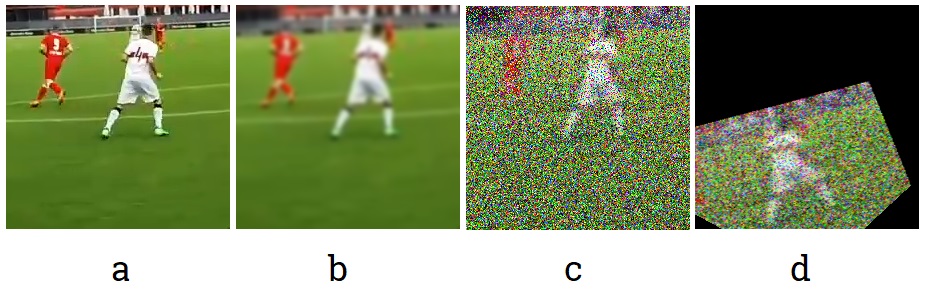}
	\caption{Augmentation pipeline. a) shows the original image cut around a fixation, b) shows the image after gaussian blur, c) shows the image after salt \& pepper noise addition, and d) shows the transformed image. }~\label{fig:augment}
\end{figure}

\subsubsection*{Data Augmentation}

For each sequence, we computed a new, modified sequence containing the same images. This means we doubled the whole data set by adding the same sequences with the same, just augmented, images. An example is shown in Figure~\ref{fig:augment}. Figure~\ref{fig:augment} a, shows an input image (image cut from the stimulus around a fixation). At first, we applied a random Gaussian blur (Figure \ref{fig:augment}, b) and salt \& pepper noise (Figure~\ref{fig:augment}, c). Afterwards, we augmented the images in a randomized manner with geometric transformations \cite{shorten2019survey} (Figure \ref{fig:augment}, d). Each image was either rotated by a random factor between -180 and 180 degrees, sheared by a random factor between -15 and 15 degrees, or both, flipped on x- or y-axis or was x- or y- translated between -80 and 80 degrees. These augmentation steps were supposed to make training the model harder in a realistic way. We assumed shear and rotation were real translational variations of the participant's head (whole field of view around fixation). This data augmentation was completed before training in an offline manner. The whole data set was augmented in 135 seconds. LSTMs usually support varying sequence lengths, however, as sequences that are much longer than typical sequences can introduce a lot of padding or discard data because of the padding or truncation of sequences, we removed an average of 20 sequences, about 2\% of all sequences. The remaining sequences were sorted by length. This led to more homogeneous padding of the input sequences.

\subsubsection*{Transfer Learning}
\begin{figure}[h]
	\centering
	\includegraphics[width=1\textwidth]{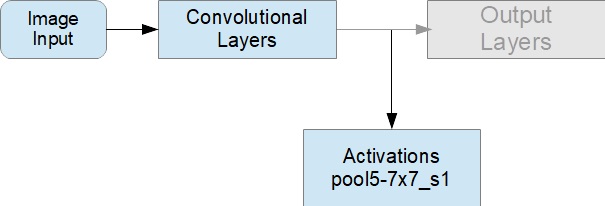}
	\caption{Transfer learning for feature selection. }~\label{fig:googlenet}
\end{figure}

To get latent spatial features in the image patches automatically, we used a convolutional neural network (CNN, GoogLeNet) as a feature extractor. The CNN was trained on ImageNet, which has about 1000 classes. Each sequence (the augmented sequences included) was fed to the CNN. We did not use the output layers, as we did not need the classification probabilities for the 1000 classes of ImageNet, but, rather, for three classes of expertise. Instead, we proceeded with transfer learning by grabbing the output of the last activation function (see Figure \ref{fig:googlenet}, the last pooling layer of the GoogLeNet network $("pool5-7x7_\textsubscript{s1}")$, and added the layers of gazePatchNet (see table \ref{tab:netModel}). We then adapted the output so that it classified our three expertise groups. By using GoogLeNet as a feature extractor, we simultaneously obtained a feature dimension reduction as our images of 224x224x3 pixels were reduced by the CNN to 1024x1 dimensions. As a result, we achieved not only shape, pattern, and object detection, but also the correct input format for an LSTM by keeping track of the input to the CNN and building sequences of related outputs (activated images). 

\subsubsection*{Training and Testing}

We trained the model in 33 runs. In each run, the samples of one subject were kept out (leave-one-out validation). The sequences of this subject were used at the end of each run to predict their class. As the model did not see the data before, it was meant to validate the predictive power of the trained model and show how it behaved with totally new data.
The data of the remaining 32 subjects were split by a ratio of 70\% / 30\%. 30\% of the data was randomly picked for testing and optimization during training. The remaining 70\% of the samples (as well as the augmented samples) were used for training the model.

\subsubsection*{Model description}

\begin{table}
	\centering
\renewcommand*{\arraystretch}{1.5}
	\begin{tcolorbox}
	\begin{tabularx}{\textwidth}{lXXc}
		\multicolumn{4}{c}{\textbf{GazePatchNet}} \\\cline{1-4}\\ 
	\cellcolor{gray!30}	& 	\cellcolor{gray!30}Name & 	\cellcolor{gray!30}Type & 	\cellcolor{gray!30}Activations  \\
		
		1 & sequence  & Sequence Input & 1024  \\
		2 &	bilstm  & BiLSTM & 1000   \\
		3 & fc-1 & Fully Connected & 100 \\
		4 & dropout & Dropout & 100   \\
		5 & fc-2 & Fully Connected & 3  \\
	\end{tabularx}
	\end{tcolorbox}
	\caption{GazePatchNet architecture.}
	\label{tab:netModel}
\end{table}

\begin{table}
	
	\centering
	\renewcommand*{\arraystretch}{1.5}
	\begin{tcolorbox}
	\begin{tabularx}{\textwidth}{lX}
			\multicolumn{2}{c}{\textbf{Training Options}} \\\cline{1-2}\\
		\cellcolor{gray!30}Parameter & \cellcolor{gray!30}Value \\
		
		MiniBatch size & 42  \\
		Learning rate & 4.4e-4\\
		L2-Regularization & 8.2e-4  \\
		Sequence length & longest sample  \\
		Shuffle & no \\
		Validation frequency & 52 \\		
		Validation patience & 6\\		
		Learning rate schedule & no\\
		Max. epochs & 30		
	\end{tabularx}
	\end{tcolorbox}
	\caption{Training Options.}	
	\label{tab:netOptions}
\end{table}

Table \ref{tab:netModel} shows the structure of the networks' layers. The sequenced activations, containing the selected features, from GoogLeNet were passed to the BiLSTM layers where the temporal relationships were calculated. To input sequences of images into the network, the first layer was a sequence input layer with the same input dimensions (1024) as the output of the activations by the CNN at the last pooling layer (GoogLeNet). As the models with gated recurrent units (GRU) and LSTM layers did not perform well in our tests (between 20-25\% lower accuracy), we chose BiLSTM layers as the next part. The BiLSTM layer had 50 hidden units (therefore 4000x1024 input weights, 4000x500 recurrent weights, and 4000x1 biases) and output only the final step. The advantage of BiLSTM layers is that they are fairly generative and take future (forward) and past (backward) states of information into account. After the BiLSTM layer, we added a fully connected layer with 13 hidden units (100x1000 weights and 100x1 biases). To prevent the model from overfitting, we added a dropout layer with a probability of not using a neuron of 50\%. As we had three classes to predict, the following fully connected layer had 3 hidden units. We took the maximum output to identify the class. To help training converge quickly, we added a softmax layer and calculated the cross-entropy loss for multi-class classification to optimize the model.

Table \ref{tab:netOptions} provides an overview of the training options. We used grid search to find an optimal hyperparameter set for the whole network \cite{feurer2019hyperparameter}. The best set consisted of a mini-batch size of 42, a low learning rate of 4.4e-4, which was not increasing during training time, an L2-regularization of 8.2e-4, to prevent overfitting and a validation frequency that was set to 52 so that the model was validated at every epoch. Validation patience of 6 seemed to be the optimal trade-off between over and underfitting. This means the training was stopped earlier if the loss on the validation set was larger than or equal to the previous smallest loss 6 times in a row. We did not shuffle training and validation data every epoch as we only wanted to use validation data to offer information about the current classification status. The maximum number of epochs for training was set to 100 as longer training results in over or underfitting. 

\subsubsection*{Metrics}

We calculated the average/median accuracy over all runs. In each run, 70\% of the samples belonged to the training set and 30\% to the validation set. We kept one participant out to test how well the model behaved on new, unseen data. As the accuracy is a metric defined by the number of correct predictions divided by the total number of predictions, we could only infer a small amount of information about the model. This was particularly because the samples of the classes used for training and validation were balanced during training, but the distribution of expert, advanced, and novice participants for testing was not. Thus, we also had to consider further performance metrics of the model. The confusion matrix is a sound metric to show the single classes' true and false positives. Similar to the confusion matrix, the following metrics were split into three classes for easy comparison. 
To gain a deeper performance insight, we showed the receiver operating characteristic (ROC) curve. A ROC curve shows the performance of a classification model at different classification thresholds. Based on the ROC curve, we simply calculated the area under the curve (AUC), which is a common single score and used for comparisons between different models usually on binary classification. Since we split the classes and computed the AUC for each, we compared which classes were predicted most successfully. A score of 1.0 described a perfect skilled model. All scores were calculated by a one-vs-all approach.

\subsection{Results}

\begin{figure}[ht]
	\centering
	\includegraphics[width=1\textwidth]{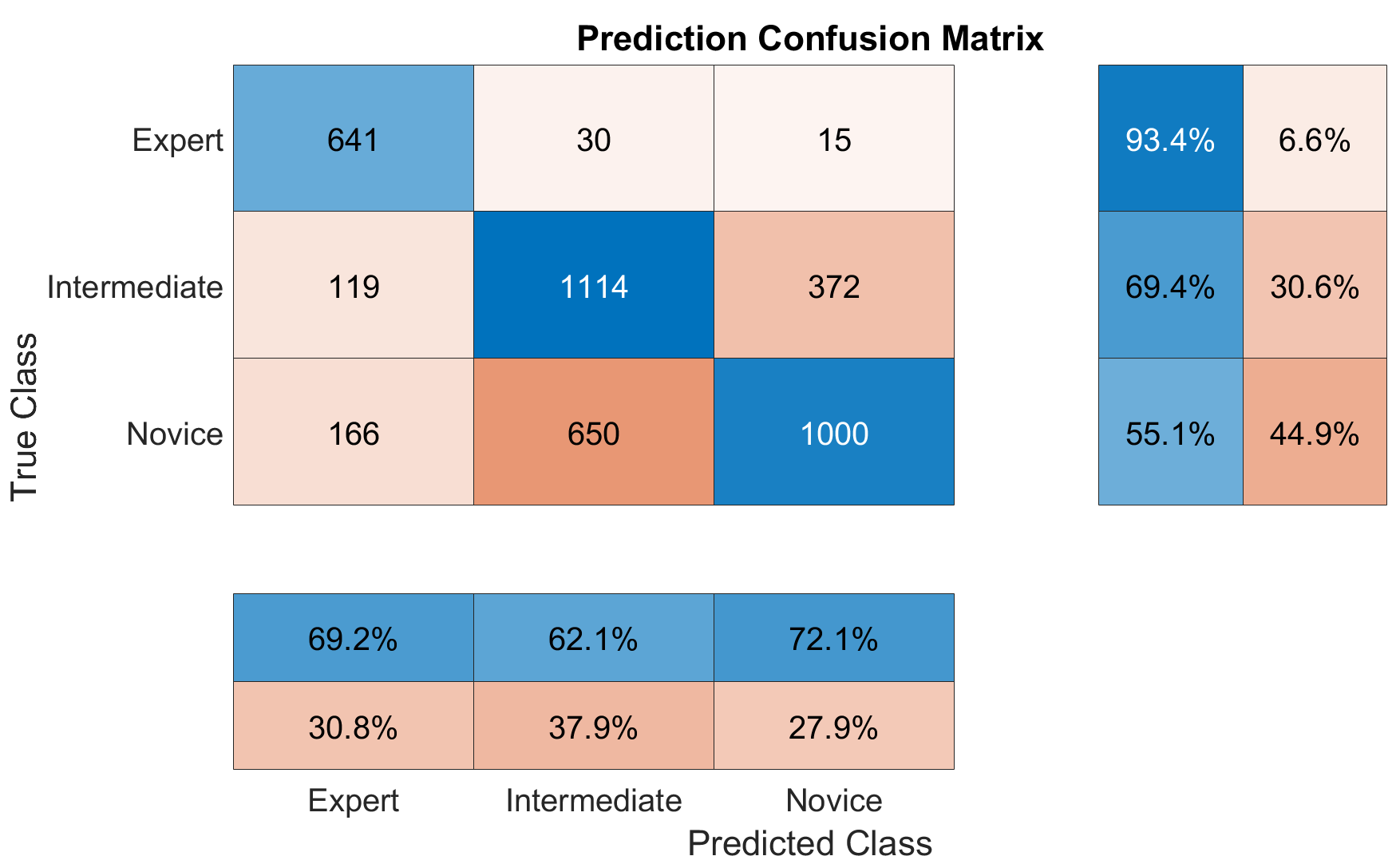}
	\caption{Confusion matrix}	~\label{figE:confMatrix}
\end{figure}

The model achieved an average accuracy of 73.11\% over 33 runs. For each run, data from one participant was kept out of training and used as test data. We looked at the data indirectly by describing one trial (one video of a participant) as one sample and classifying these samples as novice, advanced, or expert. This means that some participant samples can be detected as belonging to another group.
The distributions of the single samples to different classes can be seen nicely on the confusion matrix in Figure~\ref{figE:confMatrix}. The accuracy of predicting a novice correctly is at 55.5\%. The prediction rate of the advanced class, with an accuracy of 69.4\%, is admittedly much higher. And even higher than the advanced class, experts are predicted with an accuracy of 93.4\%.

Out of 1,816 samples of the novice class, 166 samples were predicted as belonging to the expert class and 650 to the advanced class. 1,114 samples were correctly classified as advanced. However,  about 1/3 of the advanced samples were predicted falsely, distributed with 372 on novice class and 119 on expert class. 641 of the expert samples were correctly predicted and 30 samples to advanced and 15 to novice class. \\

\begin{figure}[ht]
	\centering
	\includegraphics[width=1\textwidth]{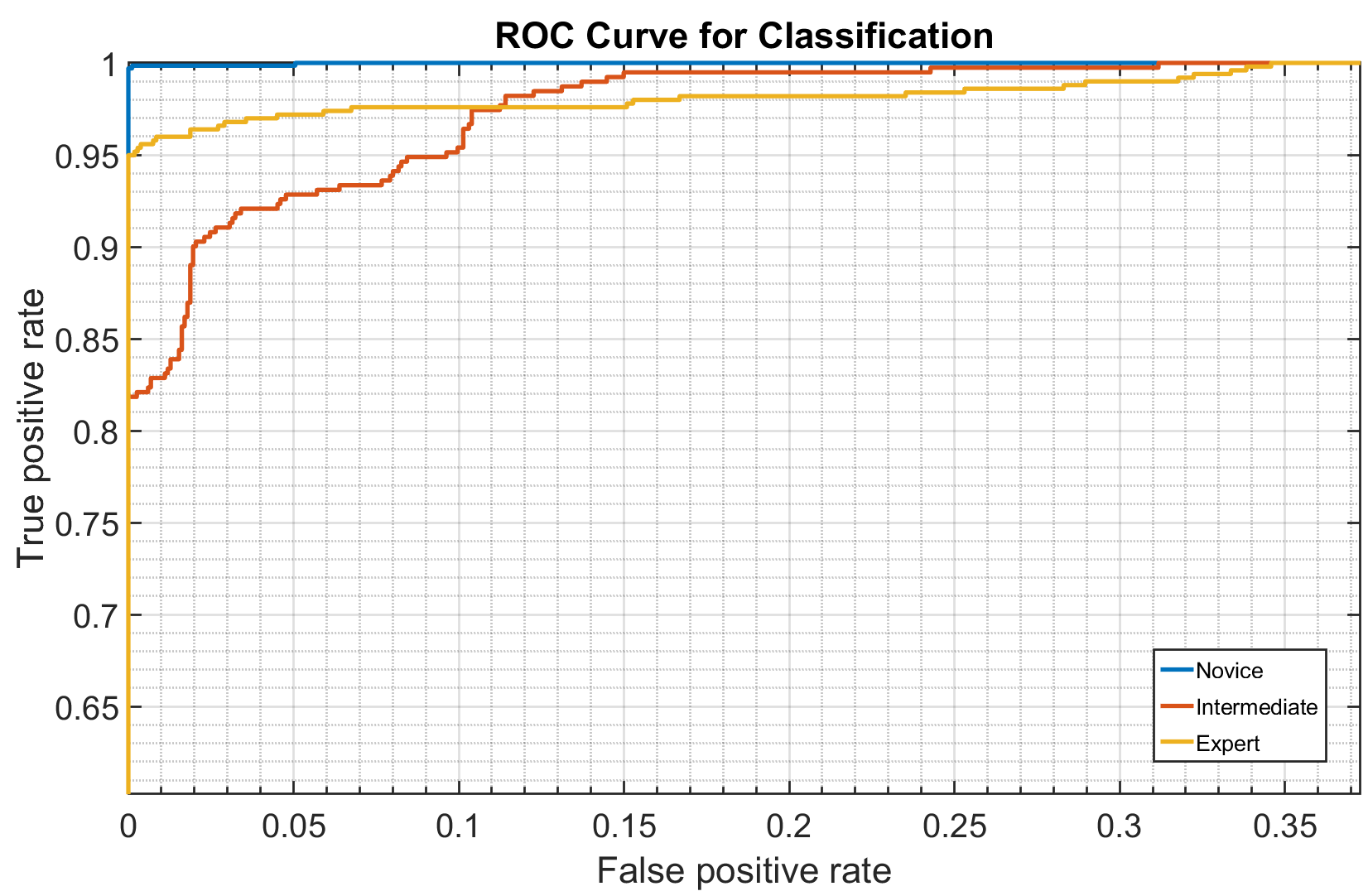}
	\caption{ROC-curve for all three classes.}~\label{fig:roc}
\end{figure}

Figure~\ref{fig:roc} shows three ROC curves with results, corresponding to the confusion matrix. The blue line represents the expert ROC curve. With an AUC of 0.951, the classification is nearly perfect. This corresponds to the confusion matrix values as well. The red line represents the advanced class that does not perform as well as the expert, but still achieves an AUC of 0.833. The yellow line shows the performance of the novice classification, which is a little bit higher than the advanced one with an AUC of 0.871.

\subsection{Discussion}

In this paper, we presented gazePatchNet, a model that, based on a by transfer learning adapted CNN for feature extraction and BiLSTM for temporal dependency identification, automates classification in the broadest sense. We recorded the gaze behavior of soccer goalkeepers during the build-up in a 360° video environment on an HMD and used their fixation image patches on the stimuli as input signals to classify three groups of expertise.
The results are promising as we can show, with a relatively small amount of data, that the combination of a CNN, transfer learning, and BiLSTM network is effective in classifying this kind of data. At least the expert and advanced classes are recognizable. However, the novices look more diverse in their behavior and therefore are much harder to predict. The model on average shows great performance, which is reflected by the average accuracy of 73.11\% and great sensitivity values visible in the ROC plot.
The differences between experts and the other groups are especially significant.

The classification of the advanced and novice class is about 20\% and 40\% respectively, lower, but advanced is still doubled when compared to chance-level. This is supposed to increase with more samples for the model to learn from. Our results are well in line with other studies on dynamic tasks, e.g. Bednarik et al.  \cite{bednarik2013computational} or Eivazi et al \cite{eivazi2017towards} who reached a classification accuracy of 66\% and 70\%, respectively, on medical applications. Both studies predicted the expertise of two skill levels. A more diverse result is found in Castner et al. \cite{castner2018scanpath}. Their study predicted the expertise of students from five different semesters alongside experts. With their one-vs-all approach, they mostly reached an accuracy of ~37\% (chance-level 20\%). 


In our model, the accuracy of predicting an expert correctly is at 93.4\% as this class is the easiest to detect. The prediction rate of the advanced class is much lower with an accuracy of 69.4\% because this class is supposed to be the hardest to detect. The accuracy, however, is nearly double the chance level with about 2/3 of the advanced samples classified correctly.
Much lower than the advanced, the novices are predicted with an accuracy of 55.1\% which is nearly twice as high as the chance level but still 15\% lower.

Out of 686 samples of the expert class, only 15 were predicted as belonging to the novice class and 30 to the advanced class. 1,114 samples were correctly classified as advanced, but about 1/3 of the advanced samples were predicted incorrectly. Although interesting, this is no surprise. It may show that the decision boundary for the advanced class does not need to be as robust as many of the advanced participants were gazing like novices and many novices as advanced according to the model. In summary of the performance of the classifications based on the ROC curves, one can state that the samples of all classes were predicted with high certainties and demonstrate the accuracy of a highly skilled predictive model. The average precision value (73.11\%), as well as the mean precision of 71.89\%, confirm the power of the model.


Looking deeper at the ROC curves, the model performs well in all classes. As samples of advanced players are often predicted as belonging to a novice class and samples of novices players as belonging to an advanced class, it may be necessary to increase the sample size, to robust the decision boundary.
In case the model predicts a sample that really belongs to the expert class, it performs this assignment with a high probability of over 93\%.

Here, the novice and advanced classes are more difficult to classify. This means that the expert group is a pretty well recognizable group. The advanced and novice groups are more heterogeneous as there are participants that have more/less experience than others. Another reason for this could be that there are missing metrics needed to divide between the two classes properly. This question is typically addressed with the availability of more data. The problem may stem from the small sample size of advanced participants as this group could be too small for the model to define robust decision boundaries. The fact that experts were barely (15 samples) predicted to be advanced shows that there are clear decision boundaries for advanced and experts.
In addition, the cognitive factor is only one of several factors that contribute to expertise. For goalkeepers, for example, it is still most important to be able to block shots on goal. If a goalkeeper can do this extremely well, he may be invited by the DFB even though he could make "worse" decisions after return passes. 
Conversely, it can also be the case with advanced participants that we have very good decision-makers, but they don't hold as many balls, which is why they are not invited by the DFB. As a result, it is very important to not just test players from different classes but to test players with the assumed highest decision-making skills. 
For the diagnosis of expertise, we aim to test the best of the very best players and compare them with other expertise groups. We need them to define an optimal behavior. Our expert players are among the 50 most successful young goalkeepers in Germany, which is reflected in the results of our model. A long-range plan is to optimize the training for young players. This work is the first step in that direction. For that, we need to know which behavior is optimal and how we can design training steps for young players to reach this optimal behavior. 


The difference in active years/training, and therefore experience, between advanced and expert participants, is much smaller and needs to be finer graded. There may be advanced players with a lot of experience that helps them to perform like experts and there may be experts that don't perform as well because they have much less experience. It is, therefore, not astonishing that some advanced samples are recognized as expert samples. If one assumes that behavior in some samples is better than others, this consequently leads to classifications distributed in different groups. It is more important that the number of classifications of higher-ranked participants into lower classes is minimized in order to depict real expertise. 

Instead of providing a description of the behavior of different classes, our model describes a pipeline to find latent features by itself. This circumvents one problem: handcrafted features. The characteristics of handcrafted features may be difficult to teach a user in the form of new behavior based on feature values. Even if the optimal set of features is found, it is difficult to incorporate the findings into a training system. Our model shows a different way of teaching a participant new behavior. As it makes more sense to be able to tell the test person what has to be observed and when and to report it visually, a model should be created that, in the best case, finds an optimal behavior. Based on such information, an optimal behavior for each class can be created and artificially extracted to create information that can be taught to users. A prerequisite will be the analysis of single scan paths, which can be accessed by looking at the fixation image patches. 

As the fixation point is currently temporally and spatially averaged, another improvement might be achieved when optimizing the input layer by using an object detection beforehand. Especially when counting in the error rate of the eye tracker and early fixations, some samples might end up directly next to an object and some directly on it. In this case, the CNN will return different shapes. By using the object as are of interest (AOI) and taking the intersection as input, this behavior can be unified as one can assume that the participant is perceiving the same object in both cases. 
The CNN can also be optimized. At the moment, this CNN is trained on ImageNet to classify about 1000 classes. By retraining the CNN on a set of 360° videos, with manually labeled teammates, opponents, goals, ball, and free spaces, the intersections of the gaze with AOIs can be advantageous and result in higher classification rates.

\subsubsection{Perspectives}

As aforementioned, the results already allow for the use of our model as a diagnostic system and as the basis for a training system. The information gathered from this work can be used to model athletes' behavior to personalize new adaptive interfaces that can understand user behaviors based on relevant user information recorded during training. For example, like Wade et al. \cite{wade2016gaze} did for intervention for individuals with autism spectrum disorders. 
With an objective way to classify the perceptual skill of a person, the first step towards a virtual reality training system (VRTS) with an adaptive design of level difficulty is achieved. With a definition of the perceptual skill of a person and the knowledge of the corresponding skill class, the choice of the difficulty of a level in a VRTS can be adapted automatically. For higher ranked users, the difficulty can be increased by pointing out fewer cues or adapting the stimulus e.g. by placing relevant information outside the foveal area (usage of peripheral vision), designing more crowded scenes (retain overview), or showing highly dynamic situations (faster perception and reaction times).
A fundamental work for such a VRTS is to enhance the model with more classes and more participants per class. More data needs to be collected to create a more robust model. A balanced data set would reveal interesting effects on recall and precision and, based on the current performance, might even increase the overall accuracy as the class with the least number of samples has the highest precision values. Different kinds of models also need to be investigated. For feature extraction, a network that is trained on human detection might provide even better results as the head/face and other parts of the human anatomy are potentially considered to be of importance. With 33 participants and an average accuracy of 73.11\% on the test set, this model is suitable to be used for this kind of classification.

In a further step, to research the applicability of our model we need to focus on adequate training scenarios. The system can, for example, already be used to create an optimal synthetic scan path. By using the knowledge discovered by our model, one can implement a generative adversarial network. This technique learns to generate new data, in our case a new scan path, with the same statistics as our training set. With enough data to train gazePatchNet to provide strong robust classes, a synthesized optimal scan path can be created. Should this be possible, it could also become relevant from a practical sports perspective to teach a certain gaze strategy obtained from the generative model. The optimal scan paths identified for each scene could be used to train the gaze behavior of athletes. The underlying hypothesis is that an improved gaze strategy leads to a more reliable recognition of cues and better decision-making based on these cues. To investigate this, however, appropriate training studies are necessary, which must provide information as to whether a) it is feasible for athletes to replace their gaze behavior, developed over years, with a foreign behavior and, if so, whether b) the modification of their gaze behavior also leads to better decision making in the lab. Then the possibility of a corresponding transfer to the field must be checked.

\subsection*{Acknowledgment}
\label{sec:ack}
This research was supported by the German Football Association (DFB). We thank our colleagues from the DFB who provided insight and expertise that greatly assisted the research. We acknowledge support from the Open Access Publishing Fund of the University of Tübingen. Enkelejda Kasneci is a member of the Machine Learning Cluster of Excellence, EXC number 2064/1  - Project number 390727645

\chapter{Gaze-Based Support Timing}

\blfootnote{This chapter is based on the following publication:\\
	
	\begin{itemize}
		\item \textbf{B.W. Hosp}, M. S. Yin, P. Haddawy, R. Watcharopas, P. Sa-ngasoongsong, and E. Kasneci (2021). "States of confusion: Eye and Head Tracking Reveal Surgeons’ Confusion during Arthroscopic Surgery". In Proceedings of the 2021 International Conference on Multimodal Interaction (ICMI ’21), October 18–22, 2021, Montréal, QC, Canada. ACM, New York, NY, USA. 
		
	\end{itemize}
}

\newpage


\section{States of Confusion: Eye and Head Tracking Reveal Surgeons’ Confusion During Arthroscopic Surgery }
\sectionmark{States of Confusion: Eye and Head Tracking Reveal Surgeons' Confusion}

\subsubsection*{Abstract}

During arthroscopic surgeries, surgeons are faced with challenges like cognitive re-projection of the 2D screen output into the 3D operating site or navigation through highly similar tissue. Training of these cognitive processes takes much time and effort for young surgeons but is necessary and crucial for their education. In this study, we want to show how to recognize states of confusion of young surgeons during arthroscopic surgery, by looking at their eye and head movements and feeding them to a machine learning model. With an accuracy of over 94\% and detection speed of 0.039 seconds, our model is a step towards online diagnostic and training systems for the perceptual-cognitive processes of surgeons during arthroscopic surgeries.

\subsection{Introduction}

Advancements in computer science have typically been a motor for new applications in fields like medicine. Next to classical imagery techniques like magnetic resonance imaging (MRI) \cite{hohne20123d} or arthroscopy \cite{ike2021arthroscopy}, nowadays, the interaction between surgeons and their patients or instruments are increasingly being investigated. There are a lot of new sources of information, e.g. about the vital parameters of the patient or new perspectives/views of the operating site, which are shown to the physicist. They are all meant to improve the work of the surgeon. However, all these new advancements come with a certain level of complexity. Surgeons need to learn how to operate and benefit from these applications. For example, in arthroscopy, the surgeon needs to transfer the 2D image on the scope output into the 3D tissue of the patient. Information is shown on the screen, but navigation takes place on the operating site with a multidimensional instrument. This translation already poses a challenge. 

Even in medical image reading, Brady et al. \cite{brady2017error} estimated that the miss rate for interpreting the results correctly, may be up to 30\% in some areas of radiology. For arthroscopy, there is no such study, but arthroscopic surgery is a much more complex procedure than image reading, as surgeons are usually under time and success pressure, while working with patients and the stimulus is constantly and dynamically changing.
Therefore, ways to teach surgeons to use these new technologies optimally, are as important as the developments of such. This is where human-computer interaction comes into play. 
Methods of human-computer interaction find their way into the world of medicine. Indeed, there are multiple goals to pursue. Besides, i.e. touchless interaction techniques \cite{mewes2017touchless}, the recognition of strategies of surgeons during an operation \cite{sodergren2010hidden} are investigated. 
The recognition of skill ~\cite{speidel2006tracking,ahmidi2010surgical,wu2021cross,yin2020study} or states of confusion \cite{stillman2000bedside,zhou2018confusion} of surgeons play a central role in interaction design, as they can help to draw a picture of a surgeons' skills and to find weak-spots that need to be focused on in training. This is done to maximize the output of surgeons and to improve their training. 
Along with confusion, often frustration or disengagement are involved, if the confusion lasts for too long \cite{d2014confusion}. Pachman et al. \cite{pachman2016eye} summarized different approaches of the last few years and show that multiple ways of detection have been tried, e.g. facial expressions \cite{zeng2008survey,mcdaniel2007facial} or learners' postures \cite{d2009automatic}. D'Mello et al. \cite{d2009automatic} postulated that models based on a single source had high error rates. Thus, later research focused on multiple sources to detect confusion but could not be fully automated, as external judges needed to be involved \cite{d2010multimodal}. Lallé et al. ~\cite{lalle2016predicting} studied various combinations to predict occurrences of confusion. They reported a 61\% prediction rate with 193 features. Thus, their system is hardly usable online since the computation of these 193 features takes too much time. Similarly, the model of Shi et al.~\cite{shi2013confusion} is hardly usable in an online setting, too, as their model is too complex and thus needs too much computation power and time. Further, they use images on a display, which might cut off environmental influences, thus, preventing the application of natural gaze behavior. Conversely, we use a soft-cadaver in a real surgical setting, which allows the application of natural gaze behavior. We further use a simple but fast and robust model for classification, which allows the usage in an online fashion. 

Most often, surgeons need both of their hands for the operation. So new information and interaction techniques need to focus on other modalities than the surgeon's hands. One way to address this is the use of eye-tracking technology. This technique can either be directly used as an interaction method \cite{mewes2017touchless} or as an information provider about the skill or current state of the surgeons themselves. 
And as these devices are getting more ubiquitous, faster, and more accurate, there are ever new possibilities to study the gaze behavior of the subject. Eye tracking can serve as a perceptual-cognitive diagnosis system. The interest in using eye tracking as a research method in medicine is growing rapidly (for an overview see Lévêque et al \cite{leveque2018state}).

There are even studies that focus on assessment of the impact of training with eye tracking, too \cite{wilson2011gaze,vine2012cheating,krupinski2013characterizing}. Wilson et al. \cite{wilson2011gaze}, i.e found significant differences in completion time when showing young surgeons a video with the gaze signal of an expert during laparoscopy, compared to only showing the plain video of the surgery or allowing a free viewing phase. There are plenty of such studies, showing that 
the findings of gaze behavior studies can even be used to optimize and/or shorten the training surgeons need to go through. 
While eye-tracking devices are getting faster and ubiquitous, they produce more data, too. On the one hand, more data means more usable information, but on the other hand, there is a rise in complexity, too. With more data, there can be more inter-dependencies that are hard to understand and handle, especially with traditional techniques like AOI intersection counts \cite{mackert2013understanding,almansa2011association,kok2015case,kelly2016development,manning2006radiologists}.
To allow the analysis of such big data to be much more complex, there is another very important advancement in computer science that has a heavy impact on medicine. Artificial intelligence is applied in a variety of applications in medicine \cite{szolovits2019artificial,holzinger2019causability,hamet2017artificial,ramesh2004artificial}. The ever-new potentials of machine learning and especially deep learning enable even more complex tasks to be solved and more data to be analyzed.

In this work, we are focusing on the analysis of data from 15 participants during arthroscopic surgery with so-called soft-cadavers. During arthroscopy, the surgeon is mainly focusing on the output of the arthroscope, which shows a plane 2D view of the arthroscopic camera inside the portal hole of the patient. Surgeons need to rely on these images, while they navigate through tissue and bones. A young surgeon with low experience may get confused during navigation since the structures look pretty similar for untrained surgeons. Expert surgeons can rely on their experience and know which visual clues they can use for navigation. In order to optimize the training of young surgeons, we introduce a real-time ready confusion detection model, that recognizes states of confusion of surgeons during arthroscopic surgeries. With the combination of eye tracking, head tracking, and machine learning methods, we present a highly accurate and fast classification model. Detections of such a model can be used to find weak spots of surgeons in real-time and signal assistive actions to be made.


\subsection{Methods}

\subsubsection*{Data collection}

We collected data of 15 surgeons who are all either members of the Orthopedic Department in the Faculty of Medicine from Mahidol University, Thailand, or in the Orthopedics Surgery Residency Program. All subjects were wearing a TobiiGlasses 2 eye tracker (running at 100 Hz) during arthroscopic surgery of the shoulder on a soft cadaver. The cadaver was placed in front of the surgeon and four feet further away we placed a 4k-screen which shows the output of the scope. During the navigation from the portal hole to the operating side, surgeons were telling verbally where they are and where they go to. They also told when they are confused. This means they can either not tell their current position inside the joint or how to continue for sure. In relation to the beginning of the operation, we measured these points of time, where the surgeon told to be unsure/confused.

\subsubsection*{Feature space}

At first, we synchronized the eye-tracking data with the timing data, by adjusting their timestamps to start at the same time relative to the start of the surgery. This allows us to find the points of time of confusion inside the eye-tracking data. In the next step, we cut out a window around every confusion point (+/- one second before and after the event). These pieces of data are considered as "confusion event" samples and the remaining data with no confusion event as "no event" samples.

Each sample contains the following features:

\begin{itemize}
	\item point of regard (x, y)
	\item pupil position (average of both eyes)
	\item pupil diameter (average of both eyes)
	\item gyroscope (x, y, z)
	\item accelerometer (x, y, z)
\end{itemize}

\subsubsection*{Classification}

To build a random forest model, we split the samples into training and test data sets. This is done in a participant-wise manner, which means, if a subject is picked to belong to the training set, all of their samples belong to the training set. We need to do this, as the model would otherwise learn person-specific, so-called idiosyncratic, features (for further information, see \cite{hosp2020eye}).  We followed two different approaches, for testing with unseen data.

The first approach follows a 2/3-strategy. We randomly pick 2/3 of the subjects for training and count the number of confusion event samples for each. Afterward, we collect the same amount of "no event" samples from the same subjects. This means for our training set we have the same amount of confusion event samples as no event samples. This firstly leads to a balanced training set (50\% confusion event samples and 50\% no event samples) and secondly, to a chance-level of 50\%, which allows easy interpretation of the results later.

In the second approach, we want to see whether cross-validation during the training would optimize the results. Thus, we split the training set data by 5-fold cross-validation, which means in every run 1/5 of the data (of the training set) is picked to validate/optimize the model, while 4/5 of the data are used for training the model. 
After each run, we use the samples of the remaining 1/3 subjects (n=5) to test the classification performance with unseen data. As we want to use our model in an online fashion, we need to test the classification accuracy (with unseen data) and the classification speed as well. We show the online computability by creating a queue, which consists of n=2000 samples. In our test, we keep reading the gaze signal and add one sample to the queue in each step, while the oldest sample is kicked out of the queue. This means at every state the queue has a total of n=2000 samples. The average of each of the features of all samples inside the queue is now computed. These values are now representing the current content of the queue, which we call delta sample. This delta sample is now given to the trained random forest model and to classify it as a "confusion sample" or "no confusion sample". To infer the average performance time, we measure the computation time of 100 single runs and calculate the average performance time.

\begin{figure}[ht]
	\centering
	\includegraphics[width=0.7\textwidth]{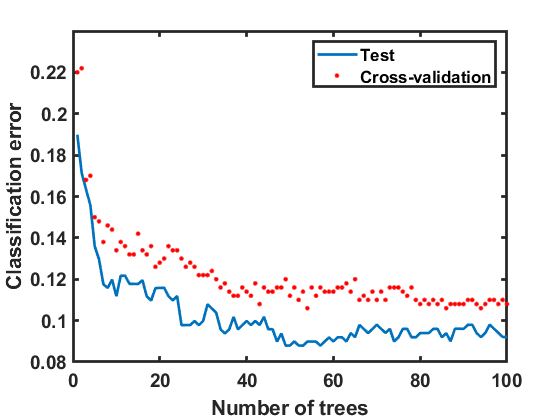}
	\caption{Validation with test data vs. validation with cross-validation as function of number of learners.}~\label{fig:testvscv}
\end{figure}

\subsection{Results}

Out of 1,266,758 samples, we have 7103 samples with a confusion event and 1,259,655 samples with no event. Out of these samples, we collect 7103 confusion samples and 7103 no confusion samples. In every run, we randomly pick 1,000 samples of both to predict their class. The other samples are used for training. We tested our approach - by randomly assigning training and testing data like the aforementioned, 100 times. The average accuracy of the random forest model is 94.2\%. According to the accuracy, the average misclassification cost/loss is 0.0595. Figure~\ref{fig:testvscv} shows the development of the loss over all runs as a function of the number of trained trees. The differences are small but noticeable. The approach with the test data set is performing a little bit better than the cross-validation approach. Test set approach reaches the best performance of the cross-validation approach (\~ 0.11) already with about 25-30 trees. The optimal loss value for the test approach is reached at ~50 trees with a misclassification cost of ~0.085.

\begin{figure}[ht]
	\centering
	\includegraphics[width=0.5\textwidth]{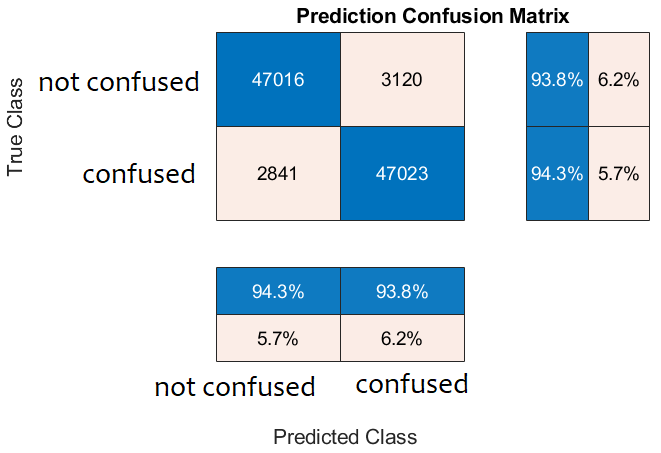}
	\caption{Confusion matrix showing number of correctly and falsely predicted samples.}	\label{figP:confMatrix}
\end{figure}

Figure~\ref{figP:confMatrix} shows the confusion matrix which contains the predictions of all 100 runs. In total, we have ~50,000 samples for each class. Of class 0 (no event), 47,016 samples out of 50,136 samples were predicted correctly and 3,120 as confusion event samples. Similarly, for class 1 (confusion event), the model predicted 47,023 samples correctly as confusion event and 2,841 samples wrongly as no event. This result is supported, by the average accuracy over all 100 runs of 94.2\%.

To measure the performance speed of the model, we measured the computing time of each of the 100 runs. On average the prediction takes 0.039 seconds. This corresponds to a frame rate of ~25 fps.

\subsection{Discussion }

\begin{figure}[ht]
	\centering
	\includegraphics[width=0.9\columnwidth]{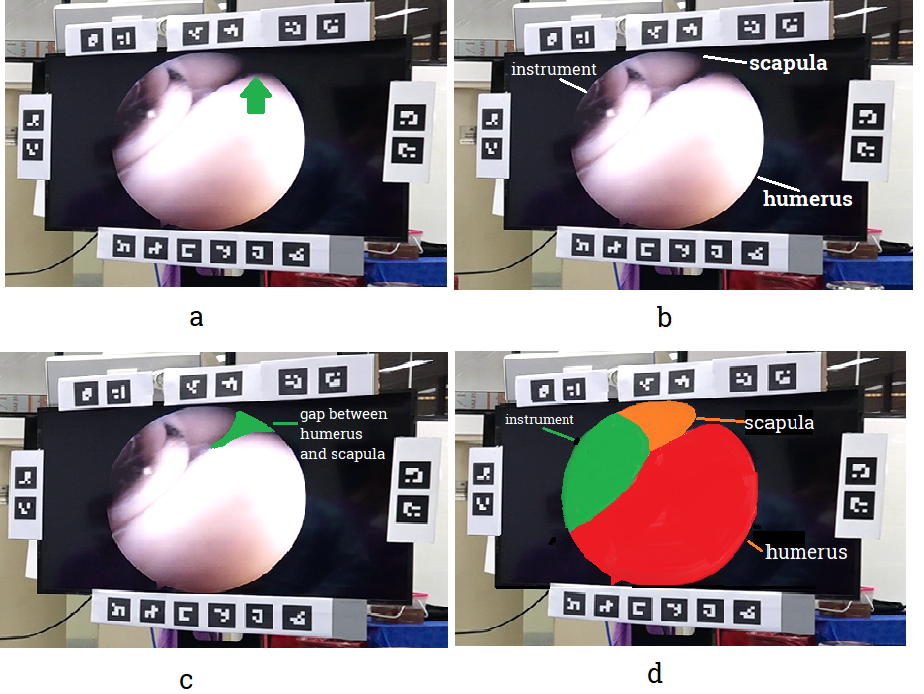}
	\caption{Different kinds of support for a confused surgeon.}	~\label{fig:support}
\end{figure}

In this work, we presented a random forest model that is able to classify states of confusion of surgeons during arthroscopic surgery of the shoulder with an accuracy of over 94.2\%, by taking only 9 features of the eye and head movement into account. In our calculations, the model was able to provide a prediction of the content of a queue containing n=2000 samples (2 seconds of samples) in 0.039 seconds. This corresponds to the temporal resolution of common head-mounted eye trackers which run at a frame rate between 25-30 fps. The speed may need to be optimized, to allow the application to higher-paced field cameras. But in the scenario of surgery, the speed is not a crucial part, rather, a high detection rate is important.		
With the detection of confusion states, one can help surgeons to proceed, either pointing out visual clues, which may be used by expert surgeons to navigate or drawing arrows on the output of the arthroscope which tells the surgeon where to navigate next. Another possible usage of the knowledge of states of confusion can be to augment the whole output by describing the scene by segmenting and labeling each bone or tissue. Or simply name the shown parts in the output. There are multiple ways of supporting the confused surgeon. Depending on the state of expertise, the level of support may be chosen, to allow different skilled surgeons, to train their different weak spots.	
The different kinds of support can be seen in Figure~\ref{fig:support}. a) shows a simple arrow, which tells the surgeon where to go next with the arthroscope. b) shows more support by naming the single party of the output, so the surgeon knows which parts are involved and may remember how to proceed. Figure~\ref{fig:support}, c shows a similar output like a), but there are only visual clues highlighted, and d) this help would provide the most support, by segmenting and coloring the single parts in different colors and naming them, accordingly.

\printbibliography[heading=bibintoc,keyword={publication},title={Publications}]

\printbibliography[heading=bibintoc,title=References]
\end{document}